\documentclass[preprint,aps,prb,showpacs,amsmath,amssymb]{revtex4}
\usepackage{graphics}
\usepackage{epsf,graphicx,pstricks,epsfig,psfrag}
\begin{document}
\title{Microscopic approach to high-temperature superconductors: \\ Pseudogap phase
}
\author{S.~Sykora and K.W.~Becker}

\affiliation{Institut f\"ur Theoretische Physik, Technische Universit\"at Dresden, 01062 Dresden, Germany}

\date{\today}



\begin{abstract}
Despite the intense theoretical and experimental effort, an understanding of the superconducting 
pairing mechanism of the high-temperature superconductors leading to an unprecedented high transition
temperature $T_c$ is still lacking.
An additional puzzle is the unknown connection between the superconducting gap 
and the  so-called pseudogap which is a central property of the 
most unusual normal state.
Angle-resolved photoemission spectroscopy (ARPES) measurements
have revealed a gap-like behavior on parts of the Fermi surface,
leaving a non-gapped segment known as Fermi arc around 
the diagonal of the Brillouin zone. Two main interpretations of the origin of the pseudogap 
have been proposed: either the pseudogap is a precursor to superconductivity, or it arises from another order competing with superconductivity. Starting from the $t$-$J$ model, in this paper
we present a microscopic approach to investigate physical properties of the
pseudogap phase in the framework of a novel renormalization scheme
called PRM. This approach is based on a stepwise elimination of high-energy transitions using unitary
transformations. We arrive at a renormalized 'free'
Hamiltonian for correlated electrons. 
The ARPES spectral function along the Fermi surface turns out to be in good agreement with experiment: 
We find well-defined excitation peaks around $\omega=0$ near the nodal direction, 
which become strongly suppressed around the antinodal point. The
origin of the pseudogap can be traced back to a suppression of spectral weight
from incoherent excitations in a small $\omega$-range around the Fermi energy. 
Therefore, both mentioned interpretations of the origin of the pseudogap can
not be held. Instead, the pseudogap is an inherent property of the unusual
normal state caused by incoherent excitations. In a subsequent paper, also the
supercunducting phase at moderate hole doping will be discussed within the PRM approach\cite{BS09}.
\end{abstract}

\maketitle

\section{Introduction}

Since the discovery of superconductivity in the cuprates \cite{BM86}, enormous
theoretical and experimental effort has been made to investigate 
the superconducting pairing mechanism which leads to an unprecedented high transition
temperature $T_c$\cite{C99}-\cite{V97}. An additional puzzle is the unknown connection 
between the superconducting gap of the superconducting phase 
and the so-called pseudogap which is a central 
property of the most unusual normal state of the cuprates. In particular, 
the pseudogap  has been subject to intense debates. 
Studies  using angle resolved photoemssion spectroscopy (ARPES) have 
revealed several key features of the pseudogap in the cuprates by 
elucidating the detailed momentum and temperature dependence\cite{N98}-\cite{CSG08}. 
It was found that the pseudogap opens on a part of the Fermi 
surface (FS) around the anti-nodal point, leaving a nongapped FS segment known as a Fermi 
arc around the nodal direction. The pseudogap also  smoothly evolves 
with decreasing temperature into the SC gap 
and was, therefore, interpreted in favor of a ``precursor
pairing'' scenario \cite{D96},\cite{L96},\cite{K08}. On the other hand, there are several 
experimental and theoretical reports which suggest a different origin for the pseudogap, 
such as caused by another order which competes with superconductivity \cite{S05}.
Superconductivity is usually understood
as an instability from a non-superconducting state. Therefore, often in  theoretical investigations, 
the starting point was either the Fermi-liquid  or the anti-ferromagnetic 
phase at large or low doping. In this paper, we take a different 
approach and only consider hole fillings, in which either a superconducting or a 
pseudogap phase is present. 

A generally accepted  model for the cuprates is the $t$-$J$ model which describes the electronic 
degrees of freedom  in the copper-oxide planes for low energies. Alternatively, one could also start
from a one-band Hubbard Hamiltonian as a minimal model. 
However, for low energy excitations, the latter model reduces to the $t$-$J$ model 
so that both models are equivalent. 
As our theoretical approach, we use a recently developed 
projector-based renormalization method which is called PRM \cite{PRM}.
The approach is based on a stepwise elimination of high-energy transitions using unitary
transformations. We thus arrive at a renormalized 'free'
Hamiltonian for correlated electrons which can describe the pseudogap phase. 
The obtained ARPES spectral function
along the Fermi surface is in good agreement with experiment: 
We find well-defined excitation peaks around $\omega=0$ near the nodal direction which are strongly suppressed around the antinodal point. The
origin of the pseudogap can be traced back to a suppression of spectral weight
of the incoherent excitations in a small $\omega$-range around the Fermi energy. 
Therefore, the usual interpretations of the pseudogap origin can not be held. Instead, the pseudogap is an inherent property of the unusual normal state caused by incoherent excitations. 

First, after a short introduction of the model in Sec.~II, it seems to be helpful, to start from a 
short outline of the basic ideas of our theoretical approach (PRM) in Sec.~III. 
A review of this approach has been given elsewhere \cite{PRM}. 
Then, in Sec.~IV,  the PRM will be applied to the $t$-$J$ model in order to
investigate the pseudogap phase at moderate hole doping. 
The final results will be discussed in Sec.~V. In a subsequent paper, the
supercunducting phase will also be discussed.

\section{Model}

A generally accepted  model for the cuprates is the $t$-$J$ model.  
In particular, in the antiferromagnetic phase at small doping, 
it has turned out that it can be used to describe the electronic degrees of freedom at low energies. 
We adopt the same model also for somewhat larger hole concentrations,
outside the antiferromagnetic phase,
where the superconducting and the pseudogap phases appear 
 \begin{eqnarray}
\label{1}
{\cal H} &=& -\sum_{ij, \sigma} t_{ij}\, \hat c_{i\sigma}^\dagger \hat c_{j\sigma} 
- \mu \sum_{i\sigma} \hat c_{i\sigma}^\dagger \hat c_{i\sigma} 
+ \sum_{ij} J_{ij} {\bf S}_i {\bf S}_j    
=: {\cal H}_t + {\cal H}_J .
\end{eqnarray}
The model consists of a hopping term ${\cal H}_t$
and an antiferromagnetic exchange ${\cal H}_J$. 
Here, $t_{ij}$ stands for the hopping matrix elements between nearest ($t$)
and next-nearest ($t'$) neighbors. $J_{ij}$ is the exchange coupling 
and $\mu$ is the chemical potential. The quantities 
\begin{eqnarray}
\label{2}
 \hat {c}_{i\sigma}^\dagger &=&  {c}_{i\sigma}^\dagger(1- n_{i,-\sigma}), \qquad 
\hat {c}_{i\sigma} =  {c}_{i\sigma}(1- n_{i,-\sigma}) 
\end{eqnarray}
are Hubbard creation and annihilation operators. They enter the model, since 
doubly occupancies of local sites are strictly forbidden due to the presence of strong 
electronic correlations. Note that 
the Hubbard operators restrict the unitary space to states with  only 
either empty or singly occupied local sites. They obey nontrivial 
anti-commutation relations 
\begin{eqnarray}
\label{3}
[\hat c_{i\sigma}^\dagger, \hat c_{j\sigma'}]_+ &=& \delta_{ij} \big( \delta_{\sigma \sigma'}
{\cal D}_\sigma(i) + \delta_{\sigma, -\sigma'} S_{i}^{\sigma} \big),
\end{eqnarray}
where the operator
\begin{eqnarray}
\label{4}
{\cal D}_\sigma (i) &=& 1 - n_{i,-\sigma}
\end{eqnarray}
can be interpreted  as a projector which projects on the local subspace at site $i$ consisting of 
either an empty or a singly occupied state with spin $\sigma$. Finally, 
$n_{i\sigma} = c_{i\sigma}^\dagger c_{i\sigma}$ is the local occupation number 
operator for spin $\sigma$, and  $S_i^{\sigma}$ is the $\sigma= \pm 1$ component of  
the local spin operator 
\begin{eqnarray}
\label{5}
 {\bf S}_i &=& \frac{1}{2} \sum_{\alpha \beta} \vec \sigma_{\alpha\beta}
\hat {c}_{i\alpha}^\dagger\hat {c}_{i\beta},
\end{eqnarray}
where $\vec \sigma_{\alpha\beta}= \sum_\nu \sigma_{\alpha\beta}^\nu\, {\bf e}_\nu$ is the 
vector formed by the Pauli spin matrices.  In Fourier notation,  the 
$t$-$J$ model \eqref{1} reads
\begin{eqnarray}
\label{6}
 {\cal H} &=&  \sum_{{\bf k}, \sigma} (\varepsilon_{\bf k} - \mu)\, \hat c_{{\bf k}\sigma}^\dagger 
\hat c_{{\bf k}\sigma} + \sum_{{\bf q}} J_{{\bf q}} {\bf S}_{\bf q} {\bf S}_{-{\bf q}} = 
{\cal H}_t +{\cal H}_J,
   \\
 \varepsilon_{\bf k} &=& - \sum_{i (\neq j)} t_{ij} e^{i{\bf k}({\bf R}_i - {\bf R}_j)},
\qquad  J_{\bf q} = \sum_{i (\neq j)} J_{ij} e^{i{\bf q}({\bf R}_i - {\bf R}_j)} .
\nonumber
\end{eqnarray}
Note that for convenience, we shall somewhat change the notation. From now on, 
all energies will be measured from the chemical potential, 
i.e., $\varepsilon_{\bf k} -\mu$  will be denoted by $\varepsilon_{\bf k}$.

\section{Projector-based renormalization method (PRM)}
\label{PRM1}

Let us start with a short introduction to the projector-based renormalization method (PRM) \cite{BHS_2002,PRM}  which we shall use as our 
theoretical tool. The general idea is as follows:
The method starts from a  decomposition of a given many-particle Hamiltonian 
\begin{eqnarray}
\label{7}
\mathcal{H} = \mathcal{H}_{0} + \mathcal{H}_{1}
\end{eqnarray}
into an unperturbed part ${\cal H}_0$ and a perturbation ${\cal H}_1$.  
In $\mathcal{H}_{1}$, no parts should be contained which
commute with $\mathcal{H}_{0}$. Therefore, 
$\mathcal{H}_{1}$ accounts for all transitions with \textit{non-zero} energies
between the eigenstates of $\mathcal{H}_{0}$. The aim of the PRM is to 
construct an effective Hamiltonian which has the same 
eigenspectrum as $\mathcal{H}$, and which can be solved. The first step is to 
construct a new renormalized Hamiltonian ${\cal H}_\lambda$ which depends on a 
given cutoff $\lambda$,
\begin{eqnarray}
  \label{8}
  \mathcal{H}_{\lambda} &=& 
 \mathcal{H}_{0,\lambda} +   \mathcal{H}_{1,\lambda},
\end{eqnarray}
with renormalized parts ${\cal H}_{0,\lambda}$ and ${\cal H}_{1,\lambda}$. Thereby,
${\cal H}_\lambda$ should have the following properties: 
(i) The eigenvalue problem of $\mathcal{H}_{0,\lambda}$ can be solved
  \begin{eqnarray*}
    \mathcal{H}_{0,\lambda} | n^{\lambda} \rangle &=& 
    E_{n}^\lambda | n^\lambda \rangle,
  \end{eqnarray*}
where $E_{n}^{\lambda}$ and $|n^{\lambda}\rangle$ are the 
renormalized eigenenergies and eigenvectors. 
(ii)
From $\mathcal{H}_{1,\lambda}$, all transition operators are eliminated
which have transition energies (with respect to  $\mathcal{H}_{0,\lambda}$)  
larger than the cutoff energy $\lambda$.  
As shown in Refs.~\cite{BHS_2002,PRM}, the renormalization step from ${\cal H}$ to 
${\cal H}_\lambda$ can be done by use of  a
unitary transformation. Therefore, the eigenspectrum of ${\cal H}_\lambda$ is the same 
as that of ${\cal H}$. 

The realization of the renormalization starts from the construction 
of  ${\cal H}_\lambda$. Here, the knowledge of the eigenvalue problem of 
${\cal H}_{0,\lambda}$ is crucial. It can be used to define generalized 
projection operators, $\sf {P}_\lambda$ and $\sf {Q}_\lambda$,
\begin{eqnarray}
  \label{9}
  \sf {P}_{\lambda} {\mathcal{A}} &=&
  \sum_{m,n} | n^{\lambda} \rangle \langle m^{\lambda} |
  \langle n^{\lambda} | \mathcal{A} | m^{\lambda}\rangle \,
 \Theta(\lambda -|E_n^\lambda -E_m^\lambda |), \nonumber\\
  \sf{Q}_{\lambda} \mathcal{A}&=& (\mathbf{1} - \sf{P}_{\lambda}) \mathcal{A},
\end{eqnarray}
which act on usual operators ${\mathcal{A}}$ of the Hilbert space.
Note that in Eq.~\eqref{9} the vectors $|n^{\lambda}\rangle$ and $|m^{\lambda}\rangle$ are 
necessarily neither low- nor high energy eigenstates of $\mathcal{H}_{0,\lambda}$. 
$\sf{P}_{\lambda}$ projects on the part of ${\mathcal{A}}$ which consists of transition
operators $| n^{\lambda}\rangle \langle m^{\lambda} |$ with excitation energies 
$| E_n^\lambda - E_m^\lambda| $ smaller than $\lambda$, whereas 
$\sf{Q}_{\lambda}$ projects on the high-energy transition operators of
${\mathcal{A}}$.

In terms of $\sf{P}_\lambda$ and 
$\sf {Q}_\lambda$, the property of ${\cal H}_\lambda$, not to allow  
transitions between eigenstates of ${\cal H}_{0,\lambda}$ with energy differences
larger than $\lambda$, reads
\begin{eqnarray}
  \label{10a}
  \sf{Q}_{\lambda} \mathcal{H}_{\lambda} = 0
 \quad \mbox{or} \quad
\mathcal{H}_{\lambda} &=& \sf{P}_{\lambda} 
  \mathcal{H}_{\lambda} \, .
\end{eqnarray}
The effective Hamiltonian ${\cal H}_\lambda$ is obtained from the 
original Hamiltonian ${\cal H}$ by use of a unitary transformation,
\begin{eqnarray}
  \label{11}
  \mathcal{H}_{\lambda} &=& 
  e^{X_{\lambda}}\; \mathcal{H}\;  e^{-X_{\lambda}},
\end{eqnarray}
where $X_\lambda$ is the generator of the unitary transformation, and the 
condition \eqref{10a} has to be fulfilled.
The renormalization procedure starts from the cutoff energy $\lambda=\Lambda$ 
of the original model ${\cal H}$ and proceeds in steps of width $\Delta \lambda$ 
to lower values of $\lambda$. Every renormalization step is performed by 
means of a new unitary transformation, 
\begin{eqnarray}
  \label{12}
  \mathcal{H}_{(\lambda-\Delta\lambda)} &=& 
  e^{X_{\lambda,\Delta\lambda}} \, \mathcal{H}_{\lambda} \,
  e^{-X_{\lambda,\Delta\lambda}}.
\end{eqnarray}
Here, the generator $X_{\lambda,\Delta \lambda}$ 
of the transformation from cutoff $\lambda$ to the 
reduced cutoff $(\lambda - \Delta \lambda)$ has to be chosen appropriately (see below). 
In this way, difference equations are derived which 
connect the parameters of ${\cal H}_\lambda$ with those of 
${\cal H}_{(\lambda - \Delta \lambda)}$. They will be called renormalization 
equations.
The limit $\lambda \rightarrow 0$ provides the desired effective Hamiltonian 
$ 
  \tilde{\cal H}= {\cal H}_{\lambda \rightarrow 0} 
  = {\cal H}_{0, \lambda \rightarrow 0}
$.
The elimination of all transitions in the original perturbation 
$\mathcal{H}_{1}$ leads to renormalized parameters in 
${\cal H}_{0,\lambda \rightarrow 0}$. Note that $\tilde{\cal H}$ 
is diagonal or at least quasi-diagonal and allows to evaluate all relevant
physical quantities. 
The final expression for $\tilde{\cal H}$ depends on the parameter values of the 
original Hamiltonian $\mathcal{H}$. Note that $\tilde{\cal H}$ and 
${\cal H} $ have, in principle,  the same eigenspectrum because both Hamiltonians are 
connected by a unitary transformation.  

What is left, is to find an appropriate expression for the generator 
$X_{\lambda, \Delta \lambda}$ of the unitary transformation which connects ${\cal H}_\lambda$ 
with ${\cal H}_{(\lambda - \Delta \lambda)}$. According to Eq.~\eqref{10a},
$X_{\lambda, \Delta \lambda}$ is fixed by the condition 
${\sf Q}_{\lambda- \Delta \lambda}{\cal H}_{\lambda -\Delta \lambda}=0$. 
As is shown in Refs.~\cite{BHS_2002,PRM},
one can find a perturbation expansion for $X_{\lambda, \Delta \lambda}$ in terms of ${\cal H}_1$. 
The lowest non-vanishing order reads 
\begin{eqnarray}
  \label{13} 
 X_{\lambda,\Delta\lambda}^{(1)} &=& 
  \frac{1}{\sf {L}_{0,\lambda}}
  \left[
    \sf {Q}_{(\lambda-\Delta\lambda)} \mathcal{H}_{1,\lambda}
  \right] + \cdots .
\end{eqnarray}
Here, ${\sf L}_{0,\lambda}$ is the Liouville operator, defined by 
the commutator ${\sf L}_{0,\lambda}{\cal A}= [{\cal H}_{0,\lambda}, {\cal A}]$, for any operator 
quantity ${\cal A}$. Note that Eq.~\eqref{13} can further be evaluated, in case the
decomposition of ${\sf Q}_{(\lambda - \Delta \lambda)}{\cal H}_{1,\lambda}$ into eigenmodes 
of ${\sf L}_{0,\lambda}$ is known. Formally written, we decompose
 \begin{eqnarray}
\label{14}
\sf {Q}_{(\lambda-\Delta\lambda)} {\cal H}_{1,\lambda} = \sum_\nu {\bf F}^\nu_{\lambda,\Delta \lambda} ,
\quad \mbox{where} \quad 
{\sf L}_{0,\lambda}\, {\bf F}^\nu _{\lambda,\Delta \lambda} &=& \omega^\nu _{\lambda,\Delta \lambda} \, 
{\bf F}^\nu _{\lambda,\Delta \lambda} \, ,
 \end{eqnarray}
so that $ X^{(1)}_{\lambda, \Delta \lambda} $ is given by
 \begin{eqnarray}
\label{14b}
  X^{(1)}_{\lambda, \Delta \lambda} &=& \sum_\nu \frac{1}{\omega^\nu _{\lambda,\Delta \lambda}}
{\bf F}^\nu _{\lambda,\Delta \lambda}. \\
&& \nonumber
 \end{eqnarray}


\section{Application to the $t$-$J$ model}
\label{tJmod}
\subsection{Renormalization ansatz}
\label{tJmod_RA}
Our aim is to apply the PRM to the $t$-$J$ model which is a generally 
accepted model for the low-energy properties of the cuprate superconductors. 
We consider a regime with moderate hole-dopings. The hole concentrations 
should be large enough for the system to be 
outside the antiferromagnetic phase but small enough to be 
 in the metallic phase. 
Our first aim is to find the decomposition of the Hamiltonian into an 'unperturbed' 
part ${\cal H}_0$ and into a 'perturbation' ${\cal H}_1$. We assume that the 
hopping element $t$ between nearest neighbors is large compared 
to the exchange coupling $J$. Therefore, ${\cal H}_t$ is the 
dominant part of the Hamiltonian in the metallic phase and should be included in  ${\cal H}_0$. 
However, also ${\cal H}_J$ has a part,  which commutes with the hopping term,   
and which will be called ${\cal H}_J^{(0)}$.   Note that this part of 
${\cal H}_J$ will not lead to transitions between the eigenstates of ${\cal H}_t$. 
Therefore,  ${\cal H}_t$ and  ${\cal H}_J^{(0)}$ together form
the unperturbed Hamiltonian ${\cal H}_0$. 
The remaining part of ${\cal H}_J$ does not 
commute with ${\cal H}_t$ and forms the perturbation ${\cal H}_1$. Thus, we can
write 
\begin{eqnarray*}
{\cal H}_0 &=& {\cal H}_t + {\cal H}_J^{(0)}, \qquad {\cal H}_1 = {\cal H}_J -{\cal H}_J^{(0)}.
\end{eqnarray*}
In the framework of the PRM, the perturbation ${\cal H}_1$ will be integrated out by use of 
a unitary transformation. In lowest order perturbation theory,
the generator of the unitary transformation $X_{\lambda,\Delta \lambda}$ is given by Eq.~\eqref{14b} 
and relies on the decomposition of ${\cal H}_J$ into the eigenmodes of ${\sf L}_0$. 
However, it will be impossible to find the exact decomposition of ${\cal H}_J$, due to the 
presence of Hubbard operators in ${\cal H}_t$.
Therefore, we have to apply approximations.
For this purpose, we start by decomposing the electronic spin operator 
\begin{eqnarray}
\label{15}
 {\bf S}_{\bf q} &=& \frac{1}{\sqrt N} \sum_{\alpha \beta} \frac{{\vec \sigma}_{\alpha \beta}}{2}
\sum_i e^{i {\bf q} {\bf R}_i} \,\hat c_{i  \alpha}^\dagger 
\hat c_{i  \beta} 
\end{eqnarray}
into eigenmodes of ${\sf L}_{t}$ instead of into eigenmodes of ${\sf L}_0$.
Here, ${\sf L}_{t}$ is the Liouville operator corresponding  
to the hopping part ${\cal H}_t$  of ${\cal H}_0$. 
The exchange ${\cal H}_J$ is given by a sum over products of 
spin operators ${\bf S}_{\bf q}\cdot {\bf S}_{-{\bf q}} $. Therefore, 
the decomposition of ${\bf S}_{\bf q}$ into eigenmodes of ${\sf L}_t$ 
can be used to find an equivalent decomposition of ${\cal H}_J$.

The easiest way to decompose $\bf S_{q}$ is to derive an equation of motion
for the time-dependent operator ${\bf S}_{\bf q}(t)$, 
where the time dependence is governed by ${\cal H}_t$,
\begin{eqnarray}
\label{16}
  {\bf S}_{\bf q}(t) = e^{i{\cal H}_t t} \, {\bf S}_{\bf q}\, 
e^{-i{\cal H}_t t} = e^{i{\sf L}_t t}\,
{\bf S}_{\bf q} .
 \end{eqnarray}
Due to Eq.~\eqref{3}, the first time derivative reads 
\begin{eqnarray}
\label{17}
\frac{d}{dt}{\bf S}_{\bf q} &=& 
-\frac{i}{\sqrt N}\sum_{\alpha \beta} \frac{\vec \sigma_{\alpha \beta}}{2}
\sum_{i \neq l} t_{il} \, e^{i{\bf q}{\bf R}_i}
(\hat c_{l\alpha}^\dagger \hat c_{i\beta} - \hat c_{i\alpha}^\dagger \hat c_{l\beta}) \\
&=&  
\frac{i}{\sqrt N} \sum_{\alpha \beta} \frac{\vec \sigma_{\alpha \beta}}{2}
\sum_{i \neq l} t_{il} 
 e^{i{\bf q}{\bf R}_i} (1 -  e^{i{\bf q}({\bf R}_l -{\bf R}_i)})\,
\hat c_{i\alpha}^\dagger \hat c_{l\beta}. \nonumber
 \end{eqnarray}
It can be interpreted as the hopping of a hole from some site $l$ to a neighboring 
site $i$ and vice versa. 
The second derivative is characterized by a twofold hopping,
\begin{eqnarray}
\label{18}
 \frac{d^2}{dt^2}{\bf S}_{\bf q} &=& - \frac{1}{\sqrt N}\sum_{i \neq l}t^2_{il}\,
(e^{i{\bf q}{\bf R}_l} -e^{i{\bf q}{\bf R}_i}) \,
( {\bf S}_l {\cal P}_0(i) - {\bf S}_i {\cal P}_0(l) )  \\
&&
  -\frac{1}{2\sqrt N} \sum_{\alpha \beta}
\sum_{i \neq j} \sum_{j(\neq i \neq l)}\, t_{il}\, t_{lj}\, 
(e^{i{\bf qR}_i} - e^{i{\bf qR}_l})
\nonumber
\\
&& \times  \left\{ \vec \sigma_{\alpha \beta} \left(
 \hat c_{j\alpha}^\dagger \,  {\cal D}_\alpha(l) 
\,\hat c_{i\beta} + \hat c_{j,-\alpha}^\dagger
S_l^\alpha \hat c_{i\beta} \right)
 + \vec \sigma_{\alpha \beta}^* \left(
\hat c_{i\beta}^\dagger \, {\cal D}_\alpha(l) \, \hat c_{j\alpha} +
 c_{i\beta}^\dagger S_l^{-\alpha} \hat c_{m,-\alpha} \right)
\right\}. \nonumber
\end{eqnarray}
It has two different contributions. The first one describes the hopping of the hole 
from $i$ back to site $l$ from which it originally came and, equivalently, the
hopping from $l$ back to $i$. The second 
term in Eq.~\eqref{18} stands for a twofold hopping away from the starting site. 

Let us discuss the first contribution to Eq.~\eqref{18} in more detail. 
The operators 
\begin{eqnarray}
\label{19}
 {\cal P}_0(i) = (1- n_{i,\uparrow})(1- n_{i, \downarrow})
\end{eqnarray}
and ${\cal P}_0(l)$ can be interpreted as local projectors on the empty state at site $i$ 
and site $l$, respectively. They
assure that the original sites $i$ and $l$ were empty before the first hop. 
Their presence results from the fact that doubly occupancies of local sites 
are strictly forbidden which is a consequence of the strong correlations 
in the $t$-$J$ model. In a further approximation, let us replace 
${\cal P}_0(i)$ and ${\cal P}_0(l)$ by their expectation values,
\begin{eqnarray}
\label{20}
 {\cal P}_0(i) \Rightarrow  \langle (1- n_{i,\uparrow})
(1- n_{i, \downarrow}) \rangle_0 =: P_0,
\end{eqnarray}
which can be interpreted as the probability for a local site to be empty.
Without the second term in Eq.~\eqref{18}, we are led to the following equation of motion 
for ${\bf S}_{\bf q}(t)$:  
\begin{eqnarray}
\label{21}
\frac{d^2}{dt^2}{\bf S}_{\bf q} &=& - \hat \omega^2_{\bf q}\, {\bf S}_{\bf q}.
\end{eqnarray}
Obviously, the differential equation \eqref{21} describes an oscillatory motion of ${\bf S}_{\bf q}(t)$ with frequency $\omega_{\bf q}$, where 
\begin{eqnarray}
\label{22}
\hat \omega^2_{\bf q} &=&
  2P_0(t^2_{{\bf q}=0}- t^2_{{\bf q}}) =  \hat \omega^2_{-{\bf q}} \geq 0 ,\quad \qquad
t^2_{\bf q} = \sum_{l (\neq i)}t_{il}^2\, e^{i{\bf q}({\bf R}_l - {\bf R}_i)} .
\end{eqnarray}
Note that the averaged projector $P_0=1-n$ also agrees with the hole 
concentration $\delta$ away from half-filling, 
i.e.~$P_0= \delta =1-n$, where $n$ is the electron filling. 

Before carrying on with the physical implications of Eqs.~\eqref{21}, \eqref{22}, let us discuss the 
influence of the  hole (or electron) hopping in Eq.~\eqref{18} to second nearest neighbors and also
to more distant sites. As long as the dynamics of ${\bf S}_{\bf q}(t)$ is alone governed
by the hopping Hamiltonian ${\cal H}_t$, all these hopping processes are important 
and would have to be taken into account.  
For instance, for a state close to half-filling outside the antiferromagnetic regime, 
a hole and a neighboring electron can
freely interchange their positions for a system governed alone by ${\cal H}_t$.
The hole can easily move through the lattice.
However, the situation is different from the case, for which 
the dynamics is governed by ${\cal H}_0 = {\cal H}_t +{\cal H}_J^{(0)}$. 
Then,  we have to decompose the perturbation ${\cal H}_1$ into eigenstates 
of ${\sf L}_0$, where ${\sf L}_0$ is the Liouville operator 
corresponding to ${\cal H}_0$.  Thus, the dynamics of $\bf S_{q}$ is  not
governed alone by the hopping Hamiltonian ${\cal H}_t$ but also by the
yet unknown commuting part ${\cal H}_J^{(0)}$ of ${\cal H}_J$. 
However, in Appendix \ref{D}, it is shown that 
local antiferromagnetic spin fluctuations due to  ${\cal H}_J^{(0)}$  restrict the hole motion 
to neighboring sites. The hopping to more distant sites is strongly suppressed by spin fluctuations. Therefore, 
the former equation of motion \eqref{21} for ${\bf S}_{\bf q}(t)$ turns out to be
a good approximation for the case that the dynamics is determined by the full
unperturbed Hamiltonian ${\cal H}_0$ including the exchange part.

\begin{figure}
 \begin{center}
    \scalebox{0.61}{
      \includegraphics*{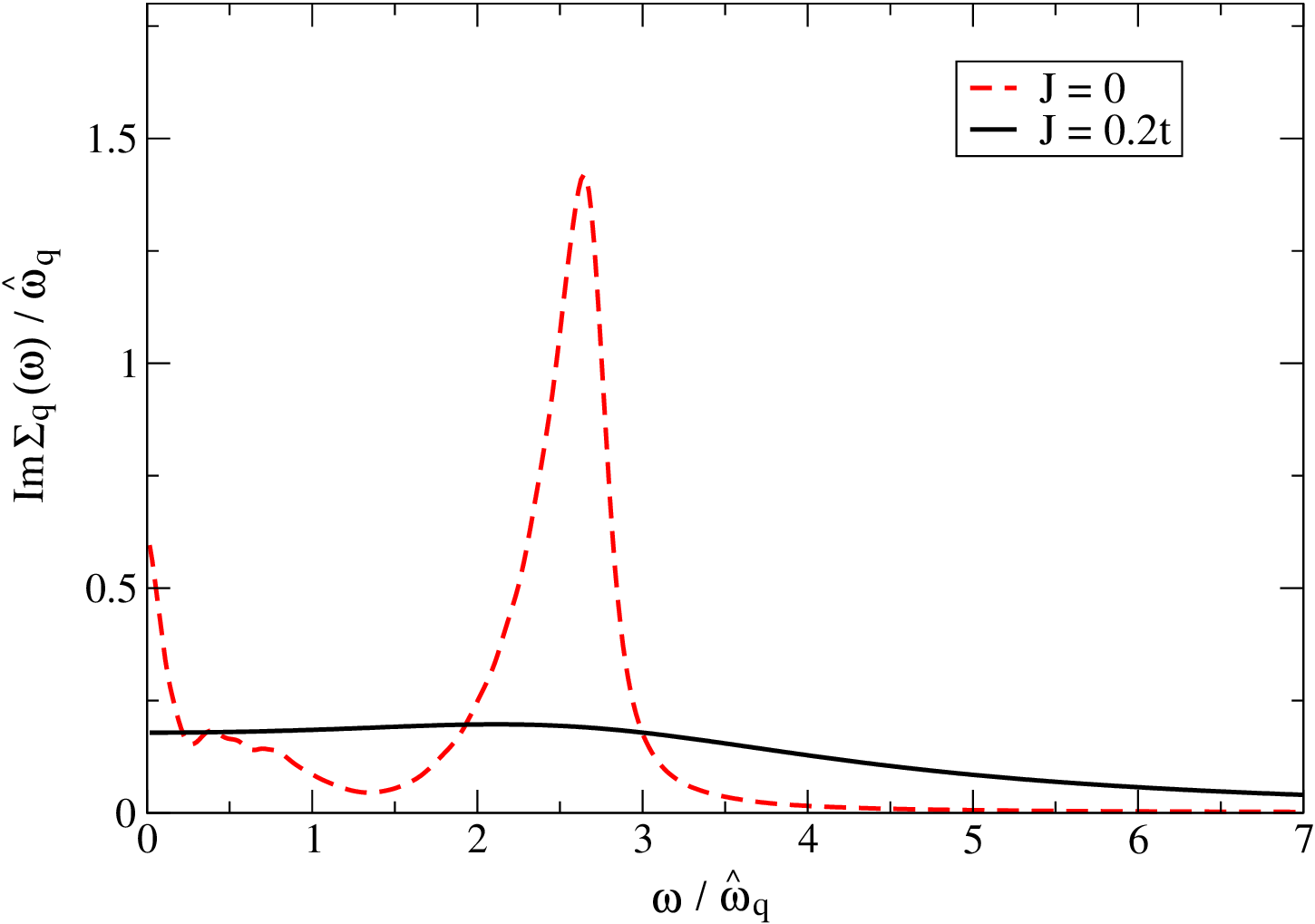}
    }
  \end{center}
  \caption{Imaginary part of the selfenergy $\Im \Sigma_{\bf q}(\omega)$ from
       Eq.~\eqref{22b} in the presence of spin fluctuations ($J =0.2t$, solid line) 
       and in the absence of spin fluctuations ($J=0$, dashed line). 
       The ${\bf q}$-vector is fixed to ${\bf q} = (\pi/20,\pi/20)$.}
\label{SE} 
\end{figure}

The arguments in Appendix \ref{D} are based on the evaluation of  
the dynamical spin susceptibility $\chi({\bf q}, \omega)$ as follows.  Using the Mori-Zwanzig projection formalism $\chi({\bf q}, \omega)$ can be written as 
\begin{eqnarray}
\label{22a}
 \chi({\bf q}, \omega)&=&  \frac{- \omega_{\bf q}^2}{\omega^2 - \omega_{\bf q}^2 - 
\omega \, \Sigma_{\bf q}(\omega)}\, \chi_{\bf q}.
\end{eqnarray}
Here, $\omega_{\bf q}^2 \approx  \hat{\omega}_{\bf q}^2$ is approximately the frequency, given in 
Eq.~\eqref{22}, and $\Sigma_{\bf q}(\omega)$ is the selfenergy. The exact expression of $\Sigma_{\bf q}(\omega)$ in terms of the Mori scalar product reads
\begin{eqnarray}
\label{22b}
 \Sigma_{\bf q}(\omega) &=& \frac{1}{ (\dot {\bf S}_{\bf q}| \dot {\bf S}_{\bf q})} \,
 ({\sf Q} \ddot {\bf S}_{\bf q}|\,
 \frac{1}{\omega -{\sf Q}{\sf L}_0 {\sf Q} - i \eta }\,  {\sf Q} \ddot {\bf S}_{\bf q}). \,
\end{eqnarray}
Here, ${\sf Q}$ is a generalized projection operator which projects perpendicular 
to ${\bf S}_{\bf q}$ and  $\dot {\bf S}_{\bf q}$ (for details see Appendix {D}). Due to construction, 
the operator  ${\sf Q}\ddot {\bf S}_{\bf q}$ in the 'bra' and 'ket' of 
Eq.~\eqref{22b} corresponds to the second line in Eq.~\eqref{18}, 
and describes a twofold hopping away from the original site. 
Therefore, the selfenergy $\Sigma_{\bf q}(\omega)$
provides information about the hopping processes between next 
nearest neighbor sites and to more distant sites. 
In Appendix \ref{D} the selfenergy $\Sigma_{\bf q}(\omega)$ is evaluated
in a  factorization approximation by including the spin fluctuations
from ${\cal H}_J^{(0)}$. The result is shown in Fig.~\ref{SE}, where 
the imaginary part of $\Sigma_{\bf q}(\omega)$ for a small ${\bf q}$-vector is 
plotted (solid line) in the presence of spin fluctuations due to ${\cal H}_J^{(0)}$.  
As is seen, $\Sigma_{\bf q}(\omega)$ is rather small and almost
$\omega$-independent over a wide frequency range. Thus, the only effect of  $\Sigma_{\bf q}(\omega)$ 
is to give rise to a small damping and lineshift of the resonances of $\chi({\bf q},\omega)$.
We have also repeated the same calculation for $\Im \Sigma_{\bf q}(\omega)$
in the absence of ${\cal H}_J^{(0)}$, i.e.~when
${\cal H}_0$ is replaced by ${\cal H}_t$ (dashed line in Fig.~\ref{SE}). 
A strong $\omega$-dependence is found 
for small ${\bf q}$-values around $\omega =0$. This shows that long reaching 
hopping processes are important in this case.
From these findings, one can conclude that the hopping to 
more distant than nearest neighbors is of minor importance as long as
the exchange part ${\cal H}_J^{(0)}$ is not neglected in ${\cal H}_0$.
 A possible explanation would be 
that local antiferromagnetic correlations are still present at moderate hole doping 
outside the antiferromagnetic phase. They lead locally to strings of spin defects  
which are well known from the hole motion in the antiferromagnetic phase.
 
Let us come back to the discussion of the  oscillation behavior in Eq.~\eqref{18} which can be 
understood as follows. When an electron hops to a neighboring site, it preferably hops back 
to the original site, since this was definitely empty after the first hop.  In contrast, 
the hopping to  
next nearest neighbor sites is energetically unfavorable due to local antiferromagnetic order. 
As will be shown in a forthcoming paper~\cite{BS09}, the proportionality  of $\hat \omega^2_{\bf q} \sim \delta$ 
turns out to be the basic feature for the understanding of the superconducting pairing mechanism 
in the cuprates.  The oscillation becomes less important for larger $\delta$ which agrees with the weakening of the superconducting phase for larger hole doping. 

The solution of Eq.~\eqref{21} is easily found,
\begin{eqnarray}
\label{23}
{\bf S}_{\bf q}(t) &=&{\bf S}_{\bf q}\,  
\cos{\hat\omega_{\bf q}t}  + \frac{1}{\hat \omega_{\bf q}}\,\dot{\bf S}_{\bf q}\,
\sin{\hat\omega_{\bf q}t} \\
&=& \frac{1}{2}
({\bf S}_{\bf q}\,  - \frac{i}{\hat \omega_{\bf q}}\,\dot{\bf S}_{\bf q})\, e^{i \hat \omega_{\bf q}\, t}
+ \frac{1}{2}
 ({\bf S}_{\bf q}\,  + \frac{i}{\hat \omega_{\bf q}}\,
\dot{\bf S}_{\bf q})\, e^{-i \hat \omega_{\bf q}\, t}, \nonumber
\end{eqnarray}
where ${\bf S}_{\bf q}= {\bf S}_{\bf q}(t=0)$ and 
$\dot{\bf S}_{\bf q}= \frac{d}{dt}{\bf S}_{\bf q}(t=0)$ was used. 
From Eq.~\eqref{23},
the decomposition of ${\bf S}_{\bf q}$ into eigenmodes of ${\sf L}_0$  
can immediately be identified,
\begin{eqnarray}
\label{24}
{\sf L}_0 \, [\frac{1}{2}({\bf S}_{\bf q}\,  \mp \frac{i}{\hat \omega_{\bf q}}\,\dot{\bf S}_{\bf q})]
&=& \pm \omega_{\bf q}\,
 [\frac{1}{2}({\bf S}_{\bf q}\,  \mp \frac{i}{\hat \omega_{\bf q}}\,\dot{\bf S}_{\bf q})],
\end{eqnarray}
which leads to the intended decomposition of  the exchange ${\cal H}_J$ as follows:
\begin{eqnarray}
\label{25}
 {\cal H}_J &=& \sum_{\bf q} J_{\bf q} \, {\bf S}_{\bf q}  \, {\bf S}_{-{\bf q}} 
= \sum_{\bf q}J_{\bf q}\, \left({\cal A}_0({\bf q}) +  {\cal A}_1({\bf q}) +  {\cal A}_1^\dagger ({\bf q})\right), 
\end{eqnarray}
where 
\begin{eqnarray}
 \label{26}
{\cal A}_0({\bf q}) &=& \frac{1}{2}\left( {\bf S}_{{\bf q}}{\bf S}_{-{\bf q}} +
\frac{1}{\hat\omega_{\bf q}^2} \dot{\bf S}_{{\bf q}}\dot{\bf S}_{-{\bf q}} 
\right),
\\
{\cal A}_1({\bf q}) &=& \frac{1}{4}\left({\bf S}_{\bf q} - \frac{i}{\hat \omega_{\bf q}}\,
 \dot{\bf S}_{\bf q}\right)
\, \left({\bf S}_{-{\bf q}} - \frac{i}{\hat \omega_{\bf q}}\, \dot{\bf S}_{-{\bf q}}\right), \nonumber
\\
{\cal A}_1^\dagger({\bf q})  &=& 
\frac{1}{4}\left({\bf S}_{\bf q} + \frac{i}{\hat \omega_{\bf q}}\,
 \dot{\bf S}_{\bf q}\right)
\, \left({\bf S}_{-{\bf q}} + \frac{i}{\hat \omega_{\bf q}}\, \dot{\bf S}_{-{\bf q}}\right), 
\nonumber
\end{eqnarray} 
and
\begin{eqnarray}
\label{27}
{\sf L}_0\,{\cal A}_0({\bf q}) &=&  0, \qquad 
{\sf L}_0\, {\cal A}_1({\bf q}) = 2 \hat \omega_{\bf q}\,  {\cal A}_1({\bf q}),   \qquad
{\sf L}_0\,{\cal A}_1^\dagger({\bf q})  = -2\hat \omega_{\bf q}\, {\cal A}_1^\dagger({\bf q})  .
\end{eqnarray}
Here, an additional approximation was used. In deriving Eqs.~\eqref{27}, 
the eigenmodes of the two 
spin operators ${\bf S_q}\cdot {\bf S}_{-{\bf q}}$ in the expression for ${\cal H}_J$ 
were taken separately from Eq.~\eqref{24}. In this way, all local  
configurations were disregarded, where two spin operators in local space 
are located on neighboring sites. Thereby, a possible hopping between the two sites
would be obstructed.  
The inclusion of these processes would need additional considerations. However, 
they would not change our results substantially.

With Eqs.~\eqref{26}, we have arrived at the intended decomposition of the $t$-$J$ model. 
The Hamiltonian 
\begin{eqnarray}
\label{27a}
 {\cal H}&=&  \sum_{{\bf k}\sigma} \varepsilon_{\bf k}
\hat c_{{\bf k}\sigma}^\dagger \hat c_{{\bf k}\sigma} +
\sum_{\bf q} J_{\bf q} {\bf S}_{\bf q} {\bf S}_{-{\bf q}}
\end{eqnarray}
can be decomposed into an 'unperturbed' part  ${\cal H}_0$ 
and into a 'perturbation'  ${\cal H}_1$. It reads
\begin{eqnarray}
\label{28}
 {\cal H}_0 &=&  {\cal H}_t + {\cal H}_{0,J} =:
\sum_{{\bf k}\sigma} \varepsilon_{\bf k} \, \hat c_{{\bf k}\sigma}^\dagger  \hat c_{{\bf k}\sigma} +
\sum_{\bf q} J_{\bf q}\, {\cal A}_0({\bf q}), \nonumber\\
{\cal H}_1 &=& \sum_{\bf q} J_{\bf q}\, \left( {\cal A}_1({\bf q}) + {\cal A}_1^\dagger({\bf q})
\right).
\end{eqnarray} 
The aim of the projector-based renormalization method (PRM) is to eliminate all transitions 
between the eigenstates of ${\cal H}_0$ which are induced  by ${\cal H}_1$.
Let us assume that all excitations with energies larger than a given cutoff $\lambda$ 
have already been eliminated. Then, the renormalized Hamiltonian ${\cal H}_\lambda$ 
should have the form
\begin{eqnarray}
 \label{28a}
{\cal H}_\lambda &=& \sum_{{\bf k}\sigma} \varepsilon_{{\bf k},\lambda}\,
\hat c_{{\bf k}\sigma}^\dagger \hat c_{{\bf k}\sigma} +
\sum_{\bf q} J_{{\bf q},\lambda} \,
{\sf P}_\lambda {\bf S}_{\bf q} {\bf S}_{-{\bf q}} \, ,
\end{eqnarray}
however, with $\lambda$-dependent prefactors $\varepsilon_{{\bf k},\lambda}$ 
and $J_{{\bf q},\lambda}$. Moreover, a projector ${\sf P}_\lambda$ \,
was introduced which acts on operator variables. It
guarantees that only transitions with excitation energies smaller  
than $\lambda$ remain from ${\bf S}_{\bf q} {\bf S}_{-{\bf q}}$.

The separation of ${\cal H}_\lambda$ 
into an unperturped part ${\cal H}_{0,\lambda}$ 
and a perturbation ${\cal H}_{1,\lambda}$ reads in analogy to Eq.~\eqref{28},
$ {\cal H}_\lambda = {\cal H}_{0,\lambda} +{\cal H}_{1,\lambda}$,
with 
\begin{eqnarray}
 {\cal H}_{0,\lambda} &=&   {\cal H}_{t,\lambda}
+\sum_{\bf q} J_{{\bf q},\lambda}\, {\cal A}_{0,\lambda}({\bf q})  
   + E_\lambda ,      \nonumber \\
{\cal H}_{1,\lambda} &=& \sum_{\bf q} J_{{\bf q},\lambda}\, \Theta(\lambda -|2\hat \omega_{{\bf q}, \lambda}|)
 \left( {\cal A}_{1,\lambda} ({\bf q}) + {\cal A}_{1,\lambda}^\dagger({\bf q})
\right),
\label{29}
\end{eqnarray} 
where we have used the $\lambda$-dependent extension of  relation \eqref{27} in order to exploit 
the properties of ${\sf P}_\lambda$. 
Note that the $\Theta$-function $\Theta(\lambda -|2\hat \omega_{{\bf q}, \lambda}|)$
in ${\cal H}_{1,\lambda}$ guarantees that only excitations with transition energies 
$|2\hat\omega_{{\bf q},\lambda}|$ smaller than $\lambda$ contribute to 
${\cal H}_{1,\lambda}$. In Eq.~\eqref{29}, 
${\cal H}_{t,\lambda}$ is the renormalized hopping term from Eq.~\eqref{28a}, 
 ${\cal H}_{t,\lambda} =
\sum_{{\bf k}\sigma} \varepsilon_{{\bf k},\lambda} \, \hat c_{{\bf k}\sigma}^\dagger  
\hat c_{{\bf k}\sigma}$.
Also, the parameters  $J_{{\bf q},\lambda}$, $ \hat 
\omega_{{\bf q}, \lambda}$  and $E_\lambda$ in Eqs.~\eqref{29} now depend on $\lambda$. 
Moreover, the new operators ${\cal A}_{\alpha,\lambda}({\bf q})$ ($\alpha=0, \pm 1$) 
depend on $\lambda$,
\begin{eqnarray}
\label{30}
{\cal A}_{0,\lambda}({\bf q}) &=& \frac{1}{2}\left( {\bf S}_{{\bf q}}{\bf S}_{-{\bf q}} +
\frac{1}{\hat\omega_{{\bf q},\lambda}^2} \dot{\bf S}_{{\bf q},\lambda}\dot{\bf S}_{-{\bf q},\lambda} 
\right), \nonumber
\\
{\cal A}_{1,\lambda}({\bf q}) &=& \frac{1}{4}\left({\bf S}_{\bf q} - \frac{i}{\hat \omega_{{\bf q},\lambda}}\,
 \dot{\bf S}_{{\bf q},\lambda} \right)
\, \left({\bf S}_{-{\bf q}} - \frac{i}{\hat \omega_{{\bf q},\lambda}}\,
  \dot{\bf S}_{-{\bf q},\lambda}\right) ,
\\
{\cal A}_{1,\lambda}^\dagger({\bf q})  &=& 
\frac{1}{4}\left({\bf S}_{\bf q} + \frac{i}{\hat \omega_{{\bf q},\lambda}}\,
 \dot{\bf S}_{{\bf q},\lambda} \right)
\, \left({\bf S}_{-{\bf q}} + \frac{i}{\hat \omega_{{\bf q},\lambda}}\, 
\dot{\bf S}_{-{\bf q},\lambda}\right) ,
\nonumber
\end{eqnarray} 
where $\hat\omega_{{\bf q},\lambda}$ and  $\dot{\bf S}_{{\bf q},\lambda}$ are defined by 
\begin{eqnarray} 
\label{30a}
 \hat\omega_{{\bf q},\lambda}^2 &=& 2P_0\,(t^2_{{\bf q}=0, \lambda} -
  t^2_{{\bf q}, \lambda}), \qquad
t^2_{{\bf q},\lambda} = \sum_{i(\neq j)}t^2_{{ij},\lambda}\, e^{i{\bf q}({\bf
  R}_i- {\bf R}_j)}, \nonumber  \\
\dot{\bf S}_{{\bf q},\lambda}&=&
\frac{i}{\hbar}[{\cal H}_{0,\lambda}, {\bf S}_{{\bf q}}] \approx
\frac{i}{\hbar}[{\cal H}_{t,\lambda}, {\bf S}_{{\bf q}}] .
\end{eqnarray}

\subsection{Generator of the unitary transformation}
To derive renormalization equations for the parameters of ${\cal H}_\lambda$, we have to 
apply the unitary transformation \eqref{12} to ${\cal H}_\lambda$ in order to eliminate 
excitations within a new energy shell between 
$\lambda$ and $\lambda - \Delta \lambda$. 
We use the lowest order expression \eqref{14b} for the new generator $X_{\lambda, \Delta \lambda}$,
\begin{eqnarray}
\label{31}
 X_{\lambda, \Delta \lambda} &=& \sum_{\bf q}\frac{J_{{\bf q},\lambda}}{2 \hat \omega_{{\bf q}, \lambda}}
\Theta_{\bf q}(\lambda, \Delta\lambda)\left({\cal A}_{1,\lambda}({\bf q}) - 
{\cal A}_{1,\lambda}^\dagger ({\bf q})  \right) .
\end{eqnarray}
Here, $\Theta_{\bf q}(\lambda, \Delta\lambda)$ denotes a product of two $\Theta$-functions,
\begin{eqnarray*}
\Theta_{\bf q}(\lambda, \Delta\lambda) &=& \Theta(\lambda -|2 \hat \omega_{{\bf q},\lambda}|)\,
\Theta\left (|2\omega_{{\bf q}, \lambda -\Delta \lambda}| -(\lambda -\Delta \lambda) \right),
\end{eqnarray*}
which confines the elimination range to excitations with 
$|2\omega_{{\bf q}, \lambda -\Delta \lambda}|$ larger than $\lambda - \Delta \lambda$
and $|2 \hat \omega_{{\bf q},\lambda}|$ smaller than $\lambda$.
Roughly speaking, for the  case of a weak $\lambda$-dependence of 
$|\omega_{{\bf q},\lambda}|$,
the elimination is restricted to all transitions within an energy  shell 
between $\lambda -\Delta \lambda$ and $\lambda$. 
With \eqref{30}, the generator $X_{\lambda, \Delta \lambda}$ can also be expressed by
\begin{eqnarray}
\label{32}
 X_{\lambda, \Delta \lambda} &=& 
-i \sum_{\bf q}\frac{J_{{\bf q},\lambda}}{ 4\hat \omega_{{\bf q}, \lambda}^2}
\Theta_{\bf q}(\lambda, \Delta\lambda)\left(
{\bf S}_{\bf q} \,\dot {\bf S}_{-{\bf q},\lambda} + 
\dot {\bf S}_{{\bf q},\lambda} \, {\bf S}_{-{\bf q}} 
 \right) .
\end{eqnarray}
In the following, we restrict ourselves to the lowest order renormalization processes.
Then, $J_{{\bf q},\lambda}$ will not be renormalized by higher 
orders in $J$, and we can use $J_{{\bf q},\lambda}= J_{\bf q}$ from the beginning.

\subsection{Renormalization equations}
The unitary transformation \eqref{12}, applied to the renormalization step between 
$\lambda$ and $\lambda - \Delta \lambda$, will 
be evaluated in perturbation theory in second order in $J_{\bf q}$,
\begin{eqnarray}
\label{33}
{\cal H}_{\lambda - \Delta \lambda} &=& e^{X_{\lambda, \Delta \lambda}}\,
{\cal H}_{\lambda} \, e^{-X_{\lambda, \Delta \lambda}} = 
{\cal H}_{\lambda - \Delta \lambda}^{(0)} + {\cal H}_{\lambda - \Delta \lambda}^{(1)} 
+{\cal H}_{\lambda - \Delta \lambda}^{(2)}  + \cdots,
\end{eqnarray}
where 
\begin{eqnarray}
\label{34}
{\cal H}_{\lambda - \Delta \lambda}^{(0)} &=& \sum_{{\bf k}\sigma}
\varepsilon_{{\bf k},\lambda} \, \hat c_{{\bf k}\sigma}^\dagger  
\hat c_{{\bf k}\sigma} + E_\lambda =
{\cal H}_{t,\lambda} + E_\lambda ,   \nonumber \\
{\cal H}_{\lambda - \Delta \lambda}^{(1)} &=&
\sum_{\bf q} J_{{\bf q}}\, {\cal A}_{0,\lambda}({\bf q}) 
+ [X_{\lambda, \Delta \lambda}, {\cal H}_{t, \lambda}]
+ 
 \sum_{\bf q} J_{{\bf q}}\, \Theta(\lambda -|2\hat \omega_{{\bf q}, \lambda}|)
 \left( {\cal A}_{1,\lambda} ({\bf q}) + {\cal A}_{1,\lambda}^\dagger({\bf q})
\right),
\nonumber \\
%
{\cal H}_{\lambda - \Delta \lambda}^{(2)} &=&
\frac{1}{2} [X_{\lambda, \Delta \lambda},  [X_{\lambda, \Delta \lambda}, {\cal H}_{t, \lambda}]\, ]
+
\sum_{\bf q} J_{{\bf q}} \,  [X_{\lambda, \Delta \lambda}, {\cal A}_{0,\lambda}({\bf q})] \nonumber \\
&& +
\sum_{\bf q} J_{{\bf q}}\, \Theta(\lambda -|2\hat \omega_{{\bf q}, \lambda}|)\,
 [ \, X_{\lambda, \Delta \lambda},
{\cal A}_{1,\lambda} ({\bf q}) + {\cal A}_{1,\lambda}^\dagger({\bf q}) \,].
\end{eqnarray}
Let us first evaluate ${\cal H}_{\lambda - \Delta \lambda}^{(2)}$ from  
second order processes.  
The commutators in Eq.~\eqref{34} are explicitly evaluated in Appendix A.  
Then, we can compare the obtained result with the formal expression for 
${\cal H}_{\lambda - \Delta \lambda}$ which has the same operator structure as 
${\cal H}_\lambda$, with  
$\lambda$ is replaced by $\lambda - \Delta \lambda$. One obtains the following 
renormalization equation from the second order contributions in $J_{\bf q}$:
\begin{eqnarray}
\label{35}
 \varepsilon_{{\bf k},\lambda - \Delta \lambda}- \varepsilon_{{\bf k},\lambda} &=&
\frac{1}{16 N}\sum_{\bf q} \frac{J_{\bf q}^2}{\hat \omega_{{\bf q}, \lambda}^4}\,
\Theta_{\bf q}(\lambda, \Delta \lambda) \,
(\varepsilon_{{\bf k}+ {\bf q}, \lambda} +  \varepsilon_{{\bf k}- {\bf q}, \lambda}
-2 \varepsilon_{{\bf k}, \lambda} )\, \langle \dot {\bf S}_{{\bf q},\lambda} \, 
\dot {\bf S}_{-{\bf q},\lambda} \rangle \nonumber \\
&& + 
\frac{3}{2N} \sum_{{\bf q}\sigma} \left(\frac{J_{\bf q}}{4 \hat \omega_{\bf q}^2}\right)^2 \,
\Theta_{\bf q}(\lambda,\Delta \lambda) \,(\varepsilon_{{\bf k},\lambda}-
\varepsilon_{{\bf k}- {\bf q},\lambda })^2 \\
&& \times \left[ \frac{1}{N} \sum_{{\bf k}'\sigma'}(2\varepsilon_{{\bf k}',\lambda } -
\varepsilon_{{\bf k}'+{\bf q},\lambda} -\varepsilon_{{\bf k}'-{\bf q},\lambda}) 
\langle \hat c_{{\bf k}'\sigma'}^\dagger  \hat c_{{\bf k}'\sigma'} \rangle 
\right] \, n_{{\bf k}-{\bf q}\alpha}^{(NL)} , \nonumber
\end{eqnarray}
where we have defined
\begin{eqnarray}
\label{35a}
 n_{{\bf k},\sigma}^{(NL)}  &=&
\langle \hat c_{{\bf k}\sigma}^\dagger \hat c_{{\bf k}\sigma} \rangle 
- \frac{1}{N} \sum_{{\bf k}'}\langle \hat c_{{\bf k}'\sigma}^\dagger \hat
c_{{\bf k}'\sigma} \rangle  
\end{eqnarray}
as non-local part of the one-particle occupation number per spin direction.
An equivalent equation also exists for $E_{\lambda -\Delta \lambda}$.
Note that in Eq.~\eqref{35} an additional factorization approximation was used in order 
to extract all terms which have the same operator structure as 
${\cal H}_\lambda$. The quantity $\langle \dot {\bf S}_{{\bf q},\lambda}  \dot {\bf S}_{-{\bf q},\lambda} \rangle$ is a correlation function of the time derivatives of ${\bf S}_{\bf q}$  which can easily 
be evaluated from Eq.~\eqref{A3}.
Note that an additional contribution 
to $\varepsilon_{{\bf k}, \lambda - \Delta \lambda}$, proportional to 
the correlation function $\langle {\bf S}_{\bf q}\cdot {\bf S}_{-{\bf q}} \rangle$, 
has been neglected. 
The remaining expectation values in Eq.~\eqref{35} have to be calculated separately. In principle, 
they should be defined with the $\lambda$-dependent Hamiltonian ${\cal H}_\lambda$, because the factorization approximation was employed 
for the renormalization step from ${\cal H}_\lambda$ to ${\cal H}_{\lambda -
\Delta \lambda}$.  However, ${\cal H}_\lambda$  still contains interactions which prevent a 
straight evaluation of $\lambda$-dependent expectation values.
The  best way to circumvent this difficulty is to calculate the expectation values with the full Hamiltonian ${\cal H}$ instead of with   ${\cal H}_\lambda$. In this case, the renormalization equations can be solved self-consistently, as will be discussed below.

Note that the renormalization \eqref{35} of $\varepsilon_{{\bf k}, \lambda}$ was evaluated from the
second order part ${\cal H}_{\lambda- \Delta \lambda}^{(2)}$  of the Hamiltonian 
\eqref{34}. Thus, we are led to
\begin{eqnarray}
\label{37}  
{\cal H}_{\lambda -\Delta \lambda} &=&  
{\cal H}_{t,\lambda- \Delta \lambda} + {\cal H}_{\lambda -\Delta \lambda}^{(1)}
 + E_{\lambda - \Delta \lambda}  ,
\end{eqnarray}
where
$ {\cal H}_{t,\lambda- \Delta \lambda} =
\sum_{{\bf k},\sigma} \varepsilon_{{\bf k},\lambda -\Delta \lambda}\,
\hat c_{{\bf k}\sigma}^\dagger \hat c_{{\bf k}\sigma} \nonumber$. What remains is to evaluate 
the renormalization part ${\cal H}_{\lambda -\Delta \lambda}^{(1)}$ 
in first order in $J_{\bf q}$ to ${\cal H}_{\lambda- \Delta \lambda}$.
First, the second term on the right hand side of Eq.~\eqref{34}
can be rewritten, since
\begin{eqnarray*}
 [X_{\lambda, \Delta \lambda}, {\cal H}_{t, \lambda}]  &=& - \sum_{\bf q}J_{{\bf q}} 
\Theta_{\bf q}(\lambda, \Delta \lambda) \, \left( {\cal A}_{1,\lambda}({\bf q}) +{\cal A}_{1,\lambda}^\dagger
({\bf q})\right).
\end{eqnarray*}
Then, by combining the second and third term, we find  
\begin{eqnarray}
\label{39}
{\cal H}_{\lambda -\Delta \lambda}^{(1)} &=&
\sum_{\bf q} J_{{\bf q}}\, {\cal A}_{0,\lambda}({\bf q})  \\
&+&
\sum_{\bf q} J_{{\bf q}}\, 
 \Theta(\lambda -|2 \hat \omega_{{\bf q},\lambda}|)\,
\Theta(\lambda - \Delta \lambda -|2 \hat \omega_{{\bf q},\lambda - \Delta \lambda}|)
 \left( {\cal A}_{1,\lambda}({\bf q}) +{\cal A}_{1,\lambda}^\dagger
({\bf q})\right). \nonumber
\end{eqnarray}
The excitation energies of ${\cal A}_{1,\lambda}({\bf q})$ and ${\cal A}_{1,\lambda}^\dagger({\bf q})$
are restricted to $|2\hat \omega_{{\bf q},\lambda}| \leq \lambda$ by the first $\Theta$-function in
Eq.~\eqref{39}. This condition is automatically fulfilled
by the second $\Theta$-function, in the case that 
$|2\hat \omega_{{\bf q},\lambda - \Delta \lambda}|$ only 
weakly depends on $\lambda$ and we can replace  $\lambda$ by $\lambda -\Delta \lambda$. 
By introducing the projector ${\sf P}_{\lambda - \Delta \lambda}$ on all 
low-energy transition operators with energies smaller than $\lambda - \Delta \lambda$, we find 
\begin{eqnarray}
\label{40}
{\cal H}_{\lambda -\Delta \lambda}^{(1)} &=&
\sum_{\bf q} J_{{\bf q}}\,{\sf P}_{\lambda - \Delta \lambda} 
\left({\cal A}_{0,\lambda}({\bf q}) 
+ {\cal A}_{1,\lambda}({\bf q}) +{\cal A}_{1,\lambda}^\dagger
({\bf q})\right) \nonumber \\
&=& 
\sum_{\bf q} J_{\bf q} {\sf P}_{\lambda - \Delta \lambda}\, 
{\bf S}_{\bf q}\cdot {\bf S}_{-{\bf q}},
\end{eqnarray}
where we have used  the representation \eqref{25} for the scalar product 
${\bf S}_{\bf q}\cdot {\bf S}_{-{\bf q}}$,
\begin{eqnarray}
\label{40a}
{\bf S}_{\bf q}\cdot {\bf S}_{-{\bf q}}
 &=&
{\cal A}_{0,\lambda}({\bf q})  + {\cal A}_{1,\lambda}({\bf q}) 
+{\cal A}_{1,\lambda}^\dagger({\bf q}) .
\end{eqnarray}
Finally, for the total Hamiltonian ${\cal H}_{\lambda - \Delta \lambda}$, we obtain
according to \eqref{37}
\begin{eqnarray}
 \label{40b}
{\cal H}_{\lambda -\Delta \lambda} &=& 
\sum_{{\bf k},\sigma} \varepsilon_{{\bf k},\lambda -\Delta \lambda}\,
\hat c_{{\bf k}\sigma}^\dagger \hat c_{{\bf k}\sigma} +
\sum_{\bf q} J_{\bf q} {\sf P}_{\lambda - \Delta \lambda}\, 
{\bf S}_{\bf q}\cdot {\bf S}_{-{\bf q}} + E_{\lambda - \Delta \lambda}.
\end{eqnarray}
Note that this expression completely agrees with the Hamiltonian 
at cutoff $\lambda$, when $\lambda$ is replaced by $\lambda -\Delta \lambda$. 
The required  decomposition into ${\cal H}_{0, \lambda - \Delta \lambda}$
and ${\cal H}_{1, \lambda - \Delta \lambda}$ is found as follows. We use again the relation  
\eqref{40a}, with   $\lambda$ is replaced by $\lambda -\Delta \lambda$, and rewrite 
${\cal H}_{\lambda -\Delta \lambda}^{(1)}$ as 
\begin{eqnarray}
{\cal H}_{\lambda -\Delta \lambda}^{(1)} &=&
\sum_{\bf q} J_{{\bf q}}\,{\sf P}_{\lambda - \Delta \lambda} 
\left({\cal A}_{0,\lambda- \Delta \lambda}({\bf q}) 
+ {\cal A}_{1,\lambda - \Delta \lambda}({\bf q}) +{\cal A}_{1,\lambda- \Delta \lambda}^\dagger
({\bf q})\right). 
\end{eqnarray}
Using again Eq.~\eqref{40}, we arrive at the renormalized Hamiltonian ${\cal H}_{\lambda - \Delta 
\lambda}= {\cal H}_{0,\lambda - \Delta \lambda}
+ {\cal H}_{1, \lambda - \Delta \lambda}$ in the following form,  
\begin{eqnarray}
 \label{41}
{\cal H}_{0,\lambda - \Delta \lambda} &=&
{\cal H}_{t,\lambda- \Delta \lambda}
+
\sum_{\bf q} J_{{\bf q}}\, {\cal A}_{0,\lambda- \Delta \lambda}({\bf q})
+ E_{\lambda - \Delta \lambda} , \nonumber \\
{\cal H}_{1, \lambda -\Delta \lambda} &=& 
\sum_{\bf q} J_{{\bf q}}\,
\Theta(\lambda -\Delta \lambda -|\hat \omega_{{\bf q}, \lambda - \Delta \lambda}|)\, 
\left({\cal A}_{1,\lambda - \Delta \lambda}({\bf q}) +{\cal A}_{1,\lambda- \Delta \lambda}^\dagger
({\bf q})\right). 
\end{eqnarray}
As expected, the renormalized Hamiltonians 
${\cal H}_{0,\lambda - \Delta \lambda}$ and  ${\cal H}_{1,\lambda - \Delta \lambda}$ 
have the same operator structure as at cutoff $\lambda$. Therefore, 
we can formulate a renormalization scheme as follows:
We start from the original $t$-$J$ model, where
the energy cutoff is denoted by  
${\lambda= \Lambda}$. Starting from a guess for the unknown expectation values,
which enter the renormalization equation \eqref{35}, 
we proceed by eliminating all  excitations in steps $\Delta \lambda$ 
from $\lambda=\Lambda$ down to $\lambda=0$.
Thereby, the parameters of the Hamiltonian change in steps according to the renormalization equation \eqref{35}. In this way, we obtain the following model at $\lambda =0$:
\begin{eqnarray}
\label{42}
 {\cal H}_{\lambda =0} &=& {\cal H}_{t, \lambda=0} + \sum_{\bf q} J_{\bf q} {\sf P}_{\lambda=0} 
{\bf S}_{\bf q} \cdot
{\bf S}_{-{\bf q}} +E_{\lambda=0} \\ 
&=& \sum_{{\bf k}\sigma} \varepsilon_{{\bf k},\lambda=0}\,
\hat c_{{\bf k} \sigma}^\dagger \, \hat c_{{\bf k} \sigma}
    +  \sum_{\bf q} J_{{\bf q}}\, {\cal A}_{{0},\lambda=0}({\bf q}) + E_{\lambda=0} .   \nonumber
\end{eqnarray}
Note that in Eq.~\eqref{42} the perturbation ${\cal H}_{1}$ is completely integrated out. Only the part 
of the exchange, which commutes with the hopping term, remains. 

Unfortunately, due to the presence of the ${\cal A}_0$-term,  the Hamiltonian ${\cal H}_{\lambda =0}$
can not be diagonalized. It does not yet allow us
to recalculate the expectation values. 
Therefore, a further approximation is necessary which consists of
a factorization of the second term 
\begin{eqnarray}
\label{43}
 \sum_{\bf q} J_{{\bf q}}\, {\cal A}_{{0},\lambda=0}({\bf q}) &=& 
\sum_{\bf q}
\frac{J_{\bf q}}{2}\left( {\bf S}_{{\bf q}}{\bf S}_{-{\bf q}} +
\frac{1}{\hat \omega_{{\bf q},\lambda=0}^2} \dot{\bf S}_{{\bf q},\lambda=0}\dot{\bf S}_{-{\bf q},\lambda=0} 
\right) .
\end{eqnarray}
According to Appendix B, ${\cal H}_{\lambda=0}$ can finally be replaced by a
modified Hamiltonian which will be denoted by
$\tilde{\cal H}^{(1)}$,
\begin{eqnarray}
 \label{44}
\tilde {\cal H}^{(1)} &=& \sum_{{\bf k}\sigma} \tilde \varepsilon_{\bf k}^{(1)}\,
\hat c_{{\bf k} \sigma}^\dagger \, \hat c_{{\bf k} \sigma}   
 +\sum_{\bf q} \frac{J_{{\bf q}}}{2}\, {\bf S}_{\bf q}\,  {\bf S}_{-{\bf q}} 
 + \tilde E^{(1)}  ,   
\end{eqnarray}
where the electron energy is modified according to
\begin{eqnarray}
\label{45}
\tilde \varepsilon_{\bf k}^{(1)} &=& 
\varepsilon_{{\bf k},\lambda=0} - \frac{1}{N} \sum_{\bf q} \frac{3J_{\bf q}}
{4 \hat \omega^2_{{\bf q},\lambda=0}} (\varepsilon_{{\bf k},\lambda=0} - 
\varepsilon_{{\bf k}- {\bf q},\lambda=0} )^2\, n_{{\bf k} -{\bf
    q},\sigma}^{(NL)},
\end{eqnarray} 
and $n_{{\bf k},\sigma}^{(NL)}$ is defined in Eq.~\eqref{35a}.
Note that the operator structure of $\tilde {\cal H}^{(1)}$ agrees 
with that of the original $t$-$J$ model of Eq.~\eqref{27a}.
However, the parameters have changed. Most important, the strength of the 
exchange coupling in Eq.~\eqref{44} is decreased by a factor $1/2$. This property allows us to start 
the whole renormalization procedure again. We consider 
the modified $t$-$J$ model of Eq.~\eqref{44} as our new initial Hamiltonian, which has to 
be renormalized again. The initial values of 
$\tilde{\cal H}^{(1)}$ at cutoff $\lambda =\Lambda$ are
$\tilde \varepsilon_{\bf k}^{(1)}$  
and $J_{\bf q}/2$. After the new renormalization cycle the exchange coupling of the new 
renormalized Hamiltonian $\tilde{\cal H}^{(2)}$ is again decreased by a factor $1/2$,
till after a sufficiently large number of renormalization cycles 
($n\rightarrow \infty$) the exchange operator completely disappears. 
Thus, we finally arrive at a 'free' model
\begin{eqnarray}
 \label{46}
\tilde {\cal H} &=& \sum_{{\bf k}\sigma} \tilde \varepsilon_{\bf k}\,
\hat c_{{\bf k} \sigma}^\dagger \, \hat c_{{\bf k} \sigma}
     + \tilde E \, ,
\end{eqnarray}
where we have introduced as new notations
$\tilde {\cal H} =  \tilde {\cal H}^{(n\rightarrow \infty)}$, $\tilde \varepsilon_{\bf k} =
\tilde \varepsilon_{\bf k}^{(n \rightarrow \infty)}$, 
and $\tilde E = \tilde E^{(n \rightarrow \infty)}$.
Note that the Hamiltonian $\tilde {\cal H}$ now
allows us  to recalculate the unknown expectation values. With the new  
 values, the whole renormalization procedure can be 
started again till, after a sufficiently large number of such overall cycles, 
the expectation values  have converged. 
The renormalization equations are solved self-consistently.
However, note that the fully renormalized Hamiltonian \eqref{46} is actually not a 'free' model.
Instead, it is still subject to strong electronic correlations which are built in  
by the presence of the Hubbard operators. 
Therefore, to evaluate the expectation values, further approximations have to be made.

\subsection{Evaluation of expectation values}
\label{expvalues}
The expectation values in Eqs.~\eqref{35} and \eqref{45} are formed with the full Hamiltonian.  
To evaluate expectation values for operator variables 
${\cal A}$, we have to apply the unitary transformation also on ${\cal A}$,
\begin{eqnarray}
\label{47}
 \langle {\cal A}\rangle &=& \frac{\mbox {Tr}\,({\cal A}\, e^{-\beta{\cal H}})}
{\mbox{Tr}\,e^{-\beta{\cal H}}} =
 \langle {\cal A}(\lambda) \rangle_{{\cal H}_\lambda} = 
\langle \tilde{\cal A} \rangle_{\tilde{\cal H}} \, ,
 \end{eqnarray}
where we have defined $
{\cal A}(\lambda) = e^{X_{\lambda}} \; {\cal A} e^{-X_{\lambda}} $ and  
${\tilde{\cal A}} = {\cal A}(\lambda \rightarrow 0)
$. Thus, additional renormalization equations for ${\cal A}(\lambda)$
have to be derived. 

As an example, let us consider the 
angle-resolved photoemission (ARPES) spectral function. It is defined by
\begin{eqnarray}
\label{48}
 { A}({\bf k}, \omega)  &=& \frac{1}{2\pi} \int_{-\infty}^{\infty}
\big< \hat c_{{\bf k}\sigma}^\dagger (-t)  \;\hat c_{{\bf k}\sigma} 
\big> \; e^{i\omega t} dt = 
\big< \hat c_{{\bf k}\sigma }^\dagger \, \delta( {\sf L} + \omega ) \;
\hat c_{{\bf k}\sigma} \big>
\end{eqnarray}
and can be rewritten by use of the dissipation-fluctuation theorem as
\begin{eqnarray}
\label{49}
 { A}({\bf k}, \omega)  &=&   \frac{1}{1+ e^{\beta \omega}} \Im  G({\bf k}, \omega) \, ,
\end{eqnarray}
where $\Im G({\bf k},\omega)$ is the dissipative part of the anti-commutator Green 
function
\begin{eqnarray*}
\Im G({\bf k}, \omega) &=& \frac{1}{2\pi} \, \int_{-\infty}^{\infty}
\big< [\hat c_{{\bf k}\sigma}^\dagger (-t)\, ,  \;\hat c_{{\bf k}\sigma}]_+ 
\big> \; e^{i\omega t} dt =
\big< [\hat c_{{\bf k}\sigma }^\dagger \, , \, \delta( {\sf L} + \omega ) \;
\hat c_{{\bf k}\sigma}]_+ \big>. \nonumber 
\end{eqnarray*}
The time dependence and the expectation value are formed 
with the full Hamiltonian ${\cal H}$, and $\sf L$ is the Liouville operator corresponding to 
${\cal H}$.
According to Eq.~\eqref{47},  the anti-commutator Green function can be expressed by
\begin{eqnarray} 
\label{50}
 {\Im  G}({\bf k}, \omega)  &=&  
\big< [\hat c_{{\bf k}\sigma }^\dagger(\lambda) \, ,\, \delta ({\sf L}_\lambda + \omega) \;
\hat c_{{\bf k}\sigma}(\lambda)]_+ \big>_\lambda \, ,
\end{eqnarray}
where now the creation and annihilation operators are also subject 
to the unitary transformation. To evaluate $A({\bf k},\omega)$,
we have to derive renormalization equations for $\hat c_{{\bf k}\sigma}(\lambda)$ and 
$\hat c_{{\bf k}\sigma}^\dagger(\lambda)$. According to Appendix \ref{B}, the following {\it ansatz} for 
$\hat c_{{\bf k}\sigma}(\lambda)$ can be used: 
\begin{eqnarray}
\label{51}
 \hat c_{{\bf k} \sigma}(\lambda) &=& u_{{\bf k}, \lambda} \hat c_{{\bf
     k}\sigma}  + \frac{1}{2N}\sum_{{\bf q k}'} v_{{\bf k}, {\bf q},\lambda} \, \frac{J_{\bf q}}{4 
\hat \omega^2_{{\bf q},\lambda}} \, 
 \sum_{\alpha \beta \gamma} (\vec \sigma_{\alpha \beta}
\cdot \vec \sigma_{\sigma \gamma}) 
 (\varepsilon_{{\bf k}',\lambda}- \varepsilon_{{\bf k}' + {\bf q},\lambda}) \, 
 \,
 \hat c^\dagger_ {{\bf k}' + {\bf q} \alpha} \ \hat c_ {{\bf k}' \beta} \
\hat c_ {{\bf k} + {\bf q} \gamma}. \nonumber \\
&&
\end{eqnarray} 
It can be justified from lowest order perturbation theory. Note that 
the $\lambda$-dependence is transferred to the 
parameters $u_{{\bf k},\lambda}$ and $v_{{\bf k},{\bf q},\lambda}$. 
Also the quantities $\hat \omega_{{\bf q},\lambda}$ and $\varepsilon_{{\bf k},\lambda}$
depend on $\lambda$. However,  
having in mind perturbation theory in $J$, this $\lambda$-dependence will be neglected
in the numerical evaluation of Sec.~\ref{NS} below.
According to  Appendix \ref{B}, the renormalization equations for 
$u_{{\bf k},\lambda}$ and  $v_{{\bf k},{\bf q},\lambda}$ read 
\begin{eqnarray}
\label{52}
  u_{{\bf k},\lambda- \Delta \lambda}^2 &=& u_{{\bf k},\lambda}^2 
-\frac{3}{2N^2} \sum_{{\bf q}{\bf k}'} \left(\frac{J_{\bf q}}{4 \hat \omega_{\bf q}^2}\right)^2
\Theta _{\bf q}(\lambda, \Delta \lambda)   (\varepsilon_{{\bf k}',\lambda} 
-\varepsilon_{{\bf k}'+{\bf q},\lambda})^2   \left\{ \left(\frac{u_{{\bf k},\lambda}}{2}\right)^2 + u_{{\bf k},\lambda}\, v_{{\bf k},{\bf q},\lambda}
\right\} \nonumber \\
&  &\times  \left\{ n_{{\bf k}'+{\bf q}} m_{{\bf k}'}+
n_{{\bf k}+{\bf q}} (D + n_{{\bf k}'} -n_{{\bf k}'+{\bf q}})
\right\} \nonumber  \\ \nonumber \\
&+& \frac{3}{4N^2} \sum_{{\bf q}{\bf q}'} \frac{J_{\bf q}}{4 \hat \omega_{\bf q}^2}\,
\frac{J_{{\bf q}'}}{4 \hat \omega_{{\bf q}'}^2} 
 (\varepsilon_{{\bf k}+{\bf q}',\lambda} -\varepsilon_{{\bf k}+{\bf q}+{\bf q}',\lambda}) \,
 (\varepsilon_{{\bf k}+{\bf q},\lambda} -\varepsilon_{{\bf k}+{\bf q}+{\bf q}',\lambda})
 \nonumber \\ 
&&\times \left\{
v_{{\bf k},{\bf q}',\lambda}\Theta _{\bf q}(\lambda, \Delta \lambda) +
v_{{\bf k},{\bf q},\lambda}\Theta _{{\bf q}'}(\lambda, \Delta \lambda)
\right\} \frac{u_{{\bf k},\lambda}} {2} \nonumber \\
&&\times \left\{
n_{{\bf k}+{\bf q}'} ( n_{{\bf k}+{\bf q}+{\bf q}'} - n_{{\bf k}+{\bf q}} -D ) -
m_{{\bf k}+{\bf q}}  n_{{\bf k}+{\bf q}+{\bf q}'} 
\right\} 
 \end{eqnarray}
and 
\begin{eqnarray}
\label{52a}
v_{{\bf k},{\bf q},\lambda- \Delta \lambda} &=& v_{{\bf k},{\bf q},\lambda}
+ u_{{\bf k},\lambda} \Theta _{\bf q}(\lambda, \Delta \lambda). 
\end{eqnarray}
The quantities $n_{{\bf k}}$ and $m_{{\bf k}}$ in Eq.~\eqref{52} are 
the ${\bf k}$-dependent occupation numbers for electrons and holes per spin direction, which are 
formed with the full Hamiltonian ${\cal H}$,
\begin{eqnarray} 
\label{53}
n_{{\bf k}} = \langle \hat c_{{\bf k}\sigma}^\dagger \hat c_{{\bf k}\sigma} \rangle,
\qquad  m_{{\bf k}} = \langle \hat c_{{\bf k}\sigma} \hat c_{{\bf k}\sigma}^\dagger \rangle.
\end{eqnarray} 
In the following, we simplify the notation by suppressing  
the spin index $\sigma$ in \eqref{53}.  
The renormalization equations \eqref{52} and \eqref{52a} for $u_{{\bf    k},\lambda}^2$ 
and $v_{{\bf k},{\bf q}, \lambda}$, 
 together with the ansatz \eqref{51} for $\hat c_{{\bf k},\sigma}(\lambda)$, enable us to evaluate 
$n_{{\bf k}}$ and  $m_{{\bf k} }$ and also the ARPES spectral function. 
With some initial guess for  $ n_{{\bf k}}$ and  $ m_{{\bf k}}$, 
we start from the parameter values of the original model at $\lambda =\Lambda$,
\begin{eqnarray}
\label{54}
 u_{{\bf k}, \Lambda} &=& 1\, , \qquad
v_{{\bf k},{\bf q}, \Lambda} =0 \, ,
\end{eqnarray}
and eliminate all excitations in steps $\Delta \lambda$ from $\lambda =\Lambda$ to $\lambda =0$.
We end up with renormalized parameters which obey
\begin{eqnarray*}
u_{{\bf k}, \lambda =0} \neq 1\, , && \qquad 
v_{{\bf k},{\bf q},\lambda=0}   \neq 0.
\end{eqnarray*}
Thus, after the renormalization, the annihilation operator 
$\hat c_{\bf k}(\lambda=0) =: \hat c_{{\bf k}\sigma}^{(1)}$ at $\lambda =0$ has the final form  
 \begin{eqnarray}
\label{55}
%
\hat c_{{\bf k}\sigma}^{(1)}
&=& u_{{\bf k},\lambda=0} 
\hat c_{{\bf k}\sigma} +
\frac{1}{2N}\sum_{{\bf q k}'} v_{{\bf k}, {\bf q},\lambda=0} \, \frac{J_{\bf q}}{4 \hat \omega^2_{\bf q}} \, 
 \sum_{\alpha \beta \gamma} (\vec \sigma_{\alpha \beta}
\cdot \vec \sigma_{\sigma \gamma}) 
(\varepsilon_{{\bf k}',\lambda=0}- \varepsilon_{{\bf k}' + {\bf q},\lambda=0}) 
 \,
\hat c^\dagger_ {{\bf k}' + {\bf q} \alpha} \hat c_ {{\bf k}' \beta}
\hat c_ {{\bf k} + {\bf q} \gamma}. \nonumber
\end{eqnarray}
As was discussed before, the Hamiltonian  
after the first renormalization $\tilde{\cal H}^{(1)}$  can not directly be used to 
recalculate the expectation values $n_{\bf k}$ and $m_{\bf k}$. 
In  $\tilde{\cal H}^{(1)}$, there is still a part of the exchange present, which is, however, 
reduced by a factor $1/2$. 
Therefore, the renormalization has to be done again by starting from   
 $\tilde{\cal H}^{(1)}$ as the new initial Hamiltonian. Similarly, $ \hat c_{{\bf k}\sigma }^{(1)}$  
can be considered as the new initial annihilation operator, i.e.,
 $ \hat c_{{\bf k}\sigma }^{(1)} = \hat c_{{\bf k}\sigma }^{(1)}(\lambda= \Lambda)$,  
 with
\begin{eqnarray*}
 u_{{\bf k},\lambda=\Lambda}^{(1)} &=& u_{{\bf k},\lambda = 0},
\hspace*{0.5cm} \, \hspace*{0.5cm}
 v_{{\bf k},{\bf q},\lambda=\Lambda}^{(1)} = v_{{\bf k},{\bf q},\lambda = 0}.
\end{eqnarray*}
After $n$ renormalization cycles, the exchange is scaled down by a factor $(1/2)^n$. 
For the renormalization equation for $u_{{\bf k},\lambda}^{(n)}$ and  
$v_{{\bf k},{\bf q}, \lambda}^{(n)}$, we obtain 
\begin{eqnarray}
\label{56}
(u_{{\bf k},\lambda- \Delta \lambda}^{(n)})^2 &=& (u_{{\bf k},\lambda}^{(n)})^2 
-\frac{3}{2N^2} \sum_{{\bf q}{\bf k}'} \left(\frac{J_{\bf q}}{4 \hat \omega_{\bf q}^2}\right)^2
\Theta _{\bf q}(\lambda, \Delta \lambda)  
(\varepsilon_{{\bf k}',\lambda} -\varepsilon_{{\bf k}'+{\bf q},\lambda})^2  \\
&& \times
 \left\{ \left(\frac{u_{{\bf k},\lambda}^{(n)}}{2^n}\right)^2 + \frac{u_{{\bf k},\lambda}^{(n)}}{2^{n-1}}\, 
v_{{\bf k},{\bf q},\lambda}^{(n)}
\right\} \, \left\{ n_{{\bf k}'+{\bf q}} m_{{\bf k}'}+
n_{{\bf k}+{\bf q}} (D + n_{{\bf k}'} -n_{{\bf k}' + {\bf q}})
\right\} \nonumber \\ \nonumber\\
&+&  
\frac{3}{4N^2} \sum_{{\bf q}{\bf q}'} \frac{J_{\bf q}}{4 \hat \omega_{\bf q}^2}\,
\frac{J_{{\bf q}'}}{4 \hat \omega_{{\bf q}'}^2} 
 (\varepsilon_{{\bf k}+{\bf q}',\lambda} -\varepsilon_{{\bf k}+{\bf q}+{\bf q}',\lambda}) \,
 (\varepsilon_{{\bf k}+{\bf q},\lambda} -\varepsilon_{{\bf k}+{\bf q}+{\bf q}',\lambda}) \, 
\nonumber \\
&& \times \left\{
v_{{\bf k},{\bf q}',\lambda}^{(n)}\Theta _{\bf q}(\lambda, \Delta \lambda) +
v_{{\bf k},{\bf q},\lambda}^{(n)}\Theta _{{\bf q}'}(\lambda, \Delta \lambda)
\right\} \frac{u_{{\bf k},\lambda}^{(n)}} {2^n} \nonumber \\
&& \times  \left\{
n_{{\bf k}+{\bf q}'} ( n_{{\bf k}+{\bf q}+{\bf q}'} - n_{{\bf k}+{\bf q}} -D ) -
m_{{\bf k}+{\bf q}}  n_{{\bf k}+{\bf q}+{\bf q}'}  \nonumber
\right\}
\end{eqnarray}
and 
\begin{eqnarray}
\label{56b}
 v_{{\bf k},{\bf q},\lambda- \Delta \lambda}^{(n)} &=& 
 v_{{\bf k},{\bf q},\lambda}^{(n)}
+ \frac{u_{{\bf k},\lambda}^{(n)}}{2^n} \Theta_{\bf q}(\lambda,\Delta \lambda) .
\end{eqnarray}
Note that the factor $1/2^n$ was incorporated in   
$v_{{\bf k}, {\bf q},\sigma}^{(n)}$,  in order to keep the shape of 
the {\it ansatz} \eqref{51} unchanged, 
 \begin{eqnarray}
\label{57}
 \hat c_{{\bf k} \sigma}^{(n)}(\lambda) &=& u_{{\bf k}, \lambda}^{(n)} \hat
 c_{{\bf k}\sigma} 
+ \frac{1}{2N}\sum_{{\bf q k}'} v_{{\bf k}, {\bf q},\lambda}^{(n)} \,
 \frac{J_{\bf q}}{4 \hat \omega^2_{\bf q}} \, 
 \sum_{\alpha \beta \gamma} (\vec \sigma_{\alpha \beta}
\cdot \vec \sigma_{\sigma \gamma}) 
(\varepsilon_{{\bf k}',\lambda}- \varepsilon_{{\bf k}' + {\bf q},\lambda}) \, 
\hat c^\dagger_ {{\bf k}' + {\bf q} \alpha} \ \hat c_ {{\bf k}' \beta} \
\hat c_ {{\bf k} + {\bf q} \gamma}  . \nonumber \\
&&
\end{eqnarray} 
For $n\rightarrow\infty$, we arrive at the fully renormalized operator
\begin{eqnarray}
\label{57a}
 \hat c_{{\bf k} \sigma}^{(n\rightarrow \infty)}(\lambda=0) &=& \tilde u_{{\bf
     k}} \hat c_{{\bf k}\sigma} 
+ \frac{1}{2N}\sum_{{\bf q k}'} \tilde{v}_{{\bf k}, {\bf q}} \, \frac{J_{\bf q}}{4 \hat \omega^2_{\bf q}} \, 
 \sum_{\alpha \beta \gamma} (\vec \sigma_{\alpha \beta}
\cdot \vec \sigma_{\sigma \gamma}) 
(\tilde{\varepsilon}_{{\bf k}'}- \tilde{\varepsilon}_{{\bf k}' + {\bf q}}) \, 
\hat c^\dagger_ {{\bf k}' + {\bf q} \alpha} \ \hat c_ {{\bf k}' \beta} \
\hat c_ {{\bf k} + {\bf q} \gamma} , \nonumber \\
&&
\end{eqnarray} 
where $ \tilde u_{\bf k}=
u_{{\bf k},\lambda = 0}^{(n \rightarrow \infty)}$ and
$ \tilde{v}_{{\bf k},{\bf q}}=
v_{{\bf k}, {\bf q}, \lambda = 0}^{(n \rightarrow \infty)}$.
Using $\tilde {\cal H}$,
the expectation values $n_{{\bf k}}$ and $m_{{\bf k}}$  as well as the spectral function 
$\Im G({\bf k},\omega)$ can be evaluated. However, due to the strong correlations in $\tilde{\cal H}$,
additional approximations will still be necessary. 

To evaluate the spectral function $\Im G({\bf k},\omega)$, we start from 
Eq.~\eqref{50} for $n\rightarrow \infty$, $\lambda = 0$
\begin{eqnarray} 
\label{58a}
 {\Im  G}({\bf k}, \omega)  &=&  
\big< [\hat c_{{\bf k}\sigma }^{(n \rightarrow \infty)\dagger}(\lambda=0) ,\, \delta (\tilde{\sf L} + \omega) \;
\hat c_{{\bf k}\sigma}^{(n\rightarrow \infty)}(\lambda=0)]_+ \big>_{\tilde{\cal H}}.
\end{eqnarray}
Here $\hat c_{{\bf k} \sigma}^{(n \rightarrow \infty)}(\lambda\rightarrow 0)$
is given by Eq.~\eqref{57a}. The time dependence and the expectation value are defined with  $\tilde{\cal H}$, and
$\tilde{\sf L}$ is the Liouville operator to  $\tilde{\cal H}$.
For a state close to half-filling, the following relation is approximately 
valid according to Appendix \ref{A}: 
 \begin{eqnarray}
\label{58b}
\tilde{\sf L} \hat c_{{\bf k} \sigma} &=& \left[\tilde{\cal H},\hat
  c_{{\bf k} \sigma}\right] = -\tilde \varepsilon_{\bf k} \,\hat c_{{\bf k} \sigma}.
\end{eqnarray}
It means, in the case that the dynamics is governed by the Hamiltonian
$\tilde{\cal H}$,  in which no magnetic interactions are present, 
a hole can move almost freely through the lattice. Using Eqs.~\eqref{57a} and \eqref{58a},
the spectral function $\Im G({\bf k},\omega)$ then reads   
\begin{eqnarray}
\label{59}
&&\Im G({\bf k},\omega) =
\tilde u_{\bf k}^2 D \, \delta(\omega - \tilde \varepsilon_{\bf k}) +
\nonumber \\
&& \quad +
\frac{3D}{2N^2} \sum_{{\bf q}{\bf q}'}
\left[
\left( \frac{J_{\bf q}\tilde v_{{\bf k},{\bf q}}}{4 \hat \omega_{\bf q}^2}  \right)^2
 ( \tilde{\varepsilon}_{{\bf k}+{\bf q}'} - \tilde{\varepsilon}_{{\bf k}+{\bf q}+{\bf q}'} 
)^2 \right. \nonumber \\
&& \quad \times \left\{ \tilde n_{{\bf k}+{\bf q}+{\bf q}'} \tilde m_{{\bf k}+{\bf q}'} +
 \tilde n_{{\bf k}+{\bf q}} (D+ \tilde n_{{\bf k}+{\bf q}'} 
- \tilde n_{{\bf k}+{\bf q}+{\bf q}'} ) \right\} 
  \\ 
&& \quad - \frac{1}{2}\frac{J_{\bf q}}{4 \hat \omega_{\bf q}^2}
\frac{J_{{\bf q}'}}{4 \hat \omega_{{\bf q}'}^2} \,
\tilde v_{{\bf k},{\bf q}}\, \tilde v_{{\bf k},{\bf q}'}\,  
( \tilde{\varepsilon}_{{\bf k}+{\bf q}'} 
- \tilde{\varepsilon}_{{\bf k}+{\bf q}+{\bf q}'} )
( \tilde{\varepsilon}_{{\bf k}+{\bf q}} 
- \tilde{\varepsilon}_{{\bf k}+{\bf q}+{\bf q}'} ) \nonumber \\
&& \quad \times \left\{ (\tilde n_{{\bf k}+{\bf q}'} -  \tilde m_{{\bf k}+{\bf q}})
\tilde  n_{{\bf k}+{\bf q}+{\bf q}'} 
- \tilde n_{{\bf k}+{\bf q}'} ( \tilde n_{{\bf k}+{\bf q}} + D)  \right\} \Bigg]  
\delta ( \omega + \tilde \varepsilon_{{\bf k}+{\bf q}+{\bf q}'} -
\tilde \varepsilon_{{\bf k}+{\bf q}'} - \tilde \varepsilon_{{\bf k}+{\bf q}} 
) . \nonumber
 \end{eqnarray} 
Note that in deriving Eq.~\eqref{59}, an additional factorization approximation was used. Thereby, an expectation 
value, formed with  six fermion operators, was replaced by a product of three two-fermion 
expectation values.  The new quantities $\tilde n_{\bf k}$ and $\tilde m_{\bf k}$  in Eq.~\eqref{59},
 \begin{eqnarray*} 
\tilde n_{\bf k} = \langle \hat c_{{\bf k}\sigma}^\dagger \hat c_{{\bf k}\sigma} 
\rangle_{\tilde{\cal H}},
\qquad  \tilde m_{\bf k} = \langle \hat c_{{\bf k}\sigma} \hat c_{{\bf k}\sigma}^\dagger 
\rangle_{\tilde {\cal H}}
\end{eqnarray*} 
are again ${\bf k}$-dependent occupation numbers for electrons and holes per spin direction, 
However, they are defined with the fully renormalized model $\tilde{\cal H}$ 
instead of with ${\cal H}$ as in Eqs.~\eqref{53}.
For $\tilde n_{\bf k}$ and $\tilde m_{\bf k}$, we use the 
Gutzwiller approximation \cite{GW}  
 \begin{eqnarray}
\label{60}
 \tilde n_{{\bf k}} &=& (D-q)  + q  \, f(\tilde \varepsilon_{\bf k}),   \\
\tilde m_{\bf k} &=& q  \, (1- f(\tilde \varepsilon_{\bf k}))  \quad 
\mbox{with} \quad q = \frac{1-n}{1-n/2} = \frac{\delta}{1-n/2} \, ,
\nonumber
\end{eqnarray}
where $f(\tilde \varepsilon_{\bf k})$ is the 
Fermi function,  $f(\tilde \varepsilon_{\bf k})= 
\Theta( - \tilde{\varepsilon}_{\bf k})$ for $T=0$. 
Note that 
$\tilde m_{{\bf k} }$ is proportional to the hole 
filling $\delta = 1 - n$. Obviously, the application of $\hat c_{{\bf k}\sigma}^\dagger$
on a Hilbert space vector is non-zero only when holes are present. 
In contrast,  $\tilde n_{{\bf k} \sigma}$ does not vanish even at half-filling.

According to \eqref{59}, the spectral function 
$\Im G({\bf k},\omega)$ consists of two parts: The first one is a 
coherent excitation of energy $\tilde \varepsilon_{\bf k}$ with the
weight $\tilde u_{\bf k}^2 D$.  The second part describes three-particle 
excitations. Also note that the sum rule 
\begin{eqnarray}
 \label{61} 
\int_{-\infty}^\infty d\omega \, \Im G({\bf k},\omega) = 
\langle [\hat c_{{\bf k}\sigma}^\dagger, \hat c_{{\bf k}\sigma}]_+ \rangle =
1 - \frac{n}{2} =D \ \ 
\end{eqnarray}
is automatically fulfilled by \eqref{59}. The sum rule is built in by the 
construction of the renormalization equations for $u_{{\bf k},\lambda}$ and
$v_{{\bf k},{\bf q}, \lambda}$ in Appendix \ref{B}.

For finite temperature, a phenomenological extension of the Gutzwiller approximation
according to \cite{SGRUV} will later be used. Here, the Fermi function is replaced by 
\begin{eqnarray}
\label{62}
 f(\tilde\varepsilon_{\bf k}) &=& \frac{1}
{1+   \exp{ [ \beta  q  \tilde{\varepsilon}_{\bf k}
/w({\bf k},n) ] } } \, ,
\end{eqnarray}
where $w({\bf k}, n)$ is a weighting function in ${\bf k}$-space. It was introduced 
in \cite{SGRUV} in order to account for an over-completeness in the Gutzwiller approximation. 
It plays the role of a ${\bf k}$-dependent effective mass and is a quantity of order 1.

Finally, note that the static expectation values $n_{\bf k}$ and $m_{\bf k}$, defined in Eq.~\eqref{53}, 
can also be evaluated from $A({\bf k},\omega)$ or $\Im G({\bf k},\omega)$:
\begin{eqnarray}
 \label{63}
n_{\bf k}&=& \int_{-\infty}^\infty A({\bf k},\omega)\, d\omega =
  \int_{-\infty}^\infty \frac{1}{1+ e^{\beta \omega}}\,  \Im G({\bf k},\omega)\, d\omega
\, , \qquad  m_{\bf k} = D -n_{\bf k}.
\end{eqnarray}

\section{Numerical evaluation for the pseudogap phase}
\label{NS}

The renormalization equations \eqref{35}, \eqref{45}, \eqref{56} and
  \eqref{56b} together with \eqref{63} form a closed system of equations,
  which  could
  be solved self-consistently. 
  However, to simplify the numerical evaluation, we calculate the expectation values  
  in Eq.~\eqref{35} and Eq.~\eqref{45}
with the renormalized Hamiltonian $ \tilde{\cal H}$ instead of with 
${\cal H}$. Within this approximation and the 
  Gutzwiller approximation \eqref{60}, 
  the renormalization equation for the energy $\varepsilon_{{\bf k},\lambda}$ 
  reads
\begin{eqnarray}
\label{63b}
 \varepsilon_{{\bf k},\lambda - \Delta \lambda}- \varepsilon_{{\bf k},\lambda} &=&
\frac{1}{16 N}\sum_{\bf q} \frac{J_{\bf q}^2}{\hat \omega_{{\bf q}, \lambda}^4}\,
\Theta_{\bf q}(\lambda, \Delta \lambda) \,
(\varepsilon_{{\bf k}+ {\bf q}, \lambda} +  \varepsilon_{{\bf k}- {\bf q}, \lambda}
-2 \varepsilon_{{\bf k}, \lambda} )\, \langle \dot {\bf S}_{\bf q} \, 
\dot {\bf S}_{-{\bf q}} \rangle \nonumber \\
&& + 
\frac{3q^2}{8N} \sum_{\bf q} \frac{J_{\bf q}^2}{\hat \omega_{{\bf q},\lambda}^4} \,
\Theta_{\bf q}(\lambda,\Delta \lambda)\,
\left[ \frac{1}{N} \sum_{{\bf k}'}(2\varepsilon_{{\bf k}',\lambda} -
\varepsilon_{{\bf k}'+{\bf q},\lambda} -\varepsilon_{{\bf k}'-{\bf q},\lambda}) 
 f_{{\bf k}'}^{(NL)}
\right]  \nonumber \\
&&  \times  (\varepsilon_{{\bf k},\lambda}- \varepsilon_{{\bf k}- {\bf q},\lambda })^2 \,
f_{{\bf k}-{\bf q}}^{(NL)} ,
\end{eqnarray}
with
\begin{eqnarray*}
\langle \dot {\bf S}_{\bf q} \, 
\dot {\bf S}_{-{\bf q}} \rangle  &=& -\frac{3q^2}{2} \, \frac{1}{N}\sum_{{\bf k}'} 
(\tilde{\varepsilon}_{{\bf k}'} - \tilde{\varepsilon}_{{\bf k}'+ {\bf q}})^2 \,
   f_{{\bf k}'}^{(NL)}  \,  f_{{\bf k}'+{\bf q}}^{(NL)}.
 \nonumber 
\end{eqnarray*}
Here, $f_{\bf k}^{(NL)}$ is the non-local part 
of the Fermi distribution,  
  $f_{\bf k}^{(NL)} = 1/(1+e^{\beta \tilde{\varepsilon}_{\bf k}}) -
 (1/N)\sum_{\bf  k}  1/(1+e^{\beta \tilde{\varepsilon}_{\bf k}})$. 
Remember that the factor $q$ as well as $\hat{\omega}_{{\bf q}, \lambda}^2$  are proportional to 
the hole concentration $\delta = 1-n$. Therefore, the renormalization contributions to 
Eq.~\eqref{63b} are almost independent of $\delta$ and turn out to be very small.
Therefore, from now on, the $\lambda$ dependence of $\varepsilon_{{\bf k},\lambda}$ 
and also of $\hat{\omega}_{{\bf q}, \lambda}$ will be neglected.

\subsection{Zero temperature results}

For the evaluation of the renormalization scheme, 
we have used a sufficiently large number 
of renormalization cycles in order to obtain self-consistency. We have considered a square 
lattice with $N=40 \times 40$ sites and a
moderate hole doping, such that the  system is outside 
the anti-ferromagnetic phase but not yet in the Fermi-liquid phase. 
Possible superconducting solutions are not considered.

The main feature of the normal state is the appearance of a pseudogap which is
experimentally observed in ARPES measurements. A small next-nearest neighbor 
hopping $t' = 0.1 t$ and an exchange constant $J= 0.2 t$ 
between nearest neighbors
are assumed.  The inclusion of a non-zero $t'$ leads to a Fermi surface (FS), as sketched in
the inset of Fig.~\ref{FS}. It closely resembles the Fermi surface of non-interacting electrons.  
The FS is determined from the condition 
$\tilde{\varepsilon}_{\bf k}=0$ for a fixed value of the electron filling $n=1-\delta$.
The temperature is set equal to $T=0$.  
Let us first concentrate on the $\omega$-dependence of the  
spectral function $\Im G({\bf k},\omega)$. In all figures, 
the symmetrized function will be plotted in order to remove the 
effects of the Fermi function
on the spectra.

\begin{figure}
  \begin{center}
    \scalebox{0.7}{
      \includegraphics*{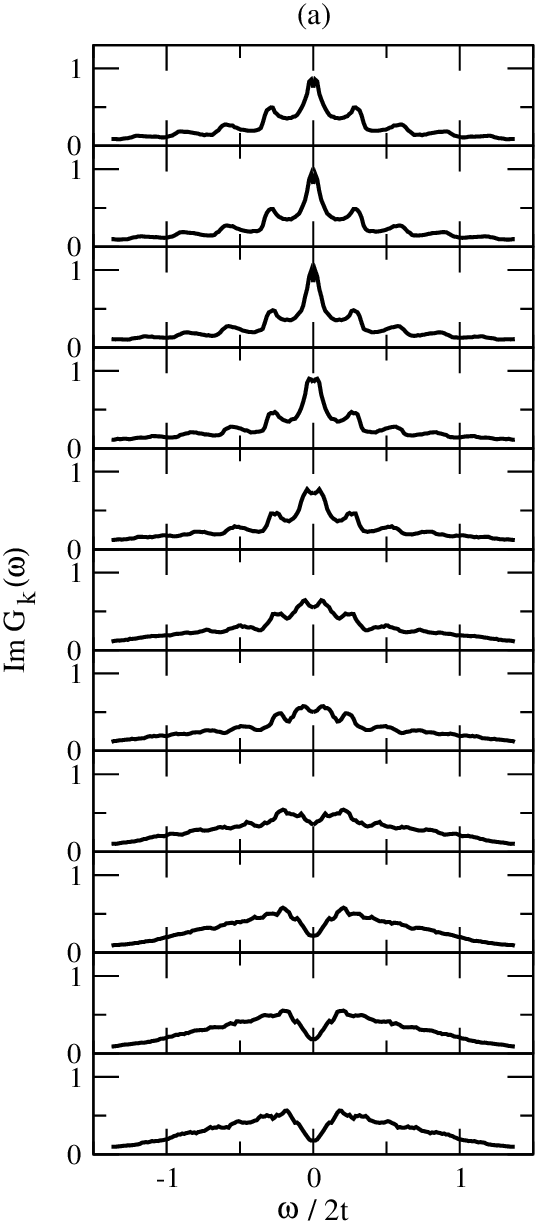}
      \includegraphics*{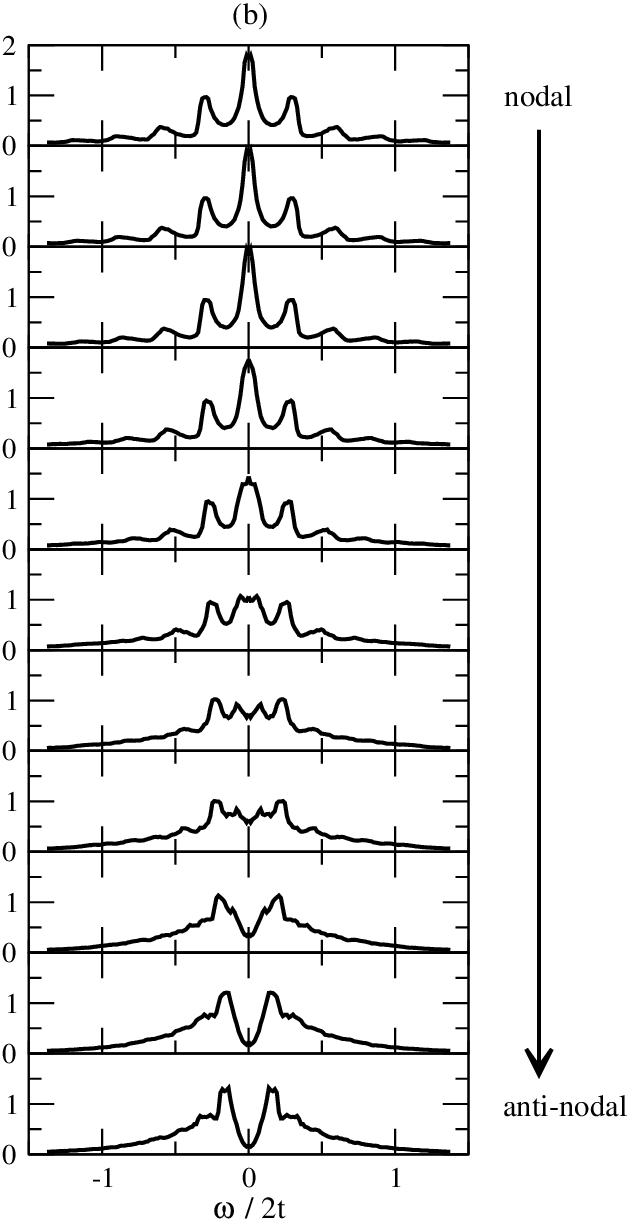}
    }
  \end{center}
  \caption{
   Symmetrized spectral function $\Im G({\bf k},\omega)$ at $T=0$ for two hole fillings
(a) $\delta=0.03$ and (b) $\delta= 0.075$
along the Fermi surface. The top $\Im G({\bf k},\omega)$
is at the node, whereas the bottom is at the anti-node, as defined in the inset of
Fig.~\ref{FS}. 
  }
  \label{Fig_1}
  \end{figure}

\begin{figure}
  \begin{center}
    \scalebox{0.55}{
      \includegraphics*{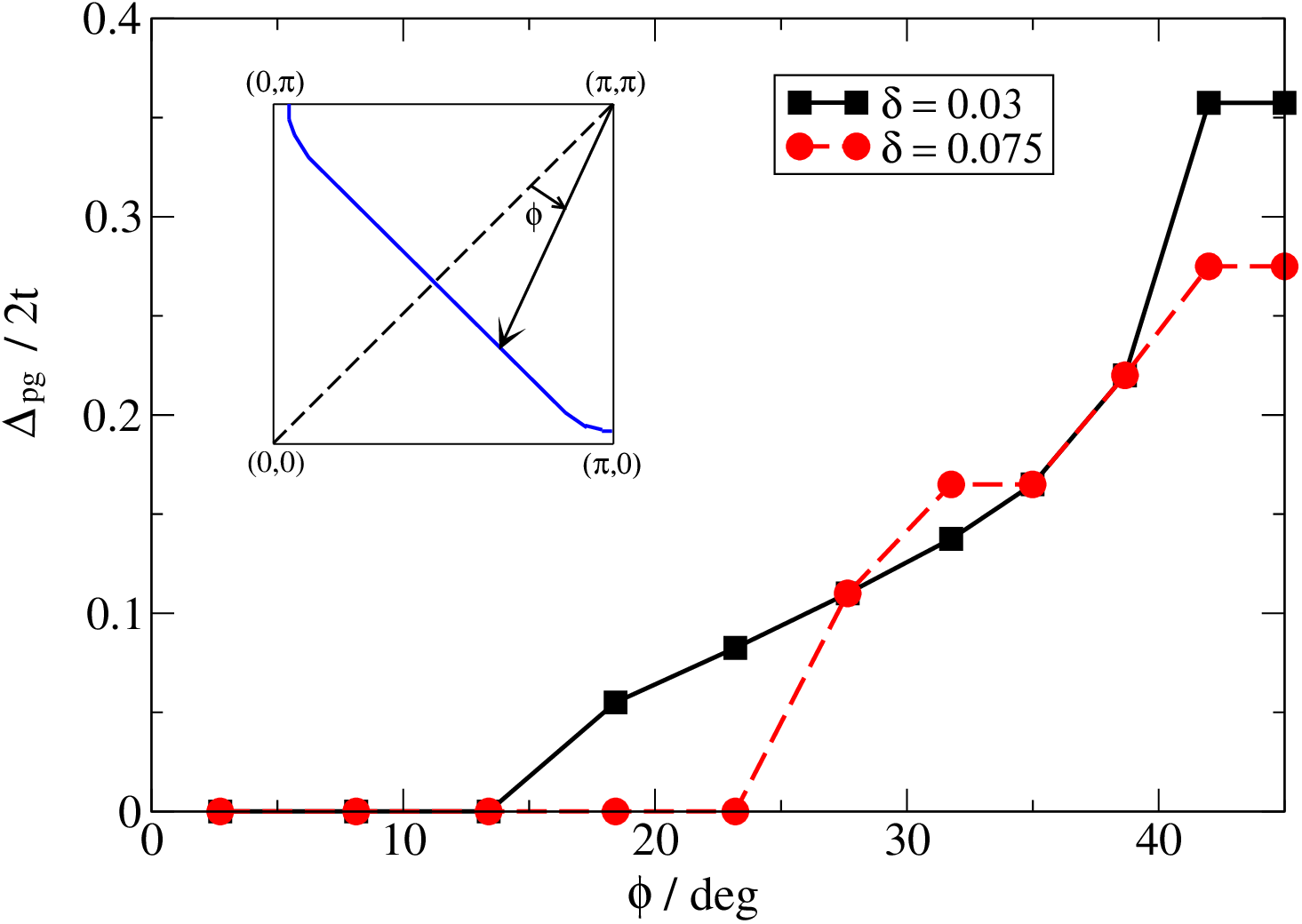}
    }
  \end{center}
  \caption{
Pseudogap size $\Delta_{pg}$ from 
  Fig.~\ref{Fig_1} as function of the Fermi surface angle $\phi$ for 
  $\delta = 0.03$ (black) and $\delta = 0.075$ (red).
  }
  \label{FS}
  \end{figure}
Fig.~\ref{Fig_1} shows the PRM result for 
$\Im G({\bf k},\omega)$  for two different hole concentrations 
in the underdoped regime (a) $\delta=0.03$  and (b) $\delta=0.075$,  for several 
$\bf k$-values on the FS between the nodal point near $(\pi/2, \pi/2)$ and the anti-nodal
near $(\pi, 0)$.  As the most important finding, one recognizes 
the opening of a pseudogap for both hole concentrations, when  
one proceeds from the nodal towards the anti-nodal direction. On a substantial part 
of the FS, the spectra show a peak-like behavior around $\omega =0$, indicating a 
Fermi arc of gapless excitations. 
Note that our analytical results show a remarkable agreement 
with findings from ARPES experiments in high-temperature superconductors
\cite{K06,TE07,K07,K08}.
Also additional peaks are found  
in the nodal direction  at lower binding energies 
which are enhanced for $\delta=0.075$.
In Fig.~\ref{FS}, the pseudogap on the FS
is shown as a function of the angle $\phi$, where $\phi$ is
defined in the inset of Fig.~\ref{FS}. The results are taken from Figs.~\ref{Fig_1}(a) and (b). 
Note that for the smaller hole filling, the length of the Fermi arc 
becomes smaller, whereas the pseudogap becomes larger. 
This behavior agrees with 
the known experimental feature of a characteristic pseudogap temperature $T^*$
which increases with decreasing hole filling\cite{K06},\cite{TL01}.

The $\omega$- and ${\bf k}$-dependence of $\Im G({\bf k},\omega)$ from 
Fig.~\ref{Fig_1} can easily be understood from equation \eqref{59},
\begin{eqnarray}
\label{63d}
\Im G({\bf k},\omega) &=&
|\tilde u_{\bf k}|^2 D \, \delta(\omega - \varepsilon_{\bf k}) \nonumber  \\ 
&+&
\frac{3D}{2N^2} \sum_{{\bf q}{\bf k}'}
\left\{
\left( \frac{J_{\bf q}}{4 \hat \omega_{\bf q}^2}\right)^2 |\tilde v_{{\bf k},{\bf q}} |^2
\, \big(  \varepsilon_{{\bf k}'} - 
 \varepsilon_{{\bf k}'+{\bf q}} \big)^2 \big( \tilde n_{{\bf k}'+{\bf q}} \tilde m_{{\bf k}'} +
 \tilde n_{{\bf k}+{\bf q}} (D+ \tilde n_{{\bf k}'} - 
\tilde n_{{\bf k}'+{\bf q}} ) \big) \right.
 \nonumber \\
&& \nonumber \\
&& \qquad + \cdots 
\Bigg\}  \, \delta \left( \omega +  \varepsilon_{{\bf k}'+{\bf q}} -
 \varepsilon_{{\bf k}'} -  \varepsilon_{{\bf k}+{\bf q}} 
\right) ,
 \end{eqnarray} 
where the dots $ + \cdots$ indicate additional terms which are less important. 
First, from the renormalization equation \eqref{52} for $u_{{\bf k},\lambda}^2$, one finds
that its original value $u_{{\bf k}}^2=1$ at $\lambda=\Lambda$ is  reduced
by renormalization contributions of order $\delta^{-2}$
according to $u_{{\bf k},\lambda -\Delta \lambda}^2- u_{{\bf k},\lambda}^2 = - \alpha_\lambda /\delta^2$. 
Thus, the weight of the coherent 
excitation $|\tilde u_{{\bf k}}|^2$ becomes small for small $\delta$, so that 
the spectral function $\Im G({\bf k},\omega)$ is dominated by the incoherent excitations in Eq.~\eqref{63d}.  
What remains is to show that the different behavior of 
$\Im G({\bf k},\omega)$ in the nodal and in the anti-nodal region can be 
understood solely from the incoherent part of Eq.~\eqref{63d}: 

First note that the dominant contribution in Eq.~\eqref{63d} at small $\omega$
arises from the small ${\bf q}$-terms in the sum over ${\bf q}$, since  in the denominator
$\hat  \omega_{\bf q}^2 \sim {\bf q}^2$. In the numerator, 
 the factor $(\varepsilon_{{\bf k}'} - \varepsilon_{{\bf k}'+{\bf q}} )^2$ is also 
proportional to 
$q^2$, so that the combined prefactor $(J_{\bf q}/ 4 \hat  \omega_{\bf q}^2 )^2
(\varepsilon_{{\bf k}'}- \varepsilon_{{\bf k}'+{\bf q}})^2$ behaves as $\sim {\bf q}^{-2}$. 
However, the small ${\bf q}$ terms do not lead to a divergency in Eq.~\eqref{63d} since the  
additional renormalization parameter  $\tilde v_{{\bf k},{\bf q}}^2$ also vanishes 
for ${\bf q} \rightarrow 0$. This behavior
can be verified by a close inspection of the renormalization equations
\eqref{52}, \eqref{52a} 
for $u_{{\bf k},\lambda}$ and $v_{{\bf k},{\bf q},\lambda}$.  
Next, let us use the small ${\bf q}$ expansion for the energy difference 
\begin{eqnarray}
\label{63e}
 \varepsilon_{{\bf k}'} - \varepsilon_{{\bf k}'+{\bf q}} = -2t \big( q_{x}
 \sin {k}_{x}' + q_{y} \sin k_{y}' \big) .
\end{eqnarray}  
The excitations from the $\delta$-function in Eq.~\eqref{63d} are given by 
\begin{eqnarray}
\label{63f}
\omega = \varepsilon_{{\bf k}'} - \varepsilon_{{\bf k}'+{\bf q}} + \varepsilon_{{\bf k}+{\bf q}}
\approx \varepsilon_{\bf k} + 2t \left\{ q_x \, (\sin k_x -\sin k_x') + q_y  \,( \sin k_y - \sin k_y')\right\},
\end{eqnarray}
which still depend on ${\bf k}'$. There is also a ${\bf k}'$-dependent factor in the numerator 
which contributes to the intensity, 
\begin{eqnarray}
\label{63g}
 (\varepsilon_{{\bf k}'} - \varepsilon_{{\bf k}'+{\bf q}})^2 = 4t^2 \big( q_{x}
 \sin {k}_{x}' + q_{y} \sin k_{y}' \big)^2  +\mathcal{O}(q^4).
\end{eqnarray}  
Now, we are able to discuss the small $\omega$-behavior 
of the spectral function $\Im G({\bf k},\omega)$, when the wave vector ${\bf k}$ is varied:

\noindent
(i) First, close to the anti-nodal point ${\bf k} = (0,\pi)$, the excitation energy \eqref{63f} 
reduces to 
\begin{eqnarray}
\label{63h}
\omega = \varepsilon_{{\bf k}'} - \varepsilon_{{\bf k}'+{\bf q}} + \varepsilon_{{\bf k}+{\bf q}}
\approx \varepsilon_{\bf k} - 2t \left( q_x \, \sin k_x' + q_y  \, \sin k_y'\right).
\end{eqnarray}
By comparing Eq.~\eqref{63h} with Eq.~\eqref{63g}, 
one realizes that the square of the frequency shift in Eq.~\eqref{63h} is
identical to the intensity factor \eqref{63g}. Thus, excitations with small shifts 
away from the Fermi surface $\varepsilon_{\bf k}=0$ also have small intensities, whereas 
those with large shifts have large intensities. This explains naturally 
the pseudogap behavior at the anti-nodal point, where a lack of intensity 
is found at $\omega=0$.

\noindent
(ii) For the nodal point near ${\bf k}=(\pi/2, \pi/2)$, the excitations have energies
\begin{eqnarray}
\label{63i}
\omega = \varepsilon_{{\bf k}'} - \varepsilon_{{\bf k}'+{\bf q}} + \varepsilon_{{\bf k}+{\bf q}}
\approx \varepsilon_{\bf k} + 2t \left\{ q_x \, (1 -\sin k_x') + q_y  \,( 1 - \sin k_y')\right\} \, ,
\end{eqnarray}
whereas the intensity factor  is again given by Eq.~\eqref{63g}. The largest 
intensity is caused by terms in the sum over ${\bf k}'$ which either belong to the region around 
${\bf k}'\approx (\pi/2, \pi/2)$ or around
${\bf k}'\approx (-\pi/2, -\pi/2)$. 
In the first case, the excitations \eqref{63i} reduces to 
$\omega \approx \varepsilon_{\bf k}$, whereas the intensity factor \eqref{63g} is given by 
$4t^2 (q_x +q_y)^2$. Thus, from 
this ${\bf k}'$-region, one obtains excitations directly at the Fermi surface. 
For the second ${\bf k}'$-region, the
excitation energies are given by  $\omega \approx \varepsilon_{\bf k} + 4t (q_x + q_y)$.
The intensity factor is the same as before. Thus, similar to the anti-nodal point,
the square of the excitation shift away from the Fermi surface $\varepsilon_{\bf k}=0$  
is proportional to the corresponding intensity. 
Therefore, from these ${\bf k}'$-terms no intensity is expected at  $\omega =0$.   
To summarize, an excitation peak at $\omega =0$ is expected 
for wave vectors ${\bf k}$ at the anti-nodal point
from the first ${\bf k}'$-regime, discussed above. 
In contrast, for wave vector ${\bf k}$ at the anti-nodal point 
a pseudogap arises. This explains the pseudogap behavior of the ARPES 
spectral function and leads to an understanding of the spectra 
of Fig.~\ref{Fig_1}.
\begin{figure}
  \begin{center}
    \scalebox{0.7}{
      \includegraphics*{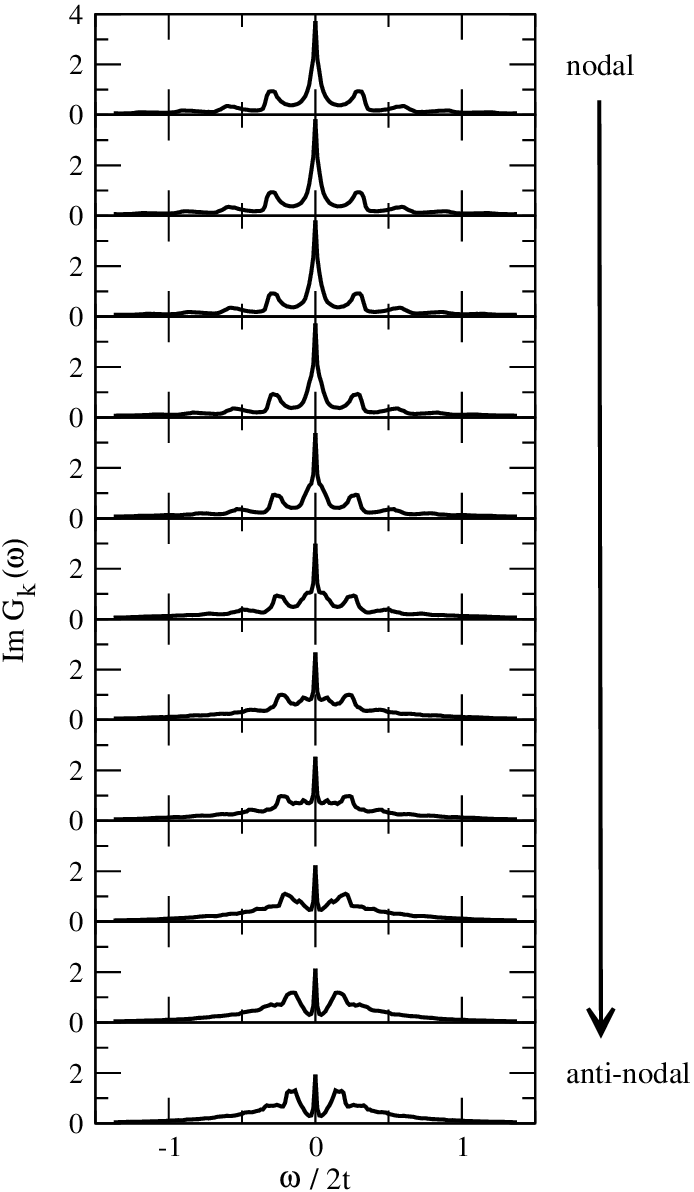}
    }
  \end{center}
  \caption{
Same quantity as in Figs.~\ref{Fig_1}(a) or (b) for a larger hole doping of 
$\delta= 0.09$.  
  }
  \label{Fig_3}
  \end{figure}
%
In  Fig.~\ref{Fig_3}, the spectral function is plotted
for a larger hole concentration $\delta= 0.09$.  The remarkable new feature is 
the occurrence of a narrow coherent excitation at $\omega =0$. Note that for this hole concentration,  
the weight $D |\tilde u_{{\bf k}}|^2$ of the coherent excitation is 
no longer negligible as in the preceding cases since the renormalization contributions 
$\sim  1/\delta^2$ to $u_{{\bf k},\lambda}^2$ are less important for larger $\delta$. 
By  increasing $\delta$, the coherent peak gains weight at the expense of the 
incoherent excitations. 
We also expect a broadening of the coherent peak due to a coupling to other degrees of 
freedom such as phonons or impurities. 

\begin{figure}
  \begin{center}
    \scalebox{0.65}{
      \includegraphics*{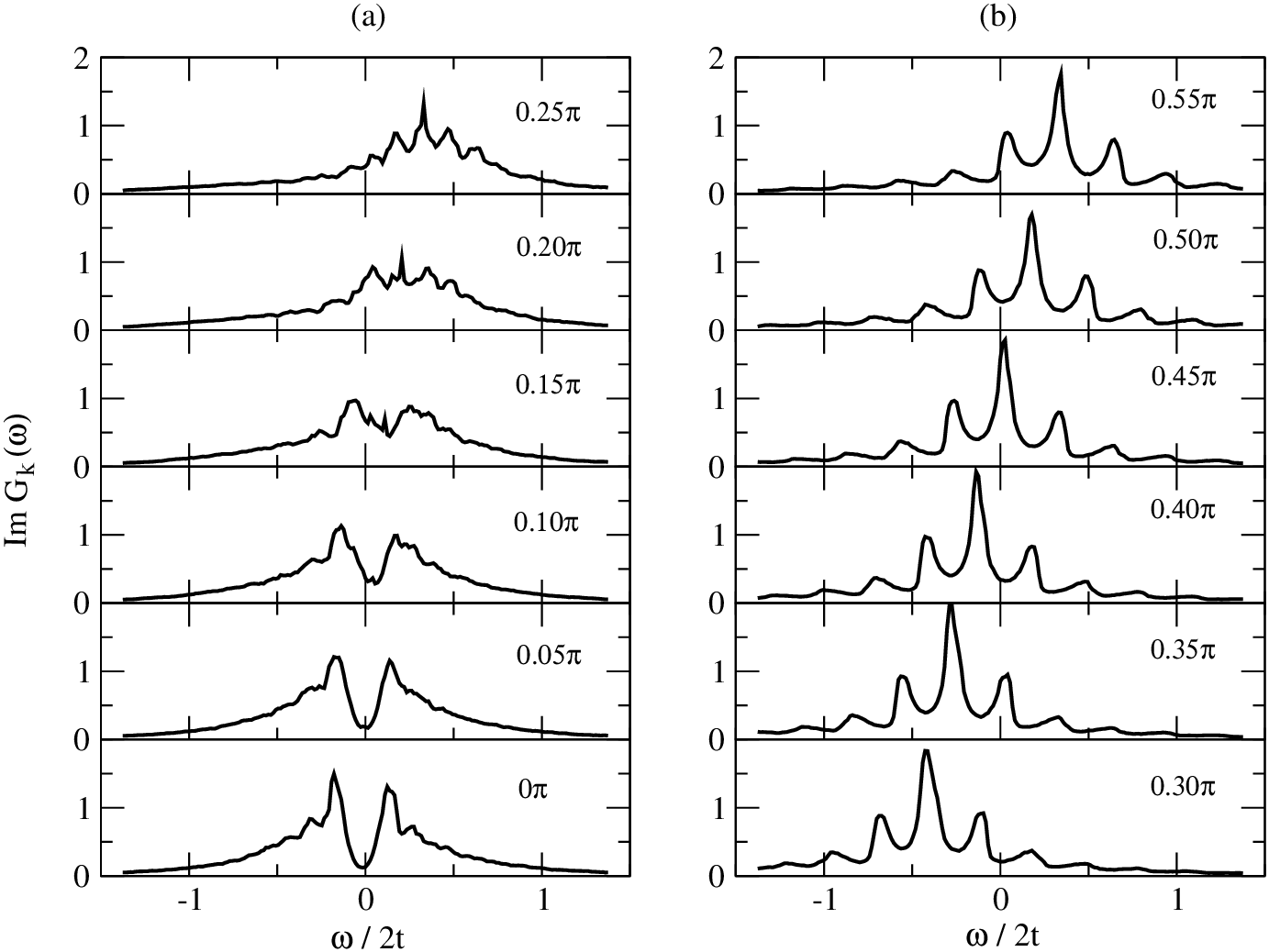}
    }
  \end{center}
  \caption{Spectral functions $\Im G({\bf k},\omega)$ for two fixed 
$k_x$ values: (a) $k_x= \pi$ and (b) $k_x= \pi/2$ and different values of $k_y$, thereby 
crossing the Fermi surface. The hole filling $\delta=0.075$ is the same as in Fig.~\ref{Fig_1}(b).
}
  \label{Fig_4}
  \end{figure}
In Figs.~\ref{Fig_4}(a) and (b), the spectral functions are 
shown 
for two different cuts in the Brillouin zone.  
In both figures, $k_x$ is  fixed and $k_y$ is varied thereby crossing the FS. 
In panel (a), where $k_x=\pi$, the cut runs along the anti-nodal region
through the FS at ${\bf k}_F\approx (\pi, 0.07 \pi)$. Note that the 
pseudogap is restricted to a small $\bf k$-range around the anti-nodal point. It 
disappears for larger $k_y$ values away from the anti-nodal point, 
in agreement with the earlier discussion on the origin of the pseudogap.
The spectra along a cut in the nodal region are shown in panel (b), where
$k_x = \pi/2$. Apart from the dominant excitation which corresponds to the gapless
excitation on the FS in Fig.~\ref{Fig_1}, 
also weaker excitations are found at lower binding energies. The complete 
peak structure is shifted almost unchanged through the FS, when $k_y$ is varied. 
The energy distance between the primary 
and the secondary peak slowly decreases by proceeding along the FS from 
the nodal point to the anti-nodal point, until finally both peaks disappear 
when the anti-nodal region is reached.
Such a double-peak structure with the same properties along the FS 
was observed in underdoped cuprate superconductors \cite{CSG08}.  
Finally, one point might still be worth mentioning. For fixed $\omega$,
the spectrum in ${\bf k}$-space is much broader than what one would expect for 
free electrons. Thus, the electron 
occupation $\langle \hat c_{{\bf k}\sigma}^\dagger 
\hat c_{{\bf k}\sigma} \rangle= \int d\omega
(1+e^{\beta \omega})^{-1} \Im G({\bf k},\omega)$ depends only weakly on ${\bf k}$. This feature is 
consistent with the former expression \eqref{60} for $\tilde n_{\bf k}$, where the 
Gutzwiller approximation was used. Remember
 that the expectation value $\tilde n_{\bf k}$  
was defined with the renormalized 
Hamiltonian $\tilde {\cal H}$ and not with ${\cal H}$.

\subsection{Finite temperature results}
\label{NSfiniteT}
Next, we discuss the influence of the temperature on the 
one-particle spectra in the normal state. For the hopping to next nearest 
neighbors, we use a somewhat larger value $t' = 0.4t$. This leads
to an enhanced curvature of the Fermi surface, as it is observed 
in most of the copper oxides superconductors. The other parameters remain unchanged. 
%
\begin{figure}
  \begin{center}
    \scalebox{0.57}{
      \includegraphics*{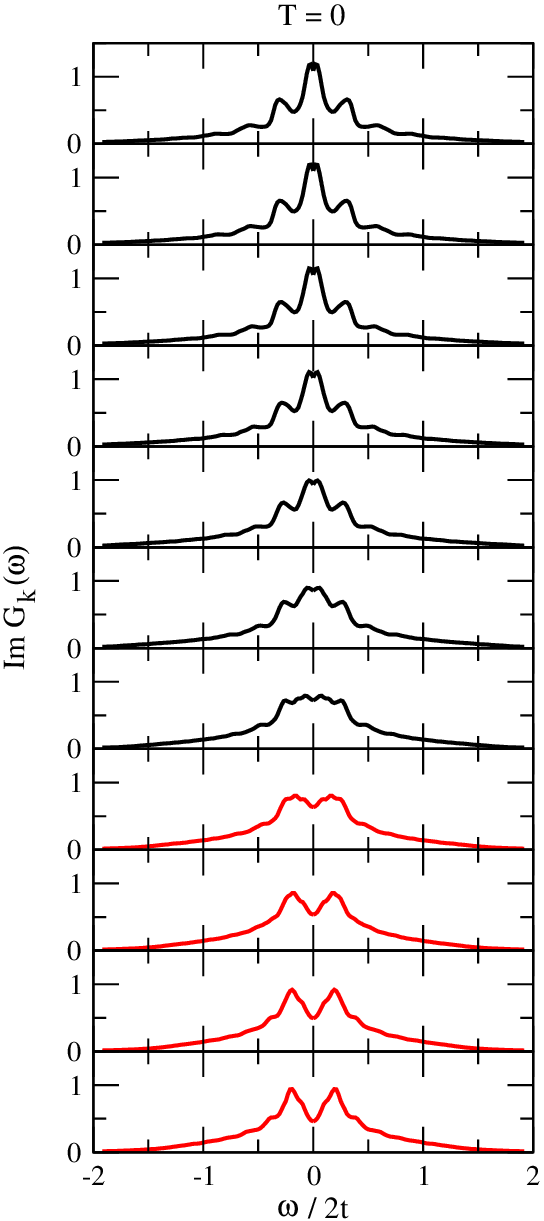}
      \includegraphics*{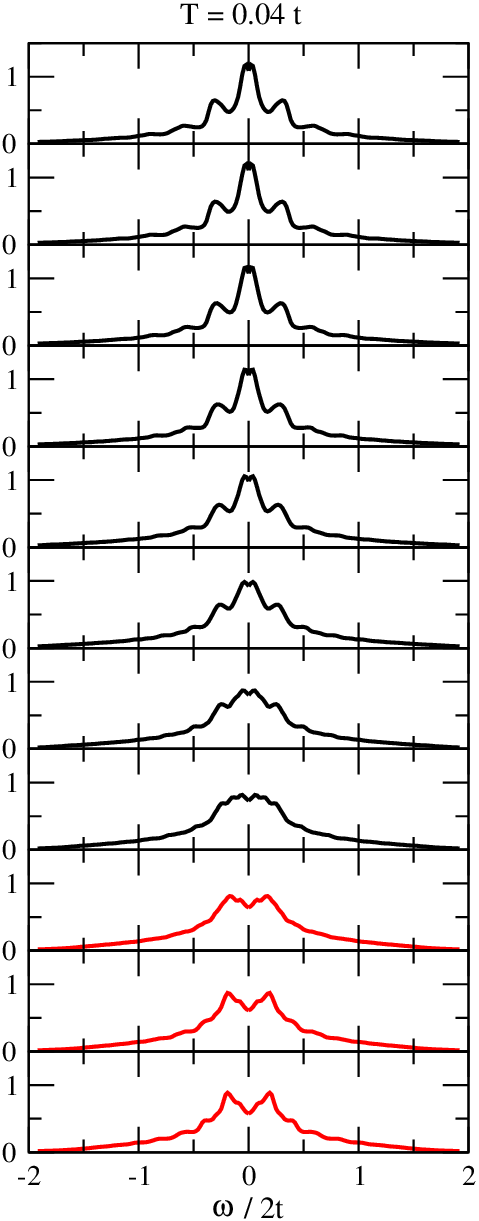}
      \includegraphics*{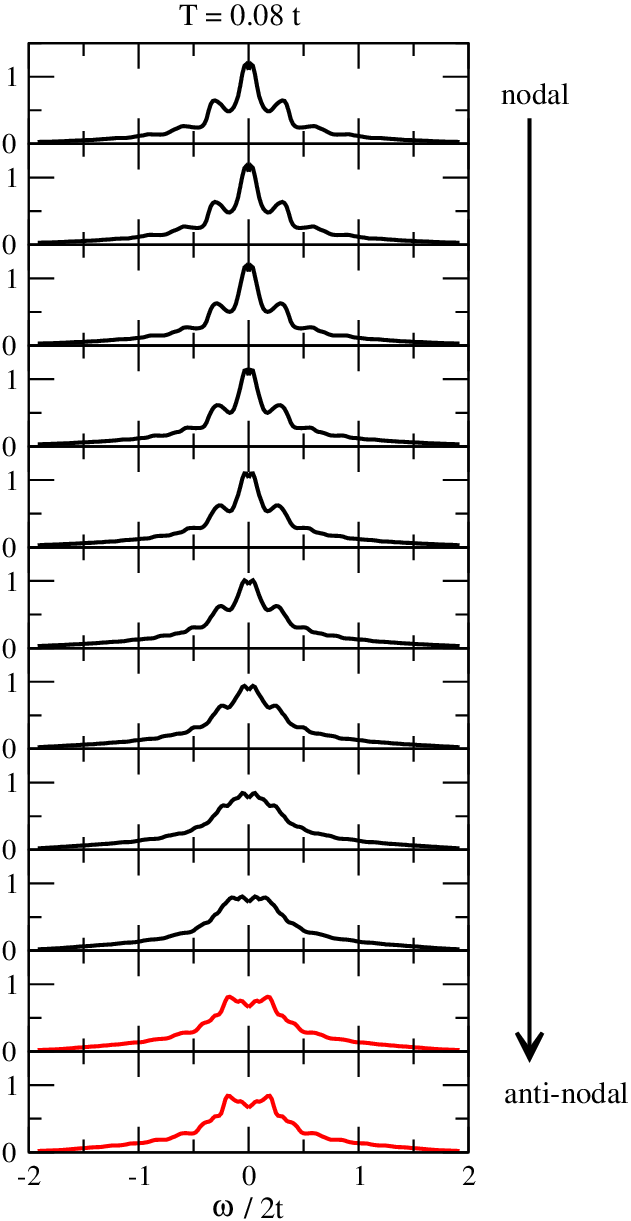}
    }
  \end{center}
  \caption{
   Symmetrized spectral function $\Im G({\bf k},\omega)$  at doping $\delta = 0.04$.
for three temperatures (a) $T=0$, (b) $T= 0.04t$ and (c) $T= 0.08t$ for ${\bf k}$-values
along the FS. The other parameters are $t' = 0.4t$
and $J = 0.2t$. The top $\Im G({\bf k},\omega)$
is at the node, whereas the bottom is at the anti-node. A possible
superconducting solution was suppressed. 
  }
  \label{Fig_5}
  \end{figure}
Fig.~\ref{Fig_5} shows the symmetrized spectral function $\Im G({\bf k},\omega)$ for
  three different temperatures (a) $T=0$, (b) $T= 0.04t$, and (c) $T=
  0.08t$. The hole concentration for all curves is $\delta = 0.04$. Possible superconducting 
  solutions are again suppressed. The results are shown  
for different ${\bf k}$-vectors on
  the Fermi surface between the nodal (top) and the anti-nodal 
  point (below). For all temperatures, a separation of the Fermi surface into two segments 
  is found, as it was already discussed in the foregoing section:
 (i) For ${\bf k}$-vectors around the nodal points, 
   $\Im G({\bf k},\omega)$ shows strong excitations at $\omega = 0$
  (black curves). They form the Fermi arc. (ii) The other segment is given by ${\bf k}$-vectors, 
  for which $\Im G({\bf k},\omega)$ shows a
  pseudogap around $\omega = 0$ (red curves). From
  Figs.~\ref{Fig_5}(a)-(c), one can see that the length of the 
  Fermi arc increases with increasing temperature. 
  This increase is equivalent to a reduction of the pseudogap region. 
  For instance, for the largest temperature $T = 0.08t$, the pseudogap is restriced to a 
  quite small region around the anti-nodal point. 
  Note that this temperature behavior is in good agreement with 
  recent ARPES experiments\cite{K06}. A comparison of the spectral functions
  at the anti-nodal point for three different temperatures (lowest curves  in
  Figs.~\ref{Fig_5}(a)-(c)) shows the influence of $T$ on the pseudogap: 
  With increasing $T$, the pseudogap is filled up
  with additional spectral weight, whereas the magnitude of the gap (i.e.~the 
  distance between the maxima on the $\omega$-axis) remains almost constant. 
  Also this temperature behavior is verified experimentally \cite{K06}. 
  A characteristic temperature $T^*$ can be defined at which the pseudogap is completely 
  filled up, and the Fermi arc extends over the whole Fermi surface. This temperature $T^*$ 
  was already introduced above and is called 
  pseudogap temperature. For the present case, $T^*$ is approximately $T^*\approx 0.1t$.
\begin{figure}
  \begin{center}
    \scalebox{0.55}{
      \includegraphics*{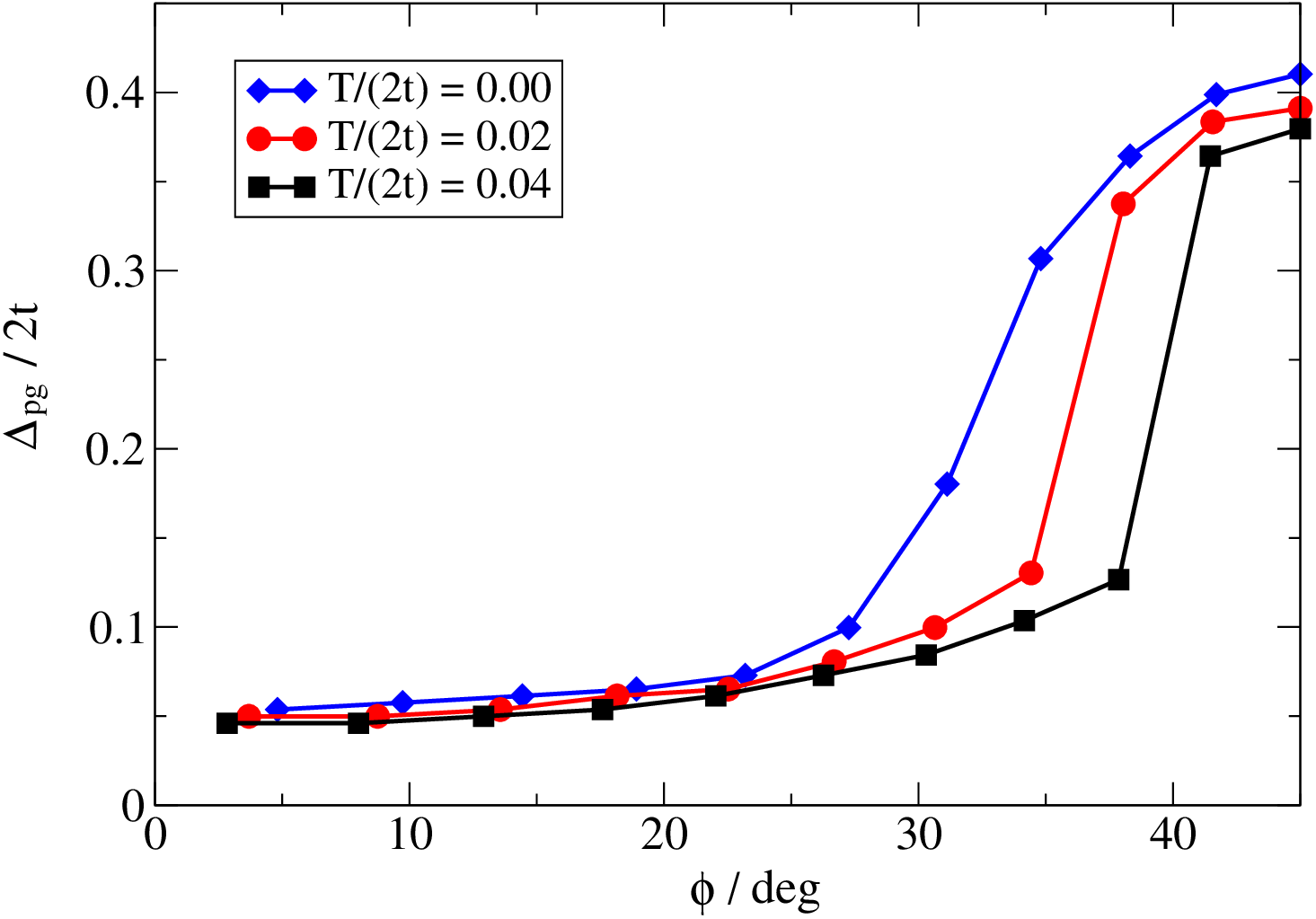}
    }
  \end{center}
  \caption{
Pseudogap  $\Delta_{pg}$ from 
  Fig.~\ref{Fig_5} as a function of the Fermi surface angle $\phi$ for 
  $T = 0$ (blue), $T = 0.04t$ (red), and $T = 0.08t$ (black).
  }
  \label{Fig_6}
  \end{figure}

  The pseudogaps, taken over from Figs.~\ref{Fig_5}(a)-(c),
  are shown in Fig.~\ref{Fig_6} for three different temperatures as function 
  of the Fermi surface angle $\phi$. 
  Note the strong increase of the pseudogap  
  at a finite Fermi angle which depends on the temperature. This particular angle marks the 
  transition
  between the Fermi arc and the pseudogap section. At $T=0$, it 
  is about 25 degrees and moves towards the anti-nodal point for higher temperatures. 
  From Fig.~\ref{Fig_6}, one may also deduce
  that the length of the Fermi arc approximately increases linearly with  $T$. 
   Also this feature is consistent with 
  ARPES experiments\cite{K06}.

\begin{figure}
  \begin{center}
    \scalebox{0.65}{
      \includegraphics*{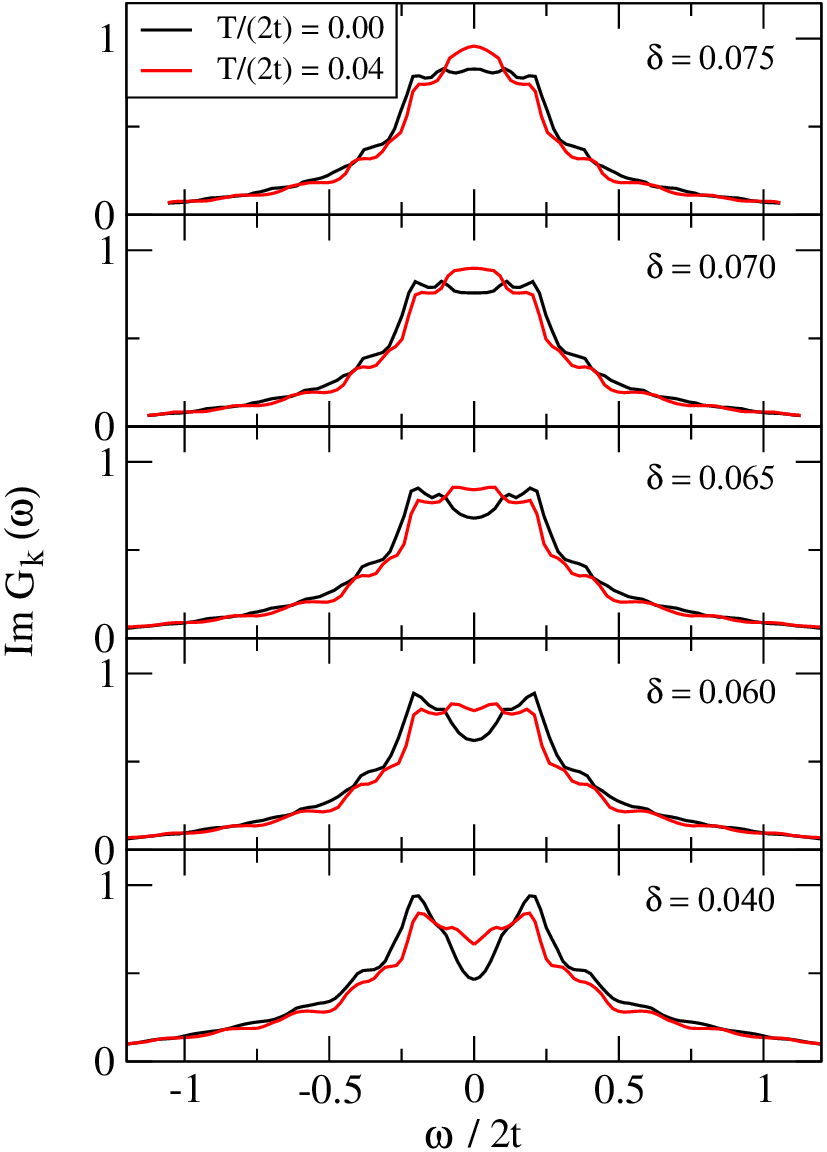}
    }
  \end{center}
  \caption{Symmetrized spectral function $\Im G({\bf k},\omega)$ for  
      fixed ${\bf k}$ value on the anti-nodal point for 
      five different hole concentrations from $\delta = 0.04$ (bottom curves) to $\delta = 0.075$
      (top curves). In each case, the results are shown for two different 
      temperatures $T=0$ (black) and $T = 0.08t$ (red). For the coherent excitations
$\sim |\tilde{u}_{\bf k}|^2$, the same broadening has been taken
for each $\delta$-value.
}
  \label{Fig_7}
  \end{figure}

To discuss the influence of $\delta$ on the 
temperature dependence, in Fig.~\ref{Fig_7}
the symmetrized spectral function $\Im G({\bf k},\omega)$
is shown as function of $\omega$ for two different temperatures  
$T=0$ (black) and $T = 0.08t$ (red) and for five different hole concentrations between $\delta =
0.04$ (bottom) and $\delta = 0.075$ (top). 
The ${\bf k}$-vector is fixed to the anti-nodal point on the FS.  
The curves for $T=0$ (black) show a decrease of the pseudogap   
with increasing hole concentration until it vanishes at 
$\delta \approx 0.075$. For the higher temperature 
$T = 0.08t$ (red), the pseudogap vanishes already at a lower hole concentration of 
$\delta \approx 0.06$. This verifies the experimentally known decrease of the  
pseudogap temperature $T^*$ with increasing hole concentration. 

\begin{figure}
  \begin{center}
    \scalebox{0.55}{
      \includegraphics*{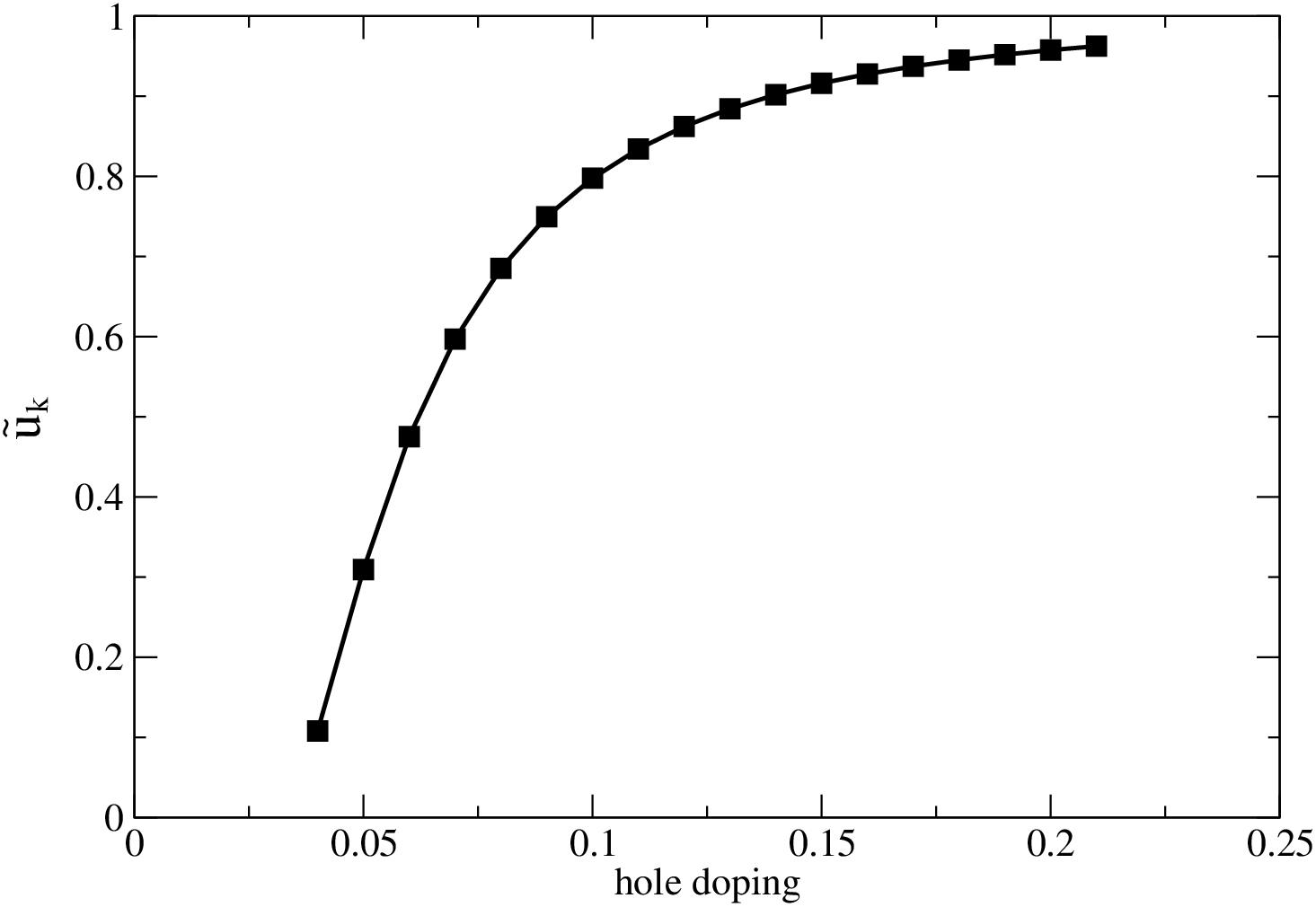}
    }
  \end{center}
  \caption{
The renormalized amplitude $\tilde u_{\bf k}$ of the coherent excitation in
  Eq.~\eqref{63d} is shown as a function of the hole concentration $\delta$. The ${\bf k}$-vector 
 is fixed to $(0,\pi)$. 
  }
  \label{Fig_8}
  \end{figure}


The doping and temperature behavior of  $\Im G({\bf k},\omega)$ 
can be understood on the basis of the former result \eqref{63d} for the spectral 
function.
First, in Fig.~\ref{Fig_8}, the parameter $\tilde{u}_{\bf k}$ is shown as a function of $\delta$
which shows a strong increase with the hole concentration.
According to the first line in Eq.~\eqref{63d}, $\tilde{u}_{\bf k}$ agrees with 
the amplitude of the coherent excitation.  Therefore,  
in Fig.~\ref{Fig_5} for instance,
the weight of the coherent excitation $\sim |\tilde{u}_{\bf k}|^2$ is negligibly small
 for the smallest hole concentration $\delta= 0.04$, 
and the spectrum is dominated by the incoherent part of Eq.~\eqref{63d}. 
In contrast, for sufficiently large $\delta$, a coherent excitation at $\omega=0$ is expected, 
when ${\bf k}$ is fixed to the Fermi surface.
This behavior is for instance realized in Fig.~\ref{Fig_3}. 
Note that an additional 
broadening of the coherent excitation should be included, 
which follows from the scattering of the 
charge carriers at additional phonons or impurities. In Fig.~\ref{Fig_7}, this
broadening was assumed to be
$T$-independent and was set equal to $0.1t$. Therefore, the following
doping behavior can be deduced from Fig.~\ref{Fig_7}: For small hole concentrations $\delta$
($\delta \ll 0.07$), the spectrum at $T=0$ is dominated by the incoherent excitations with a pronounced 
pseudogap around the anti-nodal point. For intermediate hole doping ($\delta \approx 0.07$), 
the spectrum is a superposition of a coherent and of incoherent excitations.  
Both parts are of the same order of magnitude for an intermediate doping.  
The incoherent part has still a pseudogap which is partly compensated by the 
broadening of the coherent excitation. For larger doping $\delta > 0.07$,
the spectrum mainly consists of a coherent excitation around $\omega =0$. With respect to 
temperature, the coherent excitation is almost unaffected by $T$, whereas the pseudogap is filled 
up due to the temperature-dependent shift of the Fermi surface, as will be explained below.


To understand the $T$-behavior of the spectral function, keep in mind that   
$\tilde{u}_{\bf k}$  and therefore the weight of the coherent excitation in $\Im G({\bf k},\omega)$, 
is almost independent of $T$. Moreover, the total spectral weight,
to which coherent and incoherent excitations contribute, is $T$ independent. 
This follows from the sum rule \eqref{61}, since the total electron number is fixed.
Thus, except of minor changes, the overall temperature dependence of $\Im G({\bf k},\omega)$ 
is expected to be weak.  
Instead, the main reason for the $T$-dependence can be traced back to a 
change of the Fermi surface with temperature. 
Consider a ${\bf k}$-vector on the Fermi surface at the anti-nodal point, 
 ${\bf k}_F= (\pi, k_{F}^y)$, where the $x$-component is fixed to $k_F^x=\pi$. By varying the 
temperature, one finds that the magnitude of the $y$-component $k_{F}^y$  
increases almost linearly with $T$. Due to this shift of the Fermi energy with $T$, 
also the positions of the incoherent excitations  at  $\omega=0$ are shifted. 
In this way, one understands that the pseudogap is less pronounced 
for higher temperatures, when $k_F^x$ is fixed to $k_{F}^x=\pi$. 
A similar behavior of the pseudogap was found before in Fig.~\ref{Fig_4}. There, the spectral function
is shown for fixed $k_x=\pi$ and different values of $k_y$, when the temperature is 
fixed. Also in this case, the pseudogap is suppressed for larger values of $k_y$. Finally, note that 
$k_F^y$ also strongly depends on the  nearest-neighbor hopping $t'$. 
For small $t'$, the pseudogap is more pronounced than for larger values of $t'$. 
This can be seen by comparing the spectrum in Fig.~\ref{Fig_1} (with $t'=0.1t$) 
with that  of Fig.~\ref{Fig_5}, where $t'=0.4t$.


\section{Conclusions}
In this paper, we have given a microscopic approach to the
pseudogap phase in cuprate systems at moderate hole doping. Thereby, a recently 
developed projector-based renormalization method (PRM) was applied to the $t$-$J$ model.
The pseudogap, which is found in ARPES experiments, can be traced back to 
incoherent excitations in the one-particle Green function. 
It can neither be explained by a competing order nor as a precursor of superconductivity. 
Instead, the pseudogap phase is an intrinsic property of 
the cuprates close to half-filling. In a subsequent paper\cite{BS09}, we shall show that 
a transition to a superconducting phase occurs in the formalism
either by lowering the temperature or by approaching an appropriate doping range.

\section{Acknowledgements}
We would like to acknowledge stimulating and 
enlightening discussions with J.~Fink and
A.~H\"ubsch. This work was supported by 
the DFG through the research program SFB 463.

 \begin{appendix}

\section{Derivation of the spin susceptibility $\chi({\bf q}, \omega)$}
\label{D}
The derivation of the spin susceptibility $\chi({\bf q},\omega)$ in Eq.~\eqref{22a}
for the system, described by the Hamiltonian ${\cal H}_0 = {\cal H}_t +{\cal H}_J^{(0)}$, 
is based on the Mori-Zwanzig projection formalism. This formalism allows to derive exact 
equations of motion for an appropriately chosen set of relevant operator variables 
$\{A_\alpha\}$,
 \begin{eqnarray}
\label{D1}
\frac{d}{dt}A_\alpha(t)  &=& i \sum_\beta A_\beta(t) \Omega_{\beta \alpha} -
\int_0^t \sum_\beta A_\beta (t-t')\, \Sigma_{\beta \alpha}(t') dt' + F_\alpha(t)\, ,
\end{eqnarray}
where the dynamics of the set $A_\alpha (t)$ should be governed by ${\cal H}_0$, i.e. $A_\alpha(t)=
e^{\frac{i}{\hbar}{\cal H}_0 t} A_\alpha e^{-\frac{i}{\hbar}{\cal H}_0 t}$.
The quantities $i \omega_{\alpha \beta}$, $\Sigma_{\alpha \beta}(t)$, and $F_\alpha(t)$ 
are called frequency matrix, selfenergy, and random force
\begin{align}
\label{D2}
i\Omega_{\alpha \beta} &= \sum_\gamma \chi_{\alpha \gamma}^{-1}
(A_\alpha|\dot A_\beta ), &
\Sigma_{\alpha \beta}(t) &= \sum_\gamma  \chi_{\alpha \gamma}^{-1}\, (\dot A_\gamma | {\sf Q} 
 e^{i{\sf QL}_0{\sf Q}t} {\sf Q} \dot A_\beta ), \\
F_\alpha(t) &= i \, e^{i{\sf QL}_0{\sf Q}t}{\sf Q} {\sf L}_0\, A_\alpha . \nonumber
\end{align}
Here, $\dot A_\alpha$ is the time derivative of $A_\alpha$, defined 
by $\dot A_\alpha= i \,{\sf L}_0 A_\alpha$, and $\chi_{\alpha \beta}^{-1}$ 
is the inverse of the susceptibility matrix 
$\chi_{\alpha \beta} = (A_\alpha|A_\beta)$.
In Eqs.~\eqref{D2}, we have also introduced a scalar product between 
operator quantities $A$ and $B$,
\begin{eqnarray}
\label{D3}
(A|B) &=& \int_0^\beta d\lambda \, \langle A^\dagger e^{- \,\lambda {\sf L}_0} B 
\rangle_0 \, ,
\end{eqnarray}
where the expectation value $\langle \cdots \rangle_0$ is formed with ${\cal H}_0$
and ${\sf L}_0$ is the Liouville operator, which corresponds to ${\cal H}_0$.
In $\Sigma_{\alpha \beta}(t)$ the quantity ${\sf Q}$ is a projection operator 
which projects on the subspace of all operator variables which are 
'perpendicular' to the set $\{ A_\alpha \}$, i.e. 
\begin{eqnarray}
\label{D4}
{\sf Q} &=& {\bf 1} - \sum_{\alpha} |A_\alpha) \chi_{\alpha \beta}^{-1} (A_\beta|.
\end{eqnarray}

To use the general projection formalism to derive $\chi({\bf q},\omega)$, we have to 
choose an appropriate set of relevant operator $\{ A_\alpha\}$.  In our case, this set is given 
by ${\bf S}_{\bf q}$ and its time derivative $\dot{\bf S}_{\bf q}$, i.e.
\begin{eqnarray}
\label{D5}
\{ A_\alpha \} &=& \{\,  {\bf S}_{\bf q}, \dot{ {\bf S}}_{\bf q} \, \}.
\end{eqnarray}
From the equations \eqref{D1}, one easily derives the following two equations:
\begin{eqnarray} 
\label{D6} 
\frac{d}{dt} {\bf S}_{\bf q}(t) &=& \dot {\bf S}_{\bf q}(t), \\ 
\frac{d}{dt}\dot {\bf S}_{\bf q}(t) &=& -\,\omega_{\bf q}^2 \, {\bf S}_{\bf q}(t)
- \int_0^t dt' \, \dot {\bf S}_{\bf q}(t-t') \,\Sigma_{\bf q}(t') + {\bf F}_{\bf q}(t) \, ,  \nonumber
\end{eqnarray}
where the frequency and the selfenergy in the second equation are given by
\begin{eqnarray}
\label{D7}
\omega_{\bf q}^2 &=& \frac{(\dot {\bf S}_{\bf q}|\dot {\bf S}_{\bf q})}
{({\bf S}_{\bf q}|{\bf S}_{\bf q})}, \qquad \qquad
\Sigma_{\bf q}(t) = \frac{1}{(\dot {\bf S}_{\bf q}|\dot {\bf S}_{\bf q})}
( \ddot {\bf S}_{\bf q}| {\sf Q} \, e^{i{\sf QL}_0{\sf Q}t} \, {\sf Q} \ddot {\bf S}_{\bf q} )
\end{eqnarray}
and the random force is 
${\bf F}_{\bf q}(t) =  e^{i{\sf QL}_0{\sf Q}t} {\sf Q} \ddot {\bf S}_{\bf q}$. 
The projector ${\sf Q}$ projects perpendicular to ${\bf S}_{\bf q}$ and $\dot{\bf S}_{\bf q}$.
In deriving the equations \eqref{D6}, we have also used
$
(S^\nu_{\bf q}| \dot{ S}_{\bf q}^\mu) = i \langle [{S_{\bf q}^\nu}^\dagger, 
S_{\bf q}^\mu] \rangle_0 =0
$
(for all $\nu, \mu =x,y,z$), which follows from the exact relation $(A|{\sf L}_0B)= 
\langle [A^\dagger, B]\rangle_0$.
To find the dynamical susceptibility $\chi({\bf q},\omega)$, 
we multiply both equations \eqref{D6} with the 
'bra' $({\bf S}_{\bf q}|$  and go over to the Laplace transform. Using 
$({\bf S}_{\bf q}|{\bf F}_{\bf q})=0$, we obtain 
\begin{eqnarray}
\label{D8}
 {\chi}({\bf q},\omega) &=&
 \frac{-\omega_{\bf q}^2}{\omega^2 - \omega_{\bf q}^2 - \,\omega 
\, \Sigma_{\bf q}(\omega)}  
\, \chi({\bf q}).
\end{eqnarray}
Here, $\chi({\bf q})= (S_{\bf q}|S_{\bf q})$ is the static spin susceptibility and 
$\Sigma_{\bf q}(\omega)$ is the Laplace transformed selfenergy 
 \begin{eqnarray}
\label{D9}
\Sigma_{\bf q}(\omega) &=& 
\frac{1}{(\dot {\bf S}_{\bf q}|\dot {\bf S}_{\bf q})}
( \ddot {\bf S}_{\bf q}| {\sf Q} \, \frac{1}{\omega -{\sf QL}_0{\sf Q} - i\eta} \, {\sf Q} \ddot 
{\bf S}_{\bf q} ) .
\end{eqnarray}


\noindent
To proceed, we have to evaluate the second time derivative   
$\ddot {\bf S}_{\bf q}$
\begin{eqnarray}
\label{D10}
 \ddot{\bf S}_{\bf q} &=& - \frac{1}{\sqrt N}\sum_{i \neq l}t^2_{il}\,
(e^{i{\bf q}{\bf R}_l} -e^{i{\bf q}{\bf R}_i}) \,
( \vec S_l {\cal P}_0(i) - \vec S_i {\cal P}_0(l) ) \nonumber \\
&&
  -\frac{1}{2\sqrt N} \sum_{\alpha \beta}
\sum_{i \neq j} \sum_{j(\neq i \neq l)}\, t_{il}\, t_{lj}\, 
(e^{i{\bf qR}_i} - e^{i{\bf qR}_l})
\\
&& \times  \left\{ \vec \sigma_{\alpha \beta} \left(
 \hat c_{j\alpha}^\dagger \, {\cal D}_\alpha(l)
\,\hat c_{i\beta} + \hat c_{j,-\alpha}^\dagger
S_l^\alpha \hat c_{i\beta} \right)
 + \vec \sigma_{\alpha \beta}^* \left(
\hat c_{i\beta}^\dagger \, 
 {\cal D}_\alpha(l)
\, \hat c_{j\alpha} +
 c_{i\beta}^\dagger S_l^{-\alpha} \hat c_{m,-\alpha} \right)
\right\} \, , \nonumber
\end{eqnarray}
where only the dominant part of the hopping Hamiltonian ${\cal H}_t$ was 
taken into account.
The first term on the right hand side of Eq.~\eqref{D10} enters from a twofold hopping to a 
neighboring site and back. By replacing the two projectors 
${\cal P}_0(i)$ and ${\cal P}_0(l)$ by their 
expectation values, we come back to the former equation of motion \eqref{21}.
Therefore, we can conclude that the frequency term $\omega_{\bf q}^2$, defined in  
Eq.~\eqref{D7}, agrees with the former frequency term $\hat \omega_{\bf q}^2$ from Eq.~\eqref{21},
\begin{eqnarray}
\label{D11}
\omega_{\bf q}^2 &=& \hat\omega_{\bf q}^2 =2P_0(t^2_{{\bf q}=0}- t^2_{{\bf q}}) \geq 0.
\end{eqnarray}
The second contribution in Eq.~\eqref{D10} describes a twofold hopping away 
from the starting site and agrees with the quantity ${\sf Q}\ddot{\bf S}_{\bf q}$ 
in the selfenergy,
\begin{eqnarray}
\label{D12}
{\sf Q}\ddot{\bf S}_{\bf q} &=&
 -\frac{1}{2\sqrt N} \sum_{\alpha \beta}
\sum_{i \neq j} \sum_{j(\neq i \neq l)}\, t_{il}\, t_{lj}\, 
(e^{i{\bf qR}_i} - e^{i{\bf qR}_l}) \times \\ 
&\times& {\sf Q}\,
\left\{ \vec \sigma_{\alpha \beta} \left(
 \hat c_{j\alpha}^\dagger \, {\cal D}_\alpha(l)
\,\hat c_{i\beta} + \hat c_{j,-\alpha}^\dagger
S_l^\alpha \hat c_{i\beta} \right)
 + \vec \sigma_{\alpha \beta}^* \left(
\hat c_{i\beta}^\dagger \, 
 {\cal D}_\alpha(l)
\, \hat c_{j\alpha} +
 c_{i\beta}^\dagger S_l^{-\alpha} \hat c_{m,-\alpha} \right)
\right\} .
\nonumber
\end{eqnarray}
In order to obtain a rough estimate for the selfenergy $\Sigma_{\bf q}(\omega)$,
we neglect the spin flip operators in Eq.~\eqref{D12} and 
replace the local projectors ${\cal D}_{\alpha}(i)$ and ${\cal D}_{\alpha}(l)$ as before
by their expectation value $D$. 
By introducing Fourier transformed quantities, we find
\begin{eqnarray}
\label{D13}
{\sf Q} \ddot {\bf S}_{\bf q} &=& 
\frac{D}{2\sqrt N} \sum_{\alpha \beta } \vec \sigma_{\alpha \beta}
\left( (\varepsilon_{{\bf k} +{\bf q}} - \varepsilon_k)^2 - 2(t^2_{{\bf q}=0}- t^2_{\bf q})
\right) \,{\sf Q}\,  \hat c_{{\bf k}+{\bf q}, \alpha}^\dagger \hat c_{{\bf k} \beta}  .
\end{eqnarray}
The selfenergy then reads
\begin{eqnarray}
\label{D14}
\Sigma_{\bf q}(\omega) &=& \frac{D^2}{(\dot{\vec S}_{\bf q}| \dot{\vec S}_{\bf q})}
\frac{1}{ 4 N} \sum_{{\bf k}{\bf k}'} \sum_{\alpha \beta} \sum_{\alpha' \beta'}
\vec \sigma_{\alpha\beta} \cdot \vec \sigma_{\alpha'\beta'}^* \\
&\times & [ (\varepsilon_{{\bf k}+{\bf q}} -\varepsilon_{\bf k})^2 -2(t^2_{{\bf q}=0} -t^2_{\bf q})] 
[ (\varepsilon_{{\bf k'}+{\bf q}} -\varepsilon_{\bf k'})^2 -2(t^2_{{\bf q}=0}
-t^2_{\bf q})]  \nonumber \\
& \times& (\hat c_{{\bf k}+{\bf q}, \alpha}^\dagger \hat c_{{\bf k}\beta}|
\, \frac{1}{\omega -{\sf Q}{\sf L}_0{\sf Q} - i\eta}\, {\sf Q} \,
\hat c_{{\bf k'}+{\bf q}, \alpha'}^\dagger \hat c_{{\bf k'}\beta'}) \,.
\nonumber
\end{eqnarray}
In the final step, we factorize the two-particle correlation function in Eq.~\eqref{D14}
in a product of one-particle Green functions. A straightforward calculation 
leads for the imaginary part of the selfenergy to
\begin{eqnarray}
\label{D15}
&& \Im \Sigma_{\bf q}(\omega) 
 = \frac{D^2}{(\dot{\vec S}_{\bf q}| \dot{\vec S}_{\bf q})}
\frac{3}{ 2 N} \sum_{\bf k} 
[ (\varepsilon_{{\bf k}+{\bf q}} -\varepsilon_{\bf k})^2 -2(t^2_{{\bf q}=0} -t^2_{\bf q})]^2   \, \, 
\Im M_{\bf k}({\bf q}, \omega), \\
&& \Im M_{\bf k}({\bf q}, \omega) =  
\frac{1-e^{-\beta \omega}}{\beta \omega} \frac{1}{\pi}
\int_{-\infty}^\infty d\tilde\omega \,
\frac{\Im G_{{\bf k}}^{(0)}(\omega +\tilde \omega)}
{1 + \displaystyle e^{-\beta (\omega + \tilde \omega)}}\,
\frac{\Im G_{{\bf k}+{\bf q}}^{(0)}(\tilde \omega)}{1 + 
e^{ \beta \tilde \omega}} \,. \nonumber
\end{eqnarray}
Here, $\Im G_{{\bf k}}^{(0)}(\omega)$ is the imaginary part of the one-particle
Green function, formed with the Hamiltonian ${\cal H}_0$, 
\begin{eqnarray}
\label{D16}
G_{{\bf k}}^{(0)}(\omega) &=& i \int_0^\infty dt\,
\langle [ \hat c_{{\bf k},\alpha}(t) , \hat c_{{\bf k},\alpha}^\dagger]_+ \rangle_0 \, \,
e^{-i(\omega - i\eta)t} .
\end{eqnarray}
Finally, we have to evaluate the denominator $(\dot{\vec S}_{\bf q}| \dot{\vec S}_{\bf q})$ 
of $\Sigma_{\bf q}(\omega)$. Proceeding in analogy to the evaluation of 
$\Sigma_{\bf q}(\omega)$, we find 
\begin{eqnarray}
\label{D17}
(\dot{\bf S}_{\bf q}| \dot{\bf S}_{\bf q}) &=& 
\frac{3}{2N} \sum_{\bf k} 
(\varepsilon_{{\bf k}+{\bf q}} -\varepsilon_{\bf k})^2 \,
(\hat c_{{\bf k}+{\bf q}, \alpha}^\dagger \hat c_{{\bf k}\beta}|
\hat c_{{\bf k}+{\bf q}, \alpha}^\dagger \hat c_{{\bf k}\beta}) 
\end{eqnarray}
with
 \begin{eqnarray}
\label{D18}
&& (\hat c_{{\bf k}+{\bf q}, \alpha}^\dagger \hat c_{{\bf k}\beta}|
\hat c_{{\bf k}+{\bf q}, \alpha}^\dagger \hat c_{{\bf k}\beta}) =
\int_{-\infty}^\infty d\omega \,
\frac{1-e^{-\beta \omega}}{\beta \omega} \frac{1}{\pi^2}
\int_{-\infty}^\infty d\tilde\omega \,
\frac{\Im G_{{\bf k}}^{(0)}(\omega +\tilde \omega)}
{1 + \displaystyle e^{-\beta (\omega + \tilde \omega)}}\,
\frac{\Im G_{{\bf k}+{\bf q}}^{(0)}(\tilde \omega)}{1 + 
e^{ \beta \tilde \omega}} \,. \nonumber \\
&& \nonumber 
\end{eqnarray}

\section{Factorization approximation for  $\dot{\bf S}_{{\bf q},\lambda}
\dot{\bf S}_{-{\bf q},\lambda}$}
\label{A}
The aim of this appendix is to simplify the operator product 
$\dot{\bf S}_{{\bf q},\lambda}\dot{\bf S}_{-{\bf q},\lambda}$  in the expressions for 
${\cal H}_{0,\lambda}$ and ${\cal H}_{1,\lambda}$ from Sec.~\ref{tJmod_RA},
\begin{eqnarray*}
{\cal H}_{0,\lambda} &=& \sum_{\bf q}\frac{J_{\bf q}}{2} \left( {\bf S}_{\bf q}\cdot{\bf S}_{-{\bf q}} + 
\frac{1}{\omega_{{\bf q},\lambda}}
\dot{\bf S}_{{\bf q}, \lambda} \cdot \dot{\bf S}_{-{\bf q},\lambda} \right),
\\
{\cal H}_{1,\lambda} &=& \sum_{\bf q}\frac{J_{\bf q}}{2} 
\left( {\bf S}_{\bf q}\cdot{\bf S}_{-{\bf q}}
 -\frac{1}{\omega_{{\bf q},\lambda}}
\dot{\bf S}_{{\bf q},\lambda}\cdot \dot{\bf S}_{-{\bf q},\lambda} \right).
\end{eqnarray*}
This will be done by use of a factorization approximation. 
Using for the time derivative 
\begin{eqnarray*}
\dot {\bf S}_{{\bf q},\lambda}&=& \frac{i}{2\sqrt N} \sum_{\alpha \beta} \vec \sigma_{\alpha \beta} 
\sum_{i \neq j}t_{{ij},\lambda}(e^{i {\bf q}{\bf R}_i} - e^{i {\bf q}{\bf R}_j})\, \hat c_{i \alpha}^\dagger
\hat c_{j \beta}
\end{eqnarray*}
 we first can rewrite $\dot{\bf S}_{{\bf q},\lambda} \dot{\bf S}_{-{\bf q},\lambda}$ as 
\begin{eqnarray}
\label{A1}
\dot{\bf S}_{{\bf q},\lambda}\dot{\bf S}_{-{\bf q},\lambda}
&=&  \frac{1}{4N}  
\sum_{\alpha \beta} \sum_{\gamma \delta}
(\vec \sigma_{\alpha \beta}\cdot \vec \sigma_{\delta \gamma}) 
\sum_{i \neq j}t_{{ij},\lambda}(e^{i {\bf q}{\bf R}_i} - e^{i {\bf q}{\bf R}_j}) \times \nonumber \\
&& \times\sum_{l \neq m} t_{{lm},\lambda}(e^{-i {\bf q}{\bf R}_l} - e^{-i {\bf q}{\bf R}_m})\,
\hat c_{i\alpha}^\dagger \hat c_{j\beta}\hat c_{m\delta}^\dagger \hat c_{l\gamma}.  
\end{eqnarray}
Using a factorization approximation, the four-fermion operator on the right hand side 
can be reduced to operators  
 $\hat c^\dagger_{{\bf k}\sigma} \hat c_{{\bf k}\sigma}$ which
will lead to a renormalization of $\varepsilon_{\bf k}$.
Thereby, we have to pay attention to the fact that the averaged spin operator 
vanishes ($\langle {\bf S}_i \rangle =0$) outside the antiferromagnetic regime. 
Moreover, all local indices in the four-fermion term of Eq.~\eqref{A1} should 
be different from each other. This follows from the former 
decomposition of the exchange interaction into eigenmodes of ${\sf L}_t$  
in Sec.~\ref{tJmod_RA}, where we have implicitly assumed that the 
operators $\dot {\bf S}_{{\bf q},\lambda}$ and $\dot {\bf S}_{-{\bf q},\lambda}$ 
do not overlap in the local space. Otherwise, the decomposition would be much more involved.
However, it can be shown that these 'interference' terms only make a minor impact on the 
results. For the factorization, we find
\begin{eqnarray}
\label{A2}
\dot{\bf S}_{{\bf q},\lambda}\dot{\bf S}_{-{\bf q},\lambda} &=& 
 \frac{3}{4N} 
\sum_{i \neq j}t_{{ij},\lambda}(e^{i {\bf q}{\bf R}_i} - e^{i {\bf q}{\bf R}_j})
\sum_{l \neq m} t_{{lm},\lambda}(e^{-i {\bf q}{\bf R}_l} - e^{-i {\bf q}{\bf R}_m}) \nonumber \\
 &\times&
\, \left\{
 \sum_{\alpha} \langle (\hat c_{j\beta} \hat c_{m\beta}^\dagger)_{NL} \rangle \,
(\hat c_{i\alpha}^\dagger \hat c_{l\alpha})_{NL}  
+
\sum_\beta  \langle (\hat c_{i\alpha}^\dagger \hat c_{l\alpha})_{NL} \rangle  \,
(\hat c_{j\beta} \hat c_{m\beta})_{NL} 
\right\},
\end{eqnarray}
where we have neglected an additional c-number quantity, which enters in the factorization. 
The attached subscript  in  $( \cdots )_{NL}$ on the right hand side 
indicates that the local sites of the operators inside the brackets are different from each other. 
Note that sums over spin indices in Eq.~\eqref{A1} have already been 
carried out. Fourier transforming Eq.~\eqref{A2} leads to 
\begin{eqnarray}
\label{A3}
 && 
\dot{\bf S}_{{\bf q},\lambda}\dot{\bf S}_{-{\bf q},\lambda} 
= 
 -\frac{3}{2N} \sum_{{\bf k}\sigma}
( \varepsilon_{{\bf k},\lambda} -  \varepsilon_{{\bf k}-{\bf q},\lambda} )^2
\langle (\hat c_{{\bf k}-{\bf q}\alpha}^\dagger \hat c_{{\bf k}-{\bf q}\alpha})_{NL} \rangle \,
(\hat c_{{\bf k}\sigma}^\dagger \hat c_{{\bf k}\sigma})_{NL},
\end{eqnarray}
where we have defined 
\begin{eqnarray*}
 (\hat c_{{\bf k}\sigma}^\dagger \hat c_{{\bf k}\sigma})_{NL}  &=&
\hat c_{{\bf k}\sigma}^\dagger \hat c_{{\bf k}\sigma}  
- \frac{1}{N} \sum_{{\bf k}'}\hat c_{{\bf k}'\sigma}^\dagger \hat c_{{\bf k}'\sigma}  .
\end{eqnarray*}
Using Eq.~\eqref{A3} together with Eq.~\eqref{43}, one is led 
to the renormalization result \eqref{45}   
of $\tilde{\varepsilon}_{\bf k}^{(0)}$ to first order in $J$. 

In the following, let us simplify the notation and suppress
the index $\lambda$ in $\dot {\bf S}_{{\bf q},\lambda}$, $\varepsilon_{{\bf k}, \lambda}$, and also in
$\hat \omega_{{\bf q},\lambda}$. 
With this convention, we shall use the factorization \eqref{A3} in order  
to derive  the renormalization \eqref{35} for  ${\varepsilon}_{{\bf k},\lambda}$  
in second order in $J$. We start from expression \eqref{34} for the renormalized
Hamiltonian ${\cal H}_{\lambda - \Delta \lambda}^{(2)}$ in second order
\begin{eqnarray}
 \label{A6}
{\cal H}_{\lambda - \Delta \lambda}^{(2)} &=&
\sum_{\bf q} J_{{\bf q}}\,\left\{ \Theta(\lambda -|2\hat \omega_{{\bf q}, \lambda}|)
-\frac{1}{2} \right\} \,
 [ \, X_{\lambda, \Delta \lambda},
{\cal A}_{1,\lambda} ({\bf q}) + {\cal A}_{1,\lambda}^\dagger({\bf q}) \,] 
 +\sum_{\bf q} J_{{\bf q}} \,  [X_{\lambda, \Delta \lambda}, {\cal A}_{0,\lambda}] \nonumber \\
&=&
\sum_{\bf q} J_{\bf q} \Theta_{\bf q}(\lambda, \Delta \lambda) \left( \frac{3}{4} [ X_{\lambda, \Delta \lambda}\, ,
{\bf S}_{\bf q} \cdot {\bf S}_{-{\bf q}}] + \frac{1}{4\hat \omega_{\bf q}^2} 
[ X_{\lambda, \Delta \lambda}\, ,\dot {\bf S}_{\bf q} \cdot \dot {\bf S}_{-{\bf q}}]
\right) \, ,
\end{eqnarray}
where in the first line we have already used 
$ [X_{\lambda,\Delta \lambda} \, , {\cal H}_{t,\lambda} ] =
- \sum_{\bf q}J_{\bf q}\, \Theta_{\bf q}(\lambda, \Delta \lambda)  \, 
({\cal A}_{1,\lambda}({\bf q})+ {\cal A}_{1,\lambda}^\dagger ({\bf q}) )$.
Next, we  have to evaluate the commutators of $X_{\lambda, \Delta \lambda}$ with 
${\bf S}_{\bf q}\cdot {\bf S}_{-{\bf q}}$ and 
$\dot {\bf S}_{\bf q}\cdot \dot{\bf S}_{-{\bf q}}$. Using
$[ \dot {S}_{-{\bf q}}^\nu\, , {S}_{{\bf q}}^\nu] = \frac{i}{4N} \sum_{{\bf q}\sigma}
(2\varepsilon_{\bf k}- \varepsilon_{{\bf k}+{\bf q}}-  \varepsilon_{{\bf k}-{\bf q}})\,
\hat c_{{\bf k}\sigma}^\dagger \hat c_{{\bf k}\sigma}$, 
($\nu=x,y,z$),  and Eq.~\eqref{33}, we find  
\begin{eqnarray}
\label{A7}
[X_{\lambda, \Delta \lambda}\, , {\bf S}_{\bf q}\cdot {\bf S}_{-{\bf q}}] &=&
\frac{J_{\bf q}}{4 \hat \omega_{\bf q}^2}\, \Theta_{\bf q}(\lambda, \Delta \lambda)
\left( \frac{1}{N} \sum_{{\bf k}\sigma}(2\varepsilon_{\bf k} -
\varepsilon_{{\bf k}+{\bf q}} -\varepsilon_{{\bf k}-{\bf q}}) 
\langle \hat c_{{\bf k}\sigma}^\dagger  \hat c_{{\bf k}\sigma} \rangle 
\right)\,  {\bf S}_{\bf q}\cdot {\bf S}_{-{\bf q}} \nonumber  \\
&+& 
\frac{J_{\bf q}}{4 \hat \omega_{\bf q}^2}\, \Theta_{\bf q}(\lambda, \Delta \lambda)
 \langle {\bf S}_{\bf q}\cdot {\bf S}_{-{\bf q}} \rangle \,
 \frac{1}{N} \sum_{{\bf k}\sigma}(2\varepsilon_{\bf k} -
\varepsilon_{{\bf k}+{\bf q}} -\varepsilon_{{\bf k}-{\bf q}}) \;
 \hat c_{{\bf k}\sigma}^\dagger  \hat c_{{\bf k}\sigma}, \nonumber 
\end{eqnarray}
\begin{eqnarray}
 [ X_{\lambda , \Delta \lambda}\, , \dot {\bf S}_{\bf q} \cdot \dot {\bf S}_{-{\bf q}} ]  &=&
 - \frac{J_{\bf q}}{4 \hat \omega_{\bf q}^2}\, \Theta_{\bf q}(\lambda, \Delta \lambda)
\left( \frac{1}{N} \sum_{{\bf k}\sigma}(2\varepsilon_{\bf k} -
\varepsilon_{{\bf k}+{\bf q}} -\varepsilon_{{\bf k}-{\bf q}}) 
\langle \hat c_{{\bf k}\sigma}^\dagger  \hat c_{{\bf k}\sigma} \rangle 
\right)\,  \dot {\bf S}_{\bf q}\cdot \dot {\bf S}_{-{\bf q}} \nonumber  \\
&-& 
\frac{J_{\bf q}}{4 \hat \omega_{\bf q}^2}\, \Theta_{\bf q}(\lambda, \Delta \lambda)
 \langle \dot {\bf S}_{\bf q}\cdot \dot {\bf S}_{-{\bf q}} \rangle \,
 \frac{1}{N} \sum_{{\bf k}\sigma}(2\varepsilon_{\bf k} -
\varepsilon_{{\bf k}+{\bf q}} -\varepsilon_{{\bf k}-{\bf q}}) \;
 \hat c_{{\bf k}\sigma}^\dagger  \hat c_{{\bf k}\sigma} .
\end{eqnarray}
Note that in \eqref{A7} already a factorization approximation was used. 
With the relations \eqref{A6} and \eqref{A7}, we obtain
\begin{eqnarray}
\label{A8}
{\cal H}_{\lambda - \Delta \lambda}^{(2)} &=& 3\sum_{\bf q}
(\frac{J_{\bf q}}{4 \hat \omega_{\bf q}^2})^2\, \Theta_{\bf q}(\lambda, \Delta \lambda)
\left( 
\left[
\frac{1}{N} \sum_{{\bf k}\sigma}(2\varepsilon_{\bf k} -
\varepsilon_{{\bf k}+{\bf q}} -\varepsilon_{{\bf k}-{\bf q}}) 
\langle \hat c_{{\bf k}\sigma}^\dagger  \hat c_{{\bf k}\sigma} \rangle 
\right]\,  {\bf S}_{\bf q}\cdot {\bf S}_{-{\bf q}} \right. \nonumber  \\
&& \left. +
 \langle {\bf S}_{\bf q}\cdot {\bf S}_{-{\bf q}} \rangle \,
 \frac{1}{N} \sum_{{\bf k}\sigma}(2\varepsilon_{\bf k} -
\varepsilon_{{\bf k}+{\bf q}} -\varepsilon_{{\bf k}-{\bf q}}) \;
 \hat c_{{\bf k}\sigma}^\dagger  \hat c_{{\bf k}\sigma} \right)  \\
&& \nonumber \\
&-& \sum_{\bf q}
 (\frac{J_{\bf q}}{4 \hat \omega_{\bf q}^2})^2\, \Theta_{\bf q}(\lambda, \Delta \lambda)
\left( \left[ \frac{1}{N} \sum_{{\bf k}\sigma}(2\varepsilon_{\bf k} -
\varepsilon_{{\bf k}+{\bf q}} -\varepsilon_{{\bf k}-{\bf q}}) 
\langle \hat c_{{\bf k}\sigma}^\dagger  \hat c_{{\bf k}\sigma} \rangle 
\right]\,  \dot {\bf S}_{\bf q}\cdot \dot {\bf S}_{-{\bf q}}\right.  \nonumber  \\
&& \left. + 
 \langle \dot {\bf S}_{\bf q}\cdot \dot {\bf S}_{-{\bf q}} \rangle \,
 \frac{1}{N} \sum_{{\bf k}\sigma}(2\varepsilon_{\bf k} -
\varepsilon_{{\bf k}+{\bf q}} -\varepsilon_{{\bf k}-{\bf q}}) \;
 \hat c_{{\bf k}\sigma}^\dagger  \hat c_{{\bf k}\sigma} \right).
 \nonumber  
\end{eqnarray}
In a final step, we factorize $\sim \dot {\bf S}_{\bf q}\cdot 
\dot {\bf S}_{-{\bf q}}$ according to \eqref{A3},
 \begin{eqnarray}
\label{A9}
 {\cal H}_{\lambda - \Delta \lambda}^{(2)} &=& 
3\sum_{\bf q}
(\frac{J_{\bf q}}{4 \hat \omega_{\bf q}^2})^2\, \Theta_{\bf q}(\lambda, \Delta \lambda)
\left( 
\left[
\frac{1}{N} \sum_{{\bf k}\sigma}(2\varepsilon_{\bf k} -
\varepsilon_{{\bf k}+{\bf q}} -\varepsilon_{{\bf k}-{\bf q}}) 
\langle \hat c_{{\bf k}\sigma}^\dagger  \hat c_{{\bf k}\sigma} \rangle 
\right]\,  {\bf S}_{\bf q}\cdot {\bf S}_{-{\bf q}} \right. \nonumber  \\
&& \left. \qquad +
 \langle {\bf S}_{\bf q}\cdot {\bf S}_{-{\bf q}} \rangle \,
 \frac{1}{N} \sum_{{\bf k}\sigma}(2\varepsilon_{\bf k} -
\varepsilon_{{\bf k}+{\bf q}} -\varepsilon_{{\bf k}-{\bf q}}) \;
 \hat c_{{\bf k}\sigma}^\dagger  \hat c_{{\bf k}\sigma} \right)
 \nonumber  \\
&-& \sum_{\bf q}
 (\frac{J_{\bf q}}{4 \hat \omega_{\bf q}^2})^2\, \Theta_{\bf q}(\lambda, \Delta \lambda)
 \langle \dot {\bf S}_{\bf q}\cdot \dot {\bf S}_{-{\bf q}} \rangle \,
 \frac{1}{N} \sum_{{\bf k}\sigma}(2\varepsilon_{\bf k} -
\varepsilon_{{\bf k}+{\bf q}} -\varepsilon_{{\bf k}-{\bf q}}) \;
 \hat c_{{\bf k}\sigma}^\dagger  \hat c_{{\bf k}\sigma} 
 \nonumber  \\
&+&
\frac{3}{2N} \sum_{{\bf q}\sigma} (\frac{J_{\bf q}}{4 \hat \omega_{\bf q}^2})^2 \,
\Theta_{\bf q}(\lambda,\Delta \lambda)\,
\left[ \frac{1}{N} \sum_{{\bf k}'\sigma'}(2\varepsilon_{{\bf k}'} -
\varepsilon_{{\bf k}'+{\bf q}} -\varepsilon_{{\bf k}'-{\bf q}}) 
\langle \hat c_{{\bf k}'\sigma'}^\dagger  \hat c_{{\bf k}'\sigma'} \rangle 
\right] \times \nonumber \\
&& \qquad \times \sum_{{\bf k}\sigma}   (\varepsilon_{\bf k}- \varepsilon_{{\bf k}- {\bf q} })^2
\langle (\hat c_{{\bf k}-{\bf q}\alpha}^\dagger \hat c_{{\bf k}-{\bf q}\alpha})_{NL} \rangle
\, (\hat c_{{\bf k}\sigma}^\dagger \hat c_{{\bf k}\sigma} )_{NL} .
\end{eqnarray}
From \eqref{A9}, the renormalization equatiuon \eqref{35} for
$\varepsilon_{{\bf k},\lambda - \Delta \lambda}$ can immediately be deduced.


\section{Renormalization equations  for fermion operators}
\label{B} 
The aim of this appendix is to derive the renormalization equation for the fermion operator 
$\hat c_{{\bf k}\sigma}(\lambda)=
e^{X_\lambda} \hat c_{{\bf k}\sigma} e^{-X_{\lambda}}$ in second order in $J_{\bf q}$. 
As before, we shall suppress the index $\lambda$ everywhere 
in $\dot {\bf S}_{{\bf q},\lambda}$, $\hat \omega_{{\bf q},\lambda}$, 
and  $\varepsilon_{{\bf q},\lambda}$ in order to simplify the notation.
Let us start from an {\it ansatz} for 
$\hat c_{{\bf k}\sigma}(\lambda)$ after all excitations with transition energies larger 
than $\lambda$ have been integrated out. It reads
\begin{eqnarray}
\label{B1}
 \hat c_{{\bf k} \sigma}(\lambda) &=& u_{{\bf k}, \lambda} \hat c_{{\bf k}\sigma} 
-i \sum_{\bf q} 
 \Theta(|2 \hat \omega_{\bf q}|- \lambda)\,  v_{{\bf k,q},\lambda} \,
\frac{J_{\bf q}}{4 \hat \omega^2_{\bf q}} \,
[ {\bf S_{q}} \cdot \dot {\bf S}_{-{\bf q}} + \dot {\bf S}_{-{\bf q}} \cdot {\bf S}_{{\bf q}}
, \, c_{{\bf k} \sigma}]  .
\end{eqnarray}
In Eq.~\eqref{B1}, the parameters $u_{{\bf k},\lambda}$
and $v_{{\bf k, q},\lambda}$ account for the $\lambda$-dependence. 
Note that the operator structure in Eq.~\eqref{B1} corresponds to that 
of the first order expansion for 
$\hat c_{{\bf k} \sigma}(\lambda) \approx \hat c_{{\bf k}\sigma}+ [X_{\lambda}, \hat c_{{\bf k}\sigma}]$.
Here, $X_\lambda$ has the same operator form as the generator $X_{\lambda,\Delta \lambda}$ in  
Eq.~\eqref{32}. Due to construction, the ${\bf q}$-sum in Eq.~\eqref{B1} only runs over 
$\bf q$-values with excitation energies $|2 \hat \omega_{\bf q}|$ larger than $\lambda$.
This is assured by the $\Theta$-function in Eq.~\eqref{B1}. 
For simplicity, in the following  we agree upon to incorporate the $\Theta$-function in  $v_{{\bf k,q},\lambda} $. 
Thus, we can write
\begin{eqnarray}
\label{B2}
  \hat c_{{\bf k} \sigma}(\lambda) &=& u_{{\bf k}, \lambda} \hat c_{{\bf k}\sigma} 
-i \sum_{\bf q} 
v_{{\bf k,q},\lambda} \frac{J_{\bf q}}{4 \hat \omega^2_{\bf q}} \,
\Big(\big( [ {\bf S_{q}}, c_{{\bf k} \sigma}] \cdot \dot {\bf S}_{-{\bf q}} +
\dot {\bf S}_{-{\bf q}} \cdot [ {\bf S}_{{\bf q}} , c_{{\bf k} \sigma}]  
\big) \nonumber \\
&& \hspace*{5cm}
+ \big(
{\bf S_{q}} \cdot [\dot {\bf S}_{-{\bf q}},   c_{{\bf k} \sigma}]  +
[\dot {\bf S}_{-{\bf q}},   c_{{\bf k} \sigma}] \cdot {\bf S_{q}} \,
\big) \Big).
\end{eqnarray}
For the additional renormalization from $\lambda$ to the reduced cutoff $\lambda - \Delta \lambda$, we have 
\begin{eqnarray}
\label{B3}
 && \hat c_{\bf k \sigma}(\lambda - \Delta \lambda) = e^{X_{\lambda, \Delta \lambda}}\, \hat c_{{\bf k} \sigma}(\lambda)\, 
e^{-X_{\lambda, \Delta \lambda}} =\\ 
&& \qquad = u_{{\bf k}, \lambda} 
 e^{X_{\lambda, \Delta \lambda}}\hat c_{{\bf k}\sigma} e^{-X_{\lambda, \Delta \lambda}} 
 -i \sum_{\bf q} 
v_{{\bf k,q},\lambda} \frac{J_{\bf q}}{4 \hat \omega^2_{\bf q}} \,
e^{X_{\lambda, \Delta \lambda}}
[ {\bf S_{q}} \cdot \dot {\bf S}_{-{\bf q}} + \dot {\bf S}_{-{\bf q}} \cdot {\bf S}_{{\bf q}}
, \, c_{{\bf k} \sigma}]  e^{-X_{\lambda, \Delta \lambda}}, \nonumber   
\end{eqnarray}
where $X_{\lambda, \Delta \lambda}$ is the generator from Eq.~\eqref{32},
\begin{eqnarray*}
 X_{\lambda, \Delta \lambda} &=& -i
\sum_{\bf q}\, \frac{J_{\bf q}}{4 \hat \omega_{\bf q}} 
\Theta_{\bf q}(\lambda, \Delta \lambda) 
\left( \bf S_{\bf q}\, \dot {\bf S}_{-{\bf q}} + \dot {\bf S}_{\bf q}\, {\bf S}_{-{\bf q}}
\right).
\end{eqnarray*}
First, let us expand the term $\sim u_{{\bf k},\lambda}$ in Eq.~\eqref{B3},
\begin{eqnarray}
 \label{B4}
e^{X_{\lambda, \Delta \lambda}}\, \hat c_{{\bf k} \sigma}
\, e^{-X_{\lambda, \Delta \lambda}} &=& \hat c_{{\bf k} \sigma} + [ X_{\lambda, \Delta \lambda}\, , \,
\hat c_{{\bf k} \sigma}]+
\frac{1}{2}[X_{\lambda, \Delta \lambda}, [X_{\lambda, \Delta \lambda},\hat c_{{\bf k} \sigma} ]]
+ \cdots.
\end{eqnarray}
Here, we can combine the second term in Eq.~\eqref{B3} with the second part in Eq.~\eqref{B2}, 
\begin{eqnarray}
\label{B5}
\hat c_{\bf k \sigma}(\lambda - \Delta \lambda) &=&  
( u_{{\bf k}, \lambda} + \cdots)\, \hat c_{{\bf k}\sigma}  \\
&-& i \sum_{\bf q} 
( v_{{\bf k,q},\lambda} + u_{{\bf k}, \lambda}\, \Theta_{\bf q}(\lambda, \Delta \lambda) +\cdots ) \,  
 \frac{J_{\bf q}}{4 \hat \omega^2_{\bf q}} \,
[ {\bf S_{q}} \cdot \dot {\bf S}_{-{\bf q}} + \dot {\bf S}_{-{\bf q}} \cdot {\bf S}_{{\bf q}}
, \, c_{{\bf k} \sigma}]  + \cdots , \nonumber   
\end{eqnarray}
where the dots $( +\cdots)$ mean additional contributions from higher order commutators with  $X_{\lambda, \Delta \lambda}$.
On the other hand, $\hat c_{\bf k \sigma}(\lambda - \Delta \lambda)$ should have the same form as the ansatz \eqref{B1},
when $\lambda$ is replaced by $\lambda - \Delta \lambda$,
\begin{eqnarray}
\label{B6}
  \hat c_{\bf k \sigma}(\lambda - \Delta \lambda) &=& 
u_{{\bf k}, \lambda- \Delta \lambda} \hat c_{{\bf k}\sigma} 
-i \sum_{\bf q} 
v_{{\bf k,q},\lambda - \Delta \lambda} \frac{J_{\bf q}}{4 \hat \omega^2_{\bf q}} \,
[ {\bf S_{q}} \cdot \dot {\bf S}_{-{\bf q}} + \dot {\bf S}_{-{\bf q}} \cdot {\bf S}_{{\bf q}}
, \, c_{{\bf k} \sigma}]  .
\end{eqnarray}
The comparison of Eqs.~\eqref{B6} and \eqref{B5} immediately leads to the renormalization equation \eqref{52a} 
for $v_{{\bf k, q},\lambda}$,
\begin{eqnarray}
\label{B7}
v_{{\bf k,q},\lambda - \Delta \lambda} &=&  v_{{\bf k,q},\lambda} + u_{{\bf k}, \lambda}\, \Theta_{\bf q}(\lambda, \Delta \lambda) ,
\end{eqnarray}
where we have restricted ourselves to the lowest order contributions in
$X_{\lambda, \Delta \lambda}$. Furthermore, we have exploited the very weak
$\lambda$-dependency of $\varepsilon_{{\bf k},\lambda}$ and $\hat \omega_{{\bf q},\lambda}$.

The renormalization equation for the second parameter $u_{{\bf k},\lambda}$  requires 
the evaluation of higher order commutators in Eq.~\eqref{B3}.  Alternatively, we can start
from the anti-commutator relation \eqref{3}
\begin{eqnarray*}
 [\hat c_{{\bf k}\sigma}^\dagger(\lambda), \hat c_{{\bf k}\sigma}(\lambda)]_+ = 
\frac{1}{N}\sum_i  e^{X_\lambda} {\cal D}_\sigma (i) e^{-X_{\lambda}} 
=\frac{1}{N}\sum_i {\cal D}_\sigma (i)
\end{eqnarray*}
with $ {\cal D}_\sigma (i)= 1- n_{1,-\sigma}$, where in the last relation $[X_\lambda, \sum_i 
{\cal D}_\sigma(i)] = 0$ was used. 
When we take the average, we obtain 
\begin{eqnarray}
 \langle [\hat c_{{\bf k}\sigma}^\dagger(\lambda), \hat c_{{\bf k}\sigma}(\lambda)]_+ \rangle &=& 
 \langle {\cal D}_\sigma (i) \rangle =: D.
\label{B8}
\end{eqnarray}
In order to evaluate the anti-commutator in Eq.~\eqref{B8}, we have to insert  the former ansatz \eqref{B2}
for $\hat c_{{\bf k}\sigma}(\lambda)$. Here, we make an additional approximation by taking 
into account only the two first terms in Eq.~\eqref{B2}. The remaining terms have explicit 
spin operators ${\bf S}_{\bf q}$. In the commutator of Eq.~\eqref{B8}, they lead to additional contributions with  
one or two spin operators. Outside the antiferromagnetic phase, no magnetic order is present 
and also spin correlations are weak. Therefore, it seems reasonable to neglect these terms. 
Thus,  we can approximate $ \hat c_{{\bf k}\sigma}(\lambda)$ by 
\begin{eqnarray}
\label{B9}
 \hat c_{{\bf k} \sigma}(\lambda) &=& u_{{\bf k}, \lambda} \hat c_{{\bf k}\sigma} 
-i \sum_{\bf q} 
v_{{\bf k,q},\lambda} \frac{J_{\bf q}}{4 \hat \omega^2_{\bf q}} \,
\left([ {\bf S_{q}},  c_{{\bf k} \sigma}]   \cdot \dot {\bf S}_{-{\bf q}}
+ \dot {\bf S}_{-{\bf q}}\cdot [ {\bf S_{q}},  c_{{\bf k} \sigma}] \right) \\
& =& 
u_{{\bf k}, \lambda} \hat c_{{\bf k}\sigma} 
+ \frac{1}{2N} \sum_{\bf q} 
v_{{\bf k,q},\lambda} \frac{J_{\bf q}}{4 \hat \omega^2_{\bf q}} \,
\sum_{\alpha \beta \gamma} (\vec \sigma_{\alpha \beta}
\cdot \vec \sigma_{\sigma \gamma}) 
  \sum_{{\bf k}'}  (\varepsilon_{{\bf k}'}- \varepsilon_{{\bf k}' + {\bf q}}) \, 
 \hat c^\dagger_ {{\bf k}' + {\bf q} \alpha} \ \hat c_ {{\bf k}' \beta} \
\hat c_ {{\bf k} + {\bf q} \gamma}. \nonumber 
\end{eqnarray}
Inserting Eq.~\eqref{B9} and $\hat c_{{\bf k}\sigma}^\dagger(\lambda)$ into Eq.~\eqref{B8}, we obtain
\begin{eqnarray}
 \label{B10}
D &=& |u_{{\bf k},\lambda}|^2 D  +
\frac{1}{(2N)^2}
 \sum_{{\bf q}'{\bf q}} v_{{\bf k,q}',\lambda}^*\, v_{{\bf k,q},\lambda} 
\frac{J_{{\bf q}'}}{4 \hat \omega^2_{{\bf q}'}} 
\frac{J_{\bf q}}{4 \hat \omega^2_{\bf q}} \,
\sum_{\alpha', \beta', \gamma'}
\sum_{\alpha, \beta, \gamma}
(\vec \sigma_{\beta' \alpha'}\cdot \vec \sigma_{\gamma' \sigma})
(\vec \sigma_{\alpha \beta}\cdot \vec \sigma_{\sigma \gamma}) \times \nonumber \\
&&  \times \sum_{{\bf k}',{\bf k}''}
(\varepsilon_{{\bf k}''}- \varepsilon_{{\bf k}'' + {\bf q}'}) \, 
(\varepsilon_{{\bf k}'}- \varepsilon_{{\bf k}' + {\bf q}}) \, 
\langle \, [ \,
\hat c_ {{\bf k} + {\bf q}' \gamma'}^\dagger 
\hat c_ {{\bf k}'' \beta'}^\dagger \
\hat c_ {{\bf k}'' + {\bf q}' \alpha'} \, , \,
\hat c^\dagger_ {{\bf k}' + {\bf q} \alpha} \ \hat c_ {{\bf k}' \beta} \
\hat c_ {{\bf k} + {\bf q} \gamma}  ]_+ \,
\rangle. \nonumber \\
&&
\end{eqnarray}
To find the renormalization equation for $u_{{\bf k},\lambda - \Delta \lambda}$,  we use 
the same equation, thereby replacing $\lambda$ by $\lambda - \Delta \lambda$. We then obtain
 \begin{eqnarray}
\label{B11}
D &=& |u_{{\bf k},\lambda- \Delta \lambda}|^2 D  
+ \frac{1}{(2N)^2}
 \sum_{{\bf q}'{\bf q}} 
(v_{{\bf k,q}',\lambda}^*\, + u_{{\bf k},\lambda}^* \Theta_{{\bf q}'}(\lambda, \Delta \lambda) ) \ 
( v_{{\bf k,q},\lambda} +  u_{{\bf k},\lambda} \Theta_{\bf q}(\lambda, \Delta \lambda) ) \, 
\frac{J_{{\bf q}'}}{4 \hat \omega^2_{{\bf q}'}} 
\frac{J_{\bf q}}{4 \hat \omega^2_{\bf q}}  \nonumber \\
&& \times \sum_{\alpha', \beta', \gamma'} \sum_{\alpha, \beta, \gamma}
(\vec \sigma_{\beta' \alpha'}\cdot \vec \sigma_{\gamma' \sigma})
(\vec \sigma_{\alpha \beta}\cdot \vec \sigma_{\sigma \gamma})  \\
&&  \times \sum_{{\bf k}',{\bf k}''}
(\varepsilon_{{\bf k}''}- \varepsilon_{{\bf k}'' + {\bf q}'}) \, 
(\varepsilon_{{\bf k}'}- \varepsilon_{{\bf k}' + {\bf q}}) \, 
\langle \, [ \,
\hat c_ {{\bf k} + {\bf q}' \gamma'}^\dagger 
\hat c_ {{\bf k}'' \beta'}^\dagger \
\hat c_ {{\bf k}'' + {\bf q}' \alpha'} \, , \,
\hat c^\dagger_ {{\bf k}' + {\bf q} \alpha} \ \hat c_ {{\bf k}' \beta} \
\hat c_ {{\bf k} + {\bf q} \gamma}  ]_+ \, 
\rangle \nonumber \, ,
\end{eqnarray}
where we have inserted the former renormalization result \eqref{B7} for 
$v_{{\bf k,q},\lambda- \Delta \lambda}$. Restricting ourselves to the lowest 
order contributions in $J_{\bf q}$, we can subtract Eq.~\eqref{B10} from Eq.~\eqref{B11} and obtain 
the renormalization equation which connects $u_{{\bf k},\lambda - \Delta \lambda}$ 
with $u_{{\bf k},\lambda}$, 
\begin{eqnarray}
\label{B12}
  |u_{{\bf k},\lambda- \Delta \lambda}|^2 D  &=& |u_{{\bf k},\lambda}|^2 D  
- \frac{1}{(2N)^2}  \sum_{{\bf q}'{\bf q}} 
\frac{J_{{\bf q}'}}{4 \hat \omega^2_{{\bf q}'}} 
\frac{J_{\bf q}}{4 \hat \omega^2_{\bf q}}  
\sum_{\alpha', \beta', \gamma'} \sum_{\alpha, \beta, \gamma}
(\vec \sigma_{\beta' \alpha'}\cdot \vec \sigma_{\gamma' \sigma})
(\vec \sigma_{\alpha \beta}\cdot \vec \sigma_{\sigma \gamma})  \\
&\times&
\left\{ |u_{{\bf k},\lambda}|^2 
\Theta_{{\bf q}'}(\lambda, \Delta \lambda)
\Theta_{\bf q}(\lambda, \Delta \lambda) \right. \nonumber \\
&&+ \left. ( u_{{\bf k},\lambda}\, v_{{\bf k,q}',\lambda}^*\, \Theta_{\bf q}(\lambda, \Delta \lambda)
+ u_{{\bf k},\lambda}^*\, v_{{\bf k,q},\lambda}\, \Theta_{{\bf q}'}(\lambda, \Delta \lambda))
\right\} \nonumber \\
& \times& \sum_{{\bf k}',{\bf k}''}
(\varepsilon_{{\bf k}''}- \varepsilon_{{\bf k}'' + {\bf q}'}) \, 
(\varepsilon_{{\bf k}'}- \varepsilon_{{\bf k}' + {\bf q}}) \, 
\langle \, [ \,
\hat c_ {{\bf k} + {\bf q}' \gamma'}^\dagger 
\hat c_ {{\bf k}'' \beta'}^\dagger \
\hat c_ {{\bf k}'' + {\bf q}' \alpha'} \, , \,
\hat c^\dagger_ {{\bf k}' + {\bf q} \alpha} \ \hat c_ {{\bf k}' \beta} \
\hat c_ {{\bf k} + {\bf q} \gamma}  ]_+ \,
\rangle. \nonumber
\end{eqnarray}
What remains is to evaluate the commutator in Eq.~\eqref{B12}. In a final 
factorization approximation, we find
\begin{eqnarray}
\label{B13}
|u_{{\bf k},\lambda- \Delta \lambda}|^2   &=& |u_{{\bf k},\lambda}|^2   
  -
\frac{1}{(2N)^2}  \sum_{{\bf q}} 
(\frac{J_{{\bf q}}}{4 \hat \omega^2_{{\bf q}}})^2 
\sum_{\alpha, \beta, \gamma}
|\vec \sigma_{\alpha \beta}\cdot \vec \sigma_{\sigma \gamma}|^2  \nonumber \\
&&\times
\Theta_{{\bf q}}(\lambda, \Delta \lambda)
\left\{ |u_{{\bf k},\lambda}|^2 
+ ( u_{{\bf k},\lambda}\, v_{{\bf k,q},\lambda}^*\, 
+ u_{{\bf k},\lambda}^*\, v_{{\bf k,q},\lambda}\, )
\right\} \nonumber \\
&&\times \sum_{{\bf k}'}  
(\varepsilon_{{\bf k}'}- \varepsilon_{{\bf k}' + {\bf q}})^2 \, \left\{ \,
n_{{\bf k}+{\bf q}} ( n_{{\bf k}'} +D ) + n_{{\bf k}'+{\bf q}} ( m_{{\bf k}'} - 
n_{{\bf k}+{\bf q}})  \, \right\} \nonumber \\
&+& \nonumber
\frac{1}{(2N)^2}  \sum_{{\bf q}'{\bf q}} 
\frac{J_{{\bf q}'}}{4 \hat \omega^2_{{\bf q}'}} 
\frac{J_{\bf q}}{4 \hat \omega^2_{\bf q}}  
\sum_{\alpha, \beta, \gamma}
(\vec \sigma_{\gamma \alpha}\cdot \vec \sigma_{\beta \sigma})
(\vec \sigma_{\alpha \beta}\cdot \vec \sigma_{\sigma \gamma})  \nonumber \\
&&\times
\big\{ |u_{{\bf k},\lambda}|^2 
\Theta_{{\bf q}'}(\lambda, \Delta \lambda)
\Theta_{\bf q}(\lambda, \Delta \lambda) \nonumber \\
&&+ ( u_{{\bf k},\lambda}\, v_{{\bf k,q}',\lambda}^*\, \Theta_{\bf q}(\lambda, \Delta \lambda)
+ u_{{\bf k},\lambda}^*\, v_{{\bf k,q},\lambda}\, \Theta_{{\bf q}'}(\lambda, \Delta \lambda))
\big\} \nonumber \\
&& \times
(\varepsilon_{{\bf k}+{\bf q}}- \varepsilon_{{\bf k} +{\bf q}+ {\bf q}'}) \, 
(\varepsilon_{{\bf k}+{\bf q}'}- \varepsilon_{{\bf k}+ {\bf q}' + {\bf q}}) \, 
\big\{
n_{{\bf k}+{\bf q}'}\, ( n_{{\bf k}+{\bf q}} + D) \nonumber \\
&&+
n_{{\bf k}+{\bf q}+ {\bf q}'} (m_{{\bf k}+{\bf q}} - n_{{\bf k}+{\bf q}'})
\big\}. 
\end{eqnarray}
Summing over the spin indices and exploiting that $u_{{\bf k},\lambda}$ and  $v_{{\bf k},{\bf q},\lambda}$
are real, we arrive at expression \eqref{52}.
\end{appendix}


\end{document}